\documentclass[11pt,a4paper]{article}
\pdfoutput=1
\usepackage{jheppub}
\usepackage{appendix}
\usepackage{amssymb,amsmath,amstext,amsfonts,empheq}
\usepackage{bbold}
\usepackage{hyperref}
\counterwithout{equation}{section} 
\usepackage{comment}
\usepackage{mathrsfs}
\usepackage{tikz}

\numberwithin{equation}{section}

\newcommand{\cubicdiag}{%
\begin{tikzpicture}[baseline=-0.1cm,scale=0.55]
  \draw (-.7,.65)--(0,0);
  \draw (-.7,-.65)--(0,0);
  \draw (0,0)--(.8,0);
  \fill (0,0) circle (1.7pt);
\end{tikzpicture}}

\newcommand{\taddiag}{%
\begin{tikzpicture}[baseline=-0.1cm,scale=0.55]
  \draw (0,0) circle (.62);
  \draw (.62,0)--(1.55,0);
  \fill (.62,0) circle (1.7pt);
\end{tikzpicture}}

\newcommand{\tadLdiag}{%
\begin{tikzpicture}[baseline=-0.1cm,scale=0.55]
  \draw (0,0) circle (.62);
  \draw (.62,0)--(1.55,0);
  \fill (.62,0) circle (1.7pt);
  \draw[double,double distance=1pt] (-1.65,0)-- node[midway,above] {$\zeta_L$} (-.62,0);
\end{tikzpicture}}

\newcommand{\tadLLdiag}{%
\begin{tikzpicture}[baseline=-0.1cm,scale=0.55]
  \draw (0,0) circle (.62);
  \draw (.62,0)--(1.55,0);
  \fill (.62,0) circle (1.7pt);
  \draw[double,double distance=1pt] (-1.4,.65)-- node[midway,above] {$\zeta_L$} (-.62,0);
  \draw[double,double distance=1pt] (-1.4,-.65)-- node[midway,below] {$\zeta_L$} (-.62,0);
\end{tikzpicture}}

\newcommand{\quarticBoundaryScale}{0.55}
\tikzset{
  qbd/base/.style={baseline=-0.1cm,scale=\quarticBoundaryScale,
    line width=0.4pt,line cap=round,line join=round},
  qbd/long/.style={double,double distance=1pt},
  qbd/boundary/.style={line width=0.4pt}
}
\newcommand{\quarticdiag}{%
\begin{tikzpicture}[baseline=-0.1cm,scale=0.55]
  \draw (-.7,.65)--(0,0);
  \draw (-.7,-.65)--(0,0);
  \draw (0,0)--(.7,-.65);
  \draw (0,0)--(.7,.65);
  \fill (0,0) circle (1.7pt);
\end{tikzpicture}}

\newcommand{\tadquartic}{%
\begin{tikzpicture}[baseline=-0.1cm,scale=0.55]
  \draw (0,0) circle (.62);
  \draw (0.62,0)--(1.32,-.65);
  \draw (0.62,0)--(1.32,.65);
  \fill (.62,0) circle (1.7pt);
\end{tikzpicture}}

\newcommand{\quarticonelong}{%
\begin{tikzpicture}[baseline=-0.1cm,scale=0.55]
  \draw (0,0) circle (.62);
  \draw (0.62,0)--(1.32,-.65);
  \draw (0.62,0)--(1.32,.65);
    \draw[double,double distance=1pt] (-1.65,0)-- node[midway,above] {$\zeta_L$} (-.62,0);
  \fill (.62,0) circle (1.7pt);
\end{tikzpicture}}
\newcommand{\quartictwolong}{%
\begin{tikzpicture}[baseline=-0.1cm,scale=0.55]
  \draw (0,0) circle (.62);
  \draw (0.62,0)--(1.32,-.65);
  \draw (0.62,0)--(1.32,.65);
  \draw[double,double distance=1pt] (-1.4,.65)-- node[midway,above] {$\zeta_L$} (-.62,0);
  \draw[double,double distance=1pt] (-1.4,-.65)-- node[midway,below] {$\zeta_L$} (-.62,0);
  \fill (.62,0) circle (1.7pt);
\end{tikzpicture}}
\newcommand{\qbdShortSkeleton}{%
  \draw (0,-.62) circle (.62);
  \draw (0,0)--(-.60,1.12);
  \draw (0,0)--(.60,1.12);
}

\newcommand{\quarticPSboundary}[1][]{%
\begin{tikzpicture}[qbd/base,#1]
  \qbdShortSkeleton
  \draw[qbd/boundary] (-.95,1.12)--(.95,1.12);
  \fill (0,0) circle (1.7pt);
\end{tikzpicture}}
\newcommand{\cubicPSboundary}[1][]{%
\begin{tikzpicture}[qbd/base,#1]
  \ccbShortSkeleton
  \ccbWideBoundarya
  \ccbVertices
\end{tikzpicture}}
\newcommand{\quarticBSvertexboundary}[1][]{%
\begin{tikzpicture}[qbd/base,#1]
  \qbdShortSkeleton
  \draw[qbd/long] (0,0)
    .. controls (1.05,.02) and (1.62,.22) .. (1.62,1.12);
  \draw[qbd/boundary] (-.95,1.12)--(1.92,1.12);
  \fill (0,0) circle (1.7pt);
\end{tikzpicture}}

\newcommand{\quarticBSloopboundary}[1][]{%
\begin{tikzpicture}[qbd/base,#1]
  \qbdShortSkeleton
  \draw[qbd/long] (.62,-.62)
    .. controls (1.65,-.62) and (1.62,.15) .. (1.62,1.12);
  \draw[qbd/boundary] (-.95,1.12)--(1.92,1.12);
  \fill (0,0) circle (1.7pt);
  \fill (.62,-.62) circle (1.7pt);
\end{tikzpicture}}

\newcommand{\quarticBShardboundary}[1][]{%
\begin{tikzpicture}[qbd/base,#1]
  \qbdShortSkeleton
  \draw[qbd/long] (.30,.56)
    .. controls (1.05,.36) and (1.62,.42) .. (1.62,1.12);
  \draw[qbd/boundary] (-.95,1.12)--(1.92,1.12);
  \fill (0,0) circle (1.7pt);
  \fill (.30,.56) circle (1.7pt);
\end{tikzpicture}}

\newcommand{\quarticBlobRadius}{5pt}

\newcommand{\qbdBackgroundMark}[1]{%
  \filldraw[fill=black!35,draw=black!65,line width=.35pt]
    #1 circle (\quarticBlobRadius);
}

\newcommand{\qbdRegulatorMark}[1]{%
  \begin{scope}[shift={#1}]
    \filldraw[fill=red!18,draw=red!75!black,line width=.55pt]
      (0,0) circle (\quarticBlobRadius);
    \draw[red!75!black,line width=.65pt]
      (-3pt,-3pt)--(3pt,3pt)
      (-3pt,3pt)--(3pt,-3pt);
  \end{scope}
}

\newcommand{\quarticPSconditionedboundary}[1][]{%
\begin{tikzpicture}[qbd/base,#1]
  \qbdShortSkeleton
  \draw[qbd/boundary] (-.95,1.12)--(.95,1.12);
  \qbdBackgroundMark{(-.30,.56)}
  \qbdBackgroundMark{(.30,.56)}
  \qbdBackgroundMark{(0,0)}
  \qbdBackgroundMark{(0,-1.24)}
\end{tikzpicture}}

\newcommand{\quarticPSvertexbackground}[1][]{%
\begin{tikzpicture}[qbd/base,#1]
  \qbdShortSkeleton
  \draw[qbd/long] (0,0)--(1.30,0);
  \node[above,font=\scriptsize,inner sep=1pt]
    at (.95,0) {$\zeta_L$};
  \draw[qbd/boundary] (-.95,1.12)--(.95,1.12);
  \fill (0,0) circle (1.7pt);
\end{tikzpicture}}

\newcommand{\quarticPSloopbackground}[1][]{%
\begin{tikzpicture}[qbd/base,#1]
  \qbdShortSkeleton
  \draw[qbd/long] (0,-1.24)
    .. controls (.30,-1.50) and (.90,-1.42)
    .. (1.30,-1.42);
  \node[above,font=\scriptsize,inner sep=1pt]
    at (1.0,-1.42) {$\zeta_L$};
  \draw[qbd/boundary] (-.95,1.12)--(.95,1.12);
  \fill (0,0) circle (1.7pt);
  \fill (0,-1.24) circle (1.7pt);
\end{tikzpicture}}

\newcommand{\quarticPShardbackground}[1][]{%
\begin{tikzpicture}[qbd/base,#1]
  \qbdShortSkeleton
  \draw[qbd/long] (.30,.56)--(1.30,.56);
  \node[above,font=\scriptsize,inner sep=1pt]
    at (.95,.56) {$\zeta_L$};
  \draw[qbd/boundary] (-.95,1.12)--(.95,1.12);
  \fill (0,0) circle (1.7pt);
  \fill (.30,.56) circle (1.7pt);
\end{tikzpicture}}

\newcommand{\quarticPScutoffboundary}[1][]{%
\begin{tikzpicture}[qbd/base,#1]
  \qbdShortSkeleton
  \draw[qbd/boundary] (-.95,1.12)--(.95,1.12);
  \fill (0,0) circle (1.7pt);
  \qbdRegulatorMark{(0,-1.24)}
\end{tikzpicture}}

\newcommand{\quarticBScutoffboundary}[1][]{%
\begin{tikzpicture}[qbd/base,#1]
  \qbdShortSkeleton
  \draw[qbd/long] (0,-1.24)
    .. controls (1.50,-1.44) and (1.62,-.10)
    .. (1.62,1.12);
  \draw[qbd/boundary] (-.95,1.12)--(1.92,1.12);
  \fill (0,0) circle (1.7pt);
  \qbdRegulatorMark{(0,-1.24)}
\end{tikzpicture}}

\tikzset{ccb/short/.style={draw=black,line width=.6pt}}

\newcommand{\ccbShortSkeleton}{%
  \draw[ccb/short] (0,-.51919355) circle (.72080645);
  \draw[ccb/short] (-.50,0)--(-.80,1.12);
  \draw[ccb/short] ( .50,0)--( .80,1.12);
}
\newcommand{\ccbVertices}{%
  \fill (-.50,0) circle (1.7pt);
  \fill ( .50,0) circle (1.7pt);
}
\newcommand{\ccbWideBoundary}{%
  \draw[qbd/boundary] (-.95,1.12)--(1.92,1.12);
}
\newcommand{\ccbWideBoundarya}{%
  \draw[qbd/boundary] (-.95,1.12)--(.95,1.12);
}

\newcommand{\cubicBloopboundary}[1][]{%
\begin{tikzpicture}[qbd/base,#1]
  \ccbShortSkeleton
  \draw[qbd/long] (.72080645,-.51919355)
    .. controls (1.40,-.60) and (1.62,.30) .. (1.62,1.12);
  \ccbWideBoundary
  \ccbVertices
  \fill (.72080645,-.51919355) circle (1.7pt);
\end{tikzpicture}}

\newcommand{\cubicBvertexboundary}[1][]{%
\begin{tikzpicture}[qbd/base,#1]
  \ccbShortSkeleton
  \draw[qbd/long] (.50,0)
    .. controls (1.15,.05) and (1.62,.42) .. (1.62,1.12);
  \ccbWideBoundary
  \ccbVertices
\end{tikzpicture}}

\newcommand{\cubicBhardboundary}[1][]{%
\begin{tikzpicture}[qbd/base,#1]
  \ccbShortSkeleton
  \draw[qbd/long] (.65,.56)
    .. controls (1.15,.36) and (1.62,.42) .. (1.62,1.12);
  \ccbWideBoundary
  \ccbVertices
  \fill (.65,.56) circle (1.7pt);
\end{tikzpicture}}

\newcommand{\cubicPSconditionedboundary}[1][]{%
\begin{tikzpicture}[qbd/base,#1]
  \ccbShortSkeleton
  \draw[qbd/boundary] (-.95,1.12)--(.95,1.12);
  \qbdBackgroundMark{(-.65,.56)}
  \qbdBackgroundMark{( .65,.56)}
  \qbdBackgroundMark{(-.50,0)}
  \qbdBackgroundMark{( .50,0)}
  \qbdBackgroundMark{(0,-1.24)}
\end{tikzpicture}}

\newcommand{\cubicBcutoffboundary}[1][]{%
\begin{tikzpicture}[qbd/base,#1]
  \ccbShortSkeleton
  \draw[qbd/long] (.72080645,-.51919355)
    .. controls (1.40,-.60) and (1.62,.30) .. (1.62,1.12);
  \ccbWideBoundary
  \ccbVertices
  \qbdRegulatorMark{(.72080645,-.51919355)}
\end{tikzpicture}}

\newcommand{\cubicBcutoffboundaryPower}[1][]{%
\begin{tikzpicture}[qbd/base,#1]
  \ccbShortSkeleton
  \ccbWideBoundary
  \ccbVertices
  \qbdRegulatorMark{(.72080645,-.51919355)}
\end{tikzpicture}}

\newcommand{\cubicPScutoffboundary}[1][]{%
\begin{tikzpicture}[qbd/base,#1]
  \ccbShortSkeleton
  \ccbWideBoundary
  \ccbVertices
  \qbdRegulatorMark{(.72080645,-.51919355)}
\end{tikzpicture}}
\newcommand{\ccSoftHardSkeleton}{%
  \draw[line width=.6pt] (0,0)
    .. controls (-.50,.02) and (-.65,.50) .. (-.65,1.12);
  \draw[line width=.6pt] (0,0)
    .. controls ( .50,.02) and ( .65,.50) .. ( .65,1.12);
  \draw[qbd/boundary] (-.95,1.12)--(1.92,1.12);
  \fill (0,0) circle (1.7pt);
}

\newcommand{\ccSoftBubbleSkeleton}{%
  \ccSoftHardSkeleton
  \draw[line width=.6pt] (.10,-.89) circle (.35);
  \draw[qbd/long] (0,0)
    .. controls (-.12,-.24) and (-.38,-.55) .. (-.25,-.89);
  \draw[qbd/long] (.45,-.89)
    .. controls (1.20,-.82) and (1.62,.25) .. (1.62,1.12);
}

\newcommand{\cubicBsoftboundary}[1][]{%
\begin{tikzpicture}[qbd/base,#1]
  \ccSoftBubbleSkeleton
  \fill (-.25,-.89) circle (1.7pt);
  \fill ( .45,-.89) circle (1.7pt);
\end{tikzpicture}}

\newcommand{\seagullBsoftboundary}[1][]{%
\begin{tikzpicture}[qbd/base,#1]
  \ccSoftHardSkeleton
  \draw[line width=.6pt] (0,-.89) circle (.35);
  \draw[qbd/long] (0,0)--(0,-.54);
  \draw[qbd/long] (0,-.54)
    .. controls (.85,-.65) and (1.62,.25) .. (1.62,1.12);
  \fill (0,-.54) circle (1.7pt);
\end{tikzpicture}}

\newcommand{\countertermBsoftboundary}[1][]{%
\begin{tikzpicture}[qbd/base,#1]
  \ccSoftHardSkeleton
  \draw[qbd/long] (0,0)--(0,-.70);
  \draw[qbd/long] (0,-.70)
    .. controls (.85,-.80) and (1.62,.25) .. (1.62,1.12);
  \begin{scope}[shift={(0,-.70)}]
    \draw[black,line width=.8pt]
      (-3pt,-3pt)--(3pt,3pt)
      (-3pt,3pt)--(3pt,-3pt);
  \end{scope}
\end{tikzpicture}}

\newcommand{\cutoffBsoftboundary}[1][]{%
\begin{tikzpicture}[qbd/base,#1]
  \ccSoftBubbleSkeleton
  \fill (-.25,-.89) circle (1.7pt);
  \qbdRegulatorMark{(.45,-.89)}
\end{tikzpicture}}

\tikzset{
  icseq/base/.style={
    x=0.62cm,y=0.62cm,baseline=-0.82cm,
    line width=0.35pt,line cap=round,line join=round,
    every node/.style={font=\scriptsize,inner sep=1pt}
  },
  icseq/soft/.style={double,double distance=1.1pt,line width=0.30pt},
  icseq/blob/.style={draw=black!55,fill=black!35,line width=0.25pt},
  icseq/cross/.style={draw=red,line width=0.45pt}
}

\newcommand{\icseqTadpoleSkeleton}{%
  \draw (0,-.51919355) circle[radius=.72080645];
  \draw (0,.20161290)--(0,1.12);
  \fill (0,.20161290) circle[radius=.025];
  \node[above,font=\scriptsize] at (0,1.14) {$k_1$};
}

\newcommand{\icseqHardFromCutoff}{%
  \draw (.72080645,-.51919355)--(1.18,1.12);
  \node[above,font=\scriptsize] at (1.18,1.14) {$k_2$};
}

\newcommand{\icseqSoftFromLoop}{%
  \draw[qbd/long] (-.72080645,-.51919355)--(-1.45,1.12);
  \node[above,font=\scriptsize] at (-1.45,1.14) {$p$};
}

\newcommand{\icseqSoftFromCutoff}{%
  \draw[qbd/long] (.72080645,-.51919355)--(2.02,1.12);
  \node[above,font=\scriptsize] at (2.02,1.14) {$p$};
}

\newcommand{\icseqMarkers}{%
  \qbdBackgroundMark{(-.72080645,-.51919355)}
  \qbdRegulatorMark{(.72080645,-.51919355)}
}

\newcommand{\icTadpoleConditioned}[1][]{%
\begin{tikzpicture}[qbd/base,#1]
  \icseqTadpoleSkeleton
  \draw[qbd/boundary] (-.95,1.12)--(.95,1.12);
  \icseqMarkers
\end{tikzpicture}}

\newcommand{\icTadpoleHardInsertion}[1][]{%
\begin{tikzpicture}[qbd/base,#1]
  \icseqTadpoleSkeleton
  \icseqHardFromCutoff
  \draw[qbd/boundary] (-.95,1.12)--(1.42,1.12);
  \icseqMarkers
\end{tikzpicture}}

\newcommand{\icTadpoleSoftFromLoop}[1][]{%
\begin{tikzpicture}[qbd/base,#1]
  \icseqTadpoleSkeleton
  \icseqHardFromCutoff
  \icseqSoftFromLoop
  \draw[qbd/boundary] (-1.65,1.12)--(1.42,1.12);
  \icseqMarkers
\end{tikzpicture}}

\newcommand{\icTadpoleSoftFromCutoff}[1][]{%
\begin{tikzpicture}[qbd/base,#1]
  \icseqTadpoleSkeleton
  \icseqHardFromCutoff
  \icseqSoftFromCutoff
  \draw[qbd/boundary] (-.95,1.12)--(2.22,1.12);
  \icseqMarkers
\end{tikzpicture}}

\usepackage{bm}
\usepackage{graphicx}
\usepackage{xcolor}

\usepackage{overpic}
\usepackage{subcaption}
\usepackage{braket}

\usepackage[extra]{tipa}

\newcommand{\be}{\begin{equation}}
\newcommand{\ee}{\end{equation}}
\newcommand{\bea}{\begin{eqnarray}}
\newcommand{\eea}{\end{eqnarray}}
\newcommand{\cutoff}{\Lambda}

\newcommand{\bal}{\begin{aligned}}
\newcommand{\eal}{\end{aligned}}

\newcommand{\Lt}{{\mathcal{L}^{(3)}}}

\newcommand{\Hamt}{{\mathcal{H}^{(3)}}}

\newcommand{\tz}{\tau_f}
\newcommand{\zetal}{\zeta_{L}}
\newcommand{\short}{\zeta_{S}}

\newcommand{\Mp}{M_{\rm Pl}}
\newcommand{\R}{{\zeta}}
\newcommand{\Rh}{{\zeta}}

\newcommand{\bk}{\mathbf{k}}
\newcommand{\bx}{\mathbf{x}}
\newcommand{\by}{\mathbf{y}}
\newcommand{\bK}{\mathbf{K}}
\newcommand{\bp}{\mathbf{p}}
\newcommand{\bq}{\mathbf{q}}
\definecolor{alizarin}{rgb}{0.82, 0.1, 0.26}

\newcommand{\llangleb}{\Big\langle }%\mkern-8mu \Big\langle}
\newcommand{\rrangleb}{\Big\rangle }%\mkern-8mu\Big\rangle}

\newcommand{\uv}{\text{UV}}

\newcommand{\inh}{\text{inh}}
\newcommand{\ind}{\text{ind}}

\newcommand{\expval}[1]{{\langle #1 \rangle}}

\preprint{CERN-TH-2026-303}

\title{\centering Inhomogeneous Cutoff Regularization of Loops \\[1mm]  in Cosmological Correlators}
\author[a,b,c]{Jacopo Fumagalli} \author[d]{and Keisuke Inomata} 
\affiliation[a]{Scuola Normale Superiore, Piazza dei Cavalieri 7, Pisa, 56126, Italy}

\affiliation[b]{INFN Sezione di Pisa, Largo Pontecorvo 3, 56127 Pisa, Italy}

\affiliation[c]{Departament de F\'isica Quàntica i Astrofísica and 
Institut de Ciències del Cosmos (ICC), \\
Universitat de Barcelona, Martí i Franquès 1, 08028 Barcelona, Spain}
\emailAdd{jacopo.fumagalli@sns.it}

\affiliation[d]{Theoretical Physics Department, CERN, 1211 Geneva 23, Switzerland}
\emailAdd{keisuke.inomata@cern.ch}
\abstract{We revisit the use of cutoff regularization in computing loop corrections to cosmological correlators. We argue that a cutoff corresponding to a fixed physical scale must be defined with respect to the local background geometry. 
In the presence of superhorizon perturbations, this is naturally implemented within the separate universe picture, where different local patches are regulated at the same physical scale. Expanding this cutoff in the long-wavelength background generates a series of regulator-induced contributions to loop amplitudes, which admit a diagrammatic interpretation as long-mode insertions on the corresponding loop contractions. Our cutoff prescription preserves residual spatial dilatations and volume-preserving anisotropic rescalings. It is therefore particularly relevant for loop correlators with soft external modes. We illustrate the mechanism by computing the contributions to the one-loop scalar and tensor power spectra on superhorizon scales. In both cases, the additional corrections induced by the inhomogeneous cutoff cancel the ultraviolet boundary terms arising in the ordinary diagrams. 
These calculations establish, within cutoff regularization, the late-time constancy of both power spectra and clarify its physical origin. We then investigate the squeezed one-loop scalar bispectrum and show that the additional cutoff contributions are essential for satisfying the consistency relation at finite cutoff. Finally, we discuss the role of the inhomogeneous cutoff in the late-time constancy of the one-loop bispectrum.
}
\begin{document}
\maketitle
%\newpage

\section{Introduction}
The interplay between ultraviolet (UV) regularization and symmetry has long been central to perturbative quantum field theory. Dimensional regularization owes much of its success in gauge theories to its compatibility with gauge invariance \cite{tHooft:1972tcz}. Explicit momentum cutoffs, meanwhile, provide a direct separation of momentum scales, underlying the Wilsonian description of how short-distance fluctuations contribute to effective interactions \cite{Wilson:1973jj,Polchinski:1983gv}. Their implementation requires care, since a boundary imposed on momentum integrals need not respect the symmetries of the theory. In a gravitational setting, this raises a concrete question: how should a fixed physical UV scale be implemented in an expanding and perturbed spacetime while respecting the relevant symmetries?

Loop corrections to primordial correlation functions provide a test of the
consistency and perturbative reliability of inflationary predictions, as
well as a concrete setting for quantum field theory in curved
spacetime~\cite{Weinberg:2005vy,Senatore:2009cf}. 
Recent years have seen a resurgence of interest in inflationary loop
corrections, with substantial progress on several fronts. Research on the
de Sitter wavefunction has explored how unitarity and symmetry constrain
its renormalization, with implications for primordial
correlators~\cite{Lee:2023jby,Jain:2025maa,Farren:2026hao},
while other work has developed analytic methods for cosmological loop
calculations~\cite{Benincasa:2024ptf,Pimentel:2026kqc,Henn:2026lfz}
and computed cosmological collider signatures from loops of massive
particles~\cite{Qin:2023bjk,Bodas:2025vpb,Qin:2026yuo}.
In parallel, attention has focused on perturbative control in scenarios
relevant to primordial black hole
formation, e.g.~\cite{Inomata:2022yte,Kristiano:2022maq,Riotto:2023hoz,Kristiano:2023scm,
Firouzjahi:2023aum,Motohashi:2023syh,Fumagalli:2023loc,Maity:2023qzw,Davies:2023hhn,Ballesteros:2024zdp,Frolovsky:2025qre},
alongside the related question of whether the one-loop curvature power spectrum
freezes on superhorizon scales~\cite{Fumagalli:2023zzl,Tada:2023rgp,Inomata:2024lud,
Kawaguchi:2024rsv,Fumagalli:2024jzz,Inomata:2025bqw,Inomata:2025pqa,Braglia:2025qrb}.
Other developments include the renormalization of primordial correlators
and the calculation of their finite loop
corrections~\cite{Ballesteros:2024cef,Sheikhahmadi:2024peu,Ballesteros:2025nhz,Braglia:2025cee,Fang:2026off,Ballesteros:2026fwo},
separate-universe analyses of long-wavelength
dynamics~\cite{Firouzjahi:2023ahg,Iacconi:2023ggt,Iacconi:2026uzo,Tanaka:2026zew},
the organization of loop contributions in different gauges and field
variables~\cite{Inomata:2026csq,Artigas:2026vfi,Fang:2026qhu},
and the treatment of short-mode backreaction~\cite{Fang:2025vhi,Braglia:2026fzp}.

In single-clock attractor
inflation, diffeomorphism invariance underlies cosmological consistency
relations~\cite{Maldacena:2002vr,Creminelli:2004yq,Hinterbichler:2012nm,Assassi:2012zq,Goldberger:2013rsa,Hinterbichler:2013dpa}
and plays a central role in the conservation of the adiabatic curvature
perturbation on superhorizon scales, including at loop
order. That was shown with different methods in Refs.~\cite{Pimentel:2012tw,Assassi:2012et,Senatore:2012ya}, and more recently in Refs.~\cite{Kawaguchi:2024rsv,Fumagalli:2024jzz,Ema:2026dop,Artigas:2026vfi}. These results motivate a UV cutoff prescription that respects the relevant spatial symmetries.

Following Weinberg's pioneering study of quantum corrections to cosmological correlators \cite{Weinberg:2005vy}, subsequent work emphasized the importance of distinguishing physical UV scales from comoving ones when assessing the time dependence of inflationary loop corrections \cite{Senatore:2009cf}. Keeping the UV scale fixed in physical units avoids spurious time dependence associated with an unintended evolution of the regulator's physical scale. In comoving momentum space, this is implemented by the replacement $\Lambda\longrightarrow a(t)\Lambda_{\rm UV}$, where $\Lambda_{\rm UV}$ is the fixed physical UV scale. We refer to this spatially uniform and time-dependent prescription as the \textit{homogeneous cutoff}.

In this work, we extend the cutoff prescription to the perturbed geometry, where fixing a physical UV scale also requires the comoving cutoff to depend on the local cosmological perturbations. At any given time during inflation, scalar and tensor perturbations with wavelengths much larger than the Hubble radius can be regarded, at leading order in gradients, as part of the local background geometry. This is the separate universe picture~\cite{Sasaki:1995aw,Sasaki:1998ug,Wands:2000dp}. A constant long-wavelength adiabatic scalar mode can be absorbed locally into a rescaling of spatial coordinates \cite{Weinberg:2003sw}. A long-wavelength tensor background similarly produces an anisotropic spatial rescaling \cite{Hinterbichler:2012nm,Hinterbichler:2013dpa}. Our central proposal is to regulate loop integrals by imposing the same physical UV scale in the local geometry of each patch. In this way, the UV regulator does not violate the separate universe picture. This defines the \textit{inhomogeneous cutoff} prescription 
\begin{equation}
\Lambda\,\longrightarrow\,a(t)\Lambda_{\rm UV}e^{\zeta_L}\left[\hat{k}_i\hat{k}_j(e^{-\gamma_L})^{ij}\right]^{-1/2}.
\end{equation}
Here $\zeta_L$ and $\gamma_L$ denote the long-wavelength scalar and tensor perturbations,  
where $\hat{\mathbf{k}}$ is the direction of the regulated comoving momentum.

The regulated loop integrals consequently acquire an additional dependence on the long-wavelength background through their integration boundary. Expanding this dependence and correlating it with external long-wavelength perturbations generates regulator-induced contributions to correlators with additional soft external legs. Diagrammatically, an $n$-point loop contribution can thus induce contributions to higher-point correlators through long-mode insertions on its loop contractions.
For instance, as we show explicitly at one loop, the new cutoff prescription induces an additional contribution to the one-loop power spectrum through the expansion of the regulated tadpole source in the long-wavelength background.
In practice, this mechanism follows the logic of the background-wave method for computing the leading soft limits of correlators with one or more soft external modes \cite{Creminelli:2004yq,Joyce:2014aqa}: a lower-point correlator is evaluated at fixed long-wavelength background, expanded in that background, and correlated with the external soft modes. In this language, our prescription includes the background dependence of the UV integration boundary, whose expansion generates the regulator-induced contributions.

An inhomogeneous cutoff, depending on the local inflaton background, was previously
considered in one-loop calculations of the curvature power spectrum
in spatially-flat gauge~\cite{Inomata:2025bqw,Inomata:2026csq}. In Ref.~\cite{Inomata:2025bqw}, the dependence
of the infrared boundary generates an additional backreaction
contribution that cancels a residual IR boundary term.
Ref.~\cite{Inomata:2026csq} applies analogous reasoning to the smoothing
boundary entering the $\delta N$ relation between inflaton fluctuations
and curvature perturbations. Both prescriptions are motivated
by consistency with the separate-universe picture. 
More broadly, the distinction between a fixed physical cutoff and a fixed cutoff defined with respect to a fiducial metric is familiar from the study of fluctuating geometries, particularly in Liouville gravity \cite{Knizhnik:1988ak,David:1988hj,Distler:1988jt}. 
Recently, Ref.~\cite{Cotler:2026lna} explicitly employed the associated field-dependent fiducial cutoff to study volume renormalization and its relation to stochastic and eternal inflation, with an extension beyond exact scale invariance developed in Ref.~\cite{Seo:2026ajv}. The distinction between physical and fiducial cutoffs also enters the construction of the probability measure for fluctuating reheating geometries studied in Ref.~\cite{Cotler:2026flq}. 
The present analysis focuses instead on perturbative scalar and tensor cosmological correlators, for which we develop a diagrammatic implementation of a cutoff defined with respect to the local physical geometry. We then identify the resulting cutoff-surface contributions and their implications for superhorizon conservation and soft consistency relations.

We begin by introducing the inhomogeneous cutoff in the presence of a long-wavelength scalar background in Sec.~\ref{inoscalar}. We then investigate its consequences for the late-time evolution and soft limits of one-loop cosmological correlators. The organization of the manuscript and its main results are summarized below.

\begin{itemize}

\item \textit{Scalar power spectrum.}
We first study, in Sec.~\ref{sec:one_loop_scalar_power}, the late-time evolution of the one-loop scalar power spectrum. Once the external scalar mode is well outside the horizon, it belongs to the long-wavelength sector entering the background-dependent regulator.\footnote{The cutoff prescription is independent of the correlator under consideration. Expanding its dependence on the long-wavelength background generates explicit insertions of $\zeta_L$ on regulated loop contractions. These insertions contribute when their long-wavelength fields can be contracted with the operators appearing in the calculation.}
Consistency relations and spatial dilatation symmetry organize the relevant short-mode contributions into total derivatives in loop momentum \cite{Pimentel:2012tw,Kawaguchi:2024rsv,Fumagalli:2024jzz,Iacconi:2026uzo, Artigas:2026vfi,Fang:2026qhu}. Within standard cutoff regularization with a homogeneous cutoff, these total derivatives leave a UV-divergent contribution localized at the cutoff surface. 
We show that expanding the scalar dependence of the inhomogeneous cutoff generates an additional contribution, localized on the cutoff surface, that cancels this UV boundary term exactly.

\item \textit{Tensor power spectrum.} 
In Sec.~\ref{tensorsec}, we extend the cutoff prescription to long-wavelength tensor backgrounds. Compatibility with the residual anisotropic spatial rescaling requires a direction-dependent cutoff. We derive the additional contribution generated by its dependence on the tensor background and show that it cancels the UV boundary term in the late-time evolution of the one-loop tensor power spectrum. 
The total-derivative structure underlying the soft tensor cancellation was established in Refs.~\cite{Ema:2025ftj,Fang:2025hid}, following earlier claims of scale-invariant enhancement~\cite{Ota:2022hvh,Ota:2022xni}. Our result identifies the additional regulator contribution required to cancel the resulting UV boundary term within cutoff regularization. 

These scalar and tensor cancellations constitute central results of this work, providing within cutoff regularization a proof and a geometric interpretation of one-loop conservation on superhorizon scales.

\item \textit{Squeezed bispectrum.}
In Sec.~\ref{sec:cq-squeezed-bispectrum}, we show how the inhomogeneous cutoff ensures that the one-loop squeezed scalar bispectrum satisfies the consistency relation with the power spectrum computed using the same prescription. In deriving the additional cutoff contributions required by this relation, we make explicit the connection between the inhomogeneous cutoff and the background-wave method for soft correlators, reviewed in Appendix~\ref{app:background-wave}. We then illustrate how these contributions combine with the standard in-in diagrams through an explicit example, whose detailed calculations are given in Appendix~\ref{app:cq-ordinary-inin}.

\item \textit{Late-time bispectrum.}
Finally, in Sec.~\ref{sec:bispectrum} we study the late-time evolution of the one-loop bispectrum when all external modes are superhorizon. 
We identify the additional contributions generated by the inhomogeneous cutoff and discuss their role in the conservation of the bispectrum on superhorizon scales. 
With a simplifying assumption, we see how the total derivative terms are canceled by the series expansion of the inhomogeneous cutoff in a similar manner to Sec.~\ref{sec:one_loop_scalar_power}.
\end{itemize}
\subsection*{Notation}

Throughout we use the Fourier transform convention 
\be
\hat\zeta (\bx,\tau) = \int \frac{d^3 k}{(2\pi)^3} e^{i\bk\cdot\bx}\hat{\zeta}_{\bk}(\tau).
\ee
We use a prime notation on the correlators and commutators to indicate that the Dirac delta function and $(2\pi)^3$ factors have been stripped from correlators and the commutator between two operators: 
\be\label{defdoublebracket}
\langle \hat\zeta_{\bp_1} \hat\zeta_{\bp_2} \dots \hat\zeta_{\bp_n} \rangle = (2\pi)^3\delta(\bp_1+\bp_2 + \cdots + \bp_n )\langle \hat\zeta_{\bp_1} \hat\zeta_{\bp_2} \dots \hat\zeta_{\bp_n} \rangle',
\ee
and
\begin{equation}
\left[ \hat\zeta^A_{\bp}(  \tau_1) ,\,  \hat\zeta^B_{\bp'}(  \tau) \right] =
\left( 2 \pi \right)^3 \delta (\bp + \bp' ) \left[ \hat\zeta^A_{\bp}(  \tau_1) ,\,  \hat\zeta^B_{\bp'}(  \tau) \right]',
\label{commutatornotation}
\end{equation}
where $\zeta^A, \zeta^B \in \{\zeta, \zeta', \partial_i \zeta, \cdots\}$.
We often write $\langle \hat\zeta_\bp \hat\zeta_{\bp'} \rangle' = \langle \hat\zeta_\bp \hat\zeta_{-\bp} \rangle' = P(p)$ and \begin{equation}
\left[\hat \zeta^A_{\bp}(  \tau_1) ,\,  \hat \zeta^B_{\bp'}(  \tau) \right]' = \left[ \hat\zeta^A_{\bp}(  \tau_1) ,\,  \hat \zeta^B_{-\bp}(  \tau) \right]'.
\end{equation}
We set $\Mp = 1$ and use curly letters to denote Hamiltonian and Lagrangian densities, $S = \int d\tau d^3 x \, \mathcal{L}$ and $H= \int d^3 x\, \mathcal{H}$. For brevity, we omit the hat on $\zeta$ below.
\section{Inhomogeneous cutoff regularization}\label{inoscalar}

During inflation, on scales much larger than the horizon, the curvature perturbation is approximately time-independent and the metric can be written as\footnote{We neglect for the moment the tensor long-wavelength mode that we will discuss separately in Sec.~\ref{tensorsec}. 
In terms of conformal time, defined by $d\tau=dt/a(t)$
and used below in the explicit diagrammatic calculations, the metric
reads $ds^2\simeq a^2(\tau)[-d\tau^2+e^{2\zeta_L(\mathbf x)}\delta_{ij}dx^i dx^j]$.
}
\begin{equation}
ds^2 \simeq -dt^2 + e^{2\zeta_L(\mathbf{x})} a^2(t)\,dx^i dx^i ,
\end{equation}
where $\zeta_L$ denotes the long-wavelength curvature perturbation frozen outside the horizon.
Since the spatial variation of $\zeta_L$ is negligible within a Hubble-sized region, the long-wavelength perturbation can be regarded locally as part of the background geometry. Equivalently, at leading order in the gradient expansion, it acts as a local rescaling of the spatial coordinates in the background metric~\cite{Maldacena:2002vr,Creminelli:2004yq}. Each Hubble-sized patch then evolves as an approximately homogeneous universe with a locally rescaled spatial scale factor, $
a_{\rm loc}(t,\mathbf{x}) = a(t)e^{\zeta_L(\mathbf{x})}$.
This is the basic idea of the separate universe picture.
Physical effects associated with spatial variations of the long mode arise only through gradient corrections, schematically of order
$\mathcal{O}\left(k_L^2/(a^2H^2)\right)$, where $k_L^{-1}$ denotes
a characteristic comoving length scale over which the long-wavelength
background varies. 

\begin{figure}
        \centering \includegraphics[width=0.8\columnwidth]{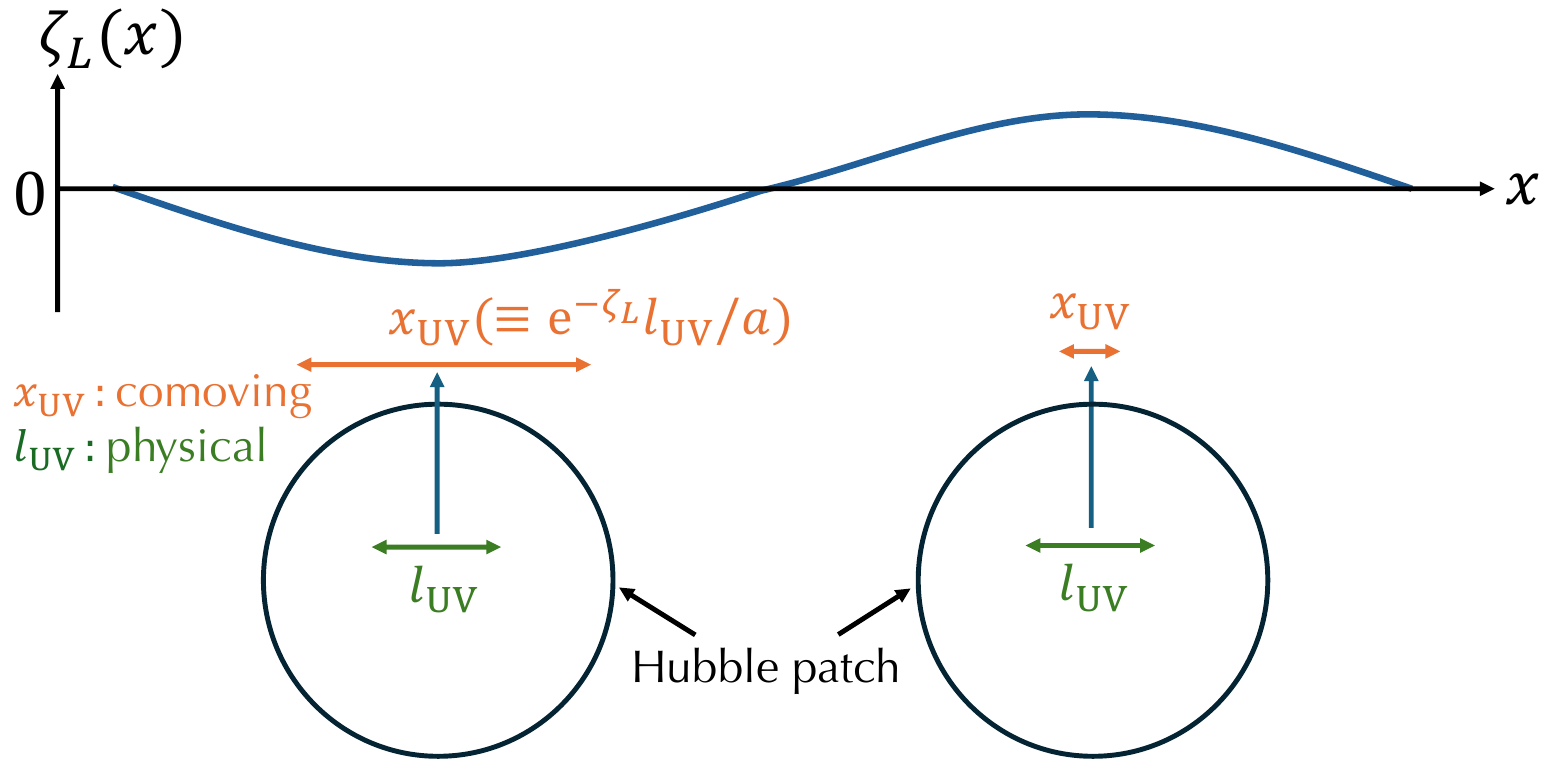}
        \caption{
        The relation between the physical cutoff $l_\uv$ for a local observer and the corresponding comoving scale $x_\uv$. 
        The left (right) Hubble patch is in the region of $\zeta_L < 0$ ($\zeta_L > 0$).
        The same physical cutoff ($l_\uv$) leads to different cutoffs for $x_\uv$ depending on the value of $\zeta_L$.
       Specifically, the left Hubble patch has $\zeta_L < 0$, which leads to $e^{-\zeta_L}>1$ and therefore leads to a larger $x_\uv$ than the right-hand one.        
        We set $a=1$ for illustrative purposes. 
        }
        \label{fig:l_uv_x_uv}
\end{figure}

We assume that the UV regulator corresponds to the same \textit{local physical cutoff scale} in each patch. See Figure~\ref{fig:l_uv_x_uv}. Let $l_{\rm UV}$ denote the physical UV cutoff distance measured by a local observer, and let $\Lambda_{\rm UV}\sim l_{\rm UV}^{-1}$ denote the corresponding physical momentum cutoff. We denote by $x_{\rm UV}$ the comoving distance corresponding to $l_{\rm UV}$. At a given time and within a given Hubble patch, $l_\uv$ and $x_\uv$ are related as
\begin{equation}
l_{\rm UV}^2 = a^2(t)e^{2\zeta_L(\mathbf{x})}x_{\rm UV}^2.
\end{equation}
This relation is understood locally within a Hubble patch, where $\zeta_L$ is approximately constant.
Thus, a regulator which is homogeneous in local physical momentum is in general inhomogeneous when expressed in terms of the global comoving momentum $k$.
Fixing the physical cutoff scale for the local observer, $l_\uv$, we find that $k_\uv$, which is the Fourier mode corresponding to the comoving coordinate $x_\uv$, depends not only on time through the scale factor, but also on the Hubble patch through the factor $e^{\zeta_L}$. 
This means that, if instead we introduce a homogeneous cutoff for the comoving momentum $k$, namely $k_\uv = a \Lambda_{\uv}$, 
then the UV cutoff scale observed by local observers, $l_\uv$, would depend on $\zeta_L$, and would therefore depend on the realization of the long-wavelength perturbation. Different Hubble patches would effectively be regulated at different physical scales.
On the contrary, imposing the same physical cutoff in each Hubble patch, consistently with the separate universe picture, leads to what we label the \textit{inhomogeneous cutoff}:
$\Lambda[\zeta_L] \equiv a(t)e^{\zeta_L(\mathbf{x})}\,\Lambda_{\uv}$, where $\Lambda_\uv$ is the UV physical cutoff scale.
That is, in terms of $k$, we need to modify the UV cutoff scale as
\be \label{ino cutoff}
\int^{k_\uv} dk \,\,\rightarrow\,\, \int^{a(t) e^{\zeta_L(\mathbf{x})}\Lambda_\uv} dk.
\ee
Expanding the regulator in powers of the long mode gives
\begin{equation}
\Lambda[\zeta_L] \equiv a \Lambda_{\rm UV}e^{\zeta_L}
=
a\Lambda_{\rm UV}
\left(
1+\zeta_L+\frac{1}{2}\zeta_L^2+\cdots
\right).
\end{equation}
The first important consequence is that the regulated contraction of two fields, $\expval{\zeta^2(\bx)_{\Lambda[\zetal]}}$, depends on the long-wavelength background. Note that, throughout this work, subscripts of correlators, such as $\Lambda[\zetal]$ for $\expval{\zeta^2(\bx)}_{\Lambda[\zetal]}$, indicate the UV cutoff that we consider for that correlator.
Expanding this dependence in powers of $\zeta_L$ from $\Lambda[\zetal]$ generates a sequence of effective interactions with an increasing number of long-mode insertions, as we will see in Sec.~\ref{subsec:ef_vertex}.
\subsection{Relation to spatial dilatation symmetry}\label{dilatationsymmetry}

The geometric construction above can also be understood through the residual spatial diffeomorphisms of comoving gauge~\cite{Hinterbichler:2012nm,Hinterbichler:2013dpa}.\footnote{Ref.~\cite{Inomata:2026csq} also uses the dilatation symmetry and discusses how the cutoff scales can be consistent with the separate universe picture in spatially-flat gauge. 
In that gauge, homogeneous cutoffs fixed in physical scales, such as the UV cutoff, are consistent with the separate universe picture simply because there is no spatial metric perturbation.
Meanwhile, cutoffs fixed in comoving scales, such as the smoothing cutoff introduced to relate the inflaton fluctuations to curvature perturbations, must be inhomogeneous in the spatially-flat gauge to be consistent with the separate universe picture because of the time dependence of the cutoff in physical scales.
} 
Consider the generators linear in the spatial coordinates,
\begin{equation}\label{residualgauge}
\xi^i=M^i_j x^j,\qquad M_{ij}=\lambda\delta_{ij}+S_{ij}+\omega_{ij},
\end{equation}
where $\lambda=M_{ii}/3$, $S_{ij}=S_{ji}$, $S_{ii}=0$, and $\omega_{ij}=-\omega_{ji}$, with
all these coefficients constant in space and time.\footnote{On the FLRW background, a time-dependent spatial diffeomorphism generates a contribution $\delta N^i=\dot\xi^i$ to the ADM shift $N^i$. For a dilatation, this is $\dot\lambda x^i$. Taking $\dot\lambda=0$ avoids this spatially growing contribution and is compatible with the standard boundary prescription used to solve the lapse and shift constraints~\cite{Urakawa:2009my,Urakawa:2010it}. Further, note that these linear generators belong to the $n=0$ sector in the classification of residual adiabatic transformations developed in Ref.~\cite{Hinterbichler:2013dpa}.}    
The trace generates an isotropic spatial dilatation, the antisymmetric part generates rotations, and the symmetric traceless part generates a volume-preserving anisotropic rescaling. The tensor realization of the latter will be discussed in Sec.~\ref{tensorsec}.

At fixed time, the action of the dilatation on the curvature perturbation is
\begin{equation}
\zeta_\lambda(\mathbf x)=\zeta(e^\lambda\mathbf x)+\lambda,\qquad \delta_\lambda\zeta=\lambda\left(1+x^i\partial_i\zeta\right).
\end{equation}
Consider a patch where $\zeta_L$ is approximately constant. 
There, the transformation gives $\zeta_L\to\zeta_L+\lambda$ for the long mode. 
Meanwhile, for the short modes, an original wave with $k$ is mapped to a new wave whose scale, in the same coordinate system, is $e^\lambda k$ through the following relation:
\begin{equation}
\zeta_\lambda(\bx)-\lambda=\zeta(e^\lambda \bx)= \int \frac{d^3 k}{(2\pi)^3} e^{i e^\lambda\bk\cdot \bx}\zeta_\bk.
\end{equation}
This effectively corresponds to $\mathbf k\to e^\lambda\mathbf k$.
The local physical momentum $k/(ae^{\zeta_L})$ is consequently unchanged. 
A local momentum cutoff compatible with this symmetry must therefore satisfy, at leading order in long-wavelength gradients,
\begin{equation}
\Lambda[\zeta_L+\lambda]=e^\lambda\Lambda[\zeta_L].
\end{equation}
Together with $\Lambda[0]=a\Lambda_{\rm UV}$, this fixes $\Lambda[\zeta_L]=ae^{\zeta_L}\Lambda_{\rm UV}$, reproducing the prescription derived geometrically above. Holding the comoving boundary at $a\Lambda_{\rm UV}$ independently of $\zeta_L$ would violate this transformation law. The inhomogeneous prescription instead allows the integration domain to transform consistently with the fields, preserving the residual spatial dilatation symmetry. Expanding this domain in the long mode generates the additional regulator contributions discussed below.

\subsection{Effective vertices and diagrammatic interpretation}
\label{subsec:ef_vertex}

Once the long mode is well outside the horizon, we treat its conserved growing-mode amplitude $\zeta_L(\mathbf x)$ as a \textit{classical stochastic background}. Short-mode contractions are evaluated quantum mechanically at fixed $\zeta_L$, and the resulting conditional correlators are subsequently averaged over the probability distribution of the long-wavelength background. This provides a separate-universe implementation of the regulator in the long-wavelength limit. To illustrate its diagrammatic consequences, we now examine how the background-dependent cutoff modifies the contraction of two fields within an arbitrary interaction vertex.

Consider, for instance, a cubic interaction $H^{(3)}$. Contracting two of its fields at the same vertex leaves a linear effective insertion, which generates a tadpole contribution when the remaining field is contracted with an external field. With a homogeneous regulator, this is the standard tadpole insertion. For the inhomogeneous cutoff, however, the regulated two-field contraction acquires an additional dependence on $\zeta_L$ through its integration boundary (see Eq.~\eqref{ino cutoff}). Expanding this dependence generates effective insertions with additional long-mode fields. To see how this works, let us write the cubic Hamiltonian as 
\begin{equation}
\label{eq:cubic_int}
\mathcal H^{(3)}=\sum_{A,B,C}\lambda_{ABC}\zeta^A \zeta^B \zeta^C,
\qquad
\zeta^I\in
\left\{
\zeta,\,
\zeta',\,
\partial_i\zeta,\ldots
\right\},
\end{equation}
where we adopt Weyl ordering for the Hamiltonian throughout this work, meaning that products of noncommuting operators are fully symmetrized. For each possible type of uncontracted leg, we define the regulated tadpole source or the amputated tadpole insertion:\footnote{
Here $\delta\mathcal H^{(3)}/\delta\zeta^A$ denotes the ordinary partial derivative of the Hamiltonian density with respect to its argument $\zeta^A$.
We adopt this notation to be aligned with Refs.~\cite{Pimentel:2012tw,Fumagalli:2024jzz}.}
\begin{equation}
\mathcal{T}_A[\zeta_L(\bx)](\tau)
\equiv
\left\langle
\frac{\delta\mathcal{H}^{(3)}}
{\delta\zeta^A}
\right\rangle_{\Lambda[\zeta_L]} .
\end{equation}
No integration by parts is implied. Since $\mathcal{H}^{(3)}$ is cubic, $\delta\mathcal{H}^{(3)}/\delta\zeta^A$ is quadratic in the perturbations. The expectation value in the equation above therefore corresponds precisely to contracting the two remaining fields at the cubic vertex while leaving the $A$-type leg uncontracted.
$\mathcal{T}_A[\zeta_L]$ is the ordinary tadpole contraction evaluated with a regulator whose position depends on the background long mode. Note again the notation $\langle\cdots\rangle_{\Lambda[\zeta_L]}$ indicates precisely that this two-field contraction is regulated using the cutoff defined with respect to the long-wavelength background in Eq.~\eqref{ino cutoff}.
For instance, consider an interaction of the form $\lambda \zeta' \zeta' \zeta$. Within a diagram having one cubic interaction, the contraction of the two $\zeta'$ would lead to 
\begin{equation}
\langle  (\Rh'(\bx,\tau_1))^2\rangle_{\Lambda[\zetal]} = \int \frac{d^3 k_1 d^3 k_2}{(2\pi)^3(2\pi)^3} e^{i (\bk_1+\bk_2)\bx}\langle \Rh'_{\bk_1}\Rh'_{\bk_2}\rangle = \int^{\Lambda[\zeta_L]} d\ln k\frac{k^3}{2\pi^2} \langle \Rh'_{\bk}\Rh'_{-\bk}\rangle',
\end{equation}
where in the last step we used the definition in Eq.~\eqref{defdoublebracket}.
In general, a regulated contraction, after angular integration, can be written  as
\begin{equation}
\label{eq:cal_T_a}
\mathcal{T}_A[\zeta_L(\bx)](\tau)
=
\int^{\Lambda[\zeta_L(\bx)]} d\ln k\,
\,T_A(k,\tau),\quad T_A(k,\tau)=\frac{k^3}{2\pi^2}\llangleb \frac{\delta \mathcal{H}^{(3)}}{\delta \zeta^A}\Big|_{\bk,-\bk}\rrangleb',
\end{equation}
where $T_A(k,\tau)$ contains the appropriate two-point contraction in momentum space, including the momentum measure and the derivatives associated with the particular structure $\zeta^A$,
and
\begin{align}
    \expval{\zeta^B \zeta^C \Big|_{\bk,-\bk}}' \equiv \expval{\zeta^B_\bk \zeta^C_{-\bk}}'.
\end{align}
Recall that $\delta \mathcal H^{(3)}/\delta \zeta^A$ is quadratic in perturbations.
The expansion of $\mathcal T_A$ takes the form
\begin{equation}
\mathcal{T}_A[\zeta_L]
=
\mathcal{T}_A^{(0)}
+
\zeta_L\,\mathcal{T}_A^{(1)}
+
\frac{1}{2}\zeta_L^2\,\mathcal{T}_A^{(2)}
+\cdots ,
\end{equation}
where, for instance,
\begin{equation}
\mathcal{T}^{(0)}_A(\tau)
=
\int^{a(\tau)\Lambda_{\rm UV}}
d\ln k\,
T_A(k,\tau),\qquad
\mathcal{T}^{(1)}_A(\tau)
\equiv
T_A(a(\tau)\Lambda_{\rm UV},\tau).
\end{equation}
After inserting this expansion back into the interaction, one obtains schematically
\begin{equation}
\mathcal I^{(3)}_{0}
=
\sum_A
\mathcal{T}_A^{(0)}\zeta^A,\quad
\mathcal I^{(3)}_{1}
=
\sum_A
\mathcal{T}_A^{(1)}
\zeta_L\zeta^A,\quad
\mathcal I^{(3)}_{2}
=
\frac{1}{2}\sum_A
\mathcal{T}_A^{(2)}
\zeta_L^2\zeta^A,\,\cdots\label{effective interaction}
\end{equation}
where $\mathcal I^{(n)}_{k}$ denotes the effective interaction Hamiltonian.
The superscript of $\mathcal I$ identifies the order of the original Hamiltonian, while the
subscript counts the long-background
insertions. At one loop, i.e. one pair contracted, the order of the effective interaction is given by $n - 2 + k$.
Despite the intuitive picture behind writing these series of effective interactions, one has to remember that $\mathcal{I}_{k
}^{(3)}$ is the result of a two-leg contraction defined conditionally on a fixed long-wavelength background. Thus, for consistency, the resulting series of $\zeta_L$ has to be introduced after the nested commutator is evaluated.

The first term in Eq.~(\ref{effective interaction}) is the ordinary homogeneous tadpole contribution. It is linear in the perturbation and is removed by the standard tadpole renormalization counterterm. The next term is qualitatively different. Although it originates from the same tadpole contraction, the dependence of the cutoff on the long-wavelength curvature perturbations converts it into an effective quadratic interaction. It can therefore contribute to a two-point function through a single vertex in-in insertion. This is the term relevant for the one-loop power spectrum studied below. 
Diagrammatically, we may represent the series of effective interactions induced by the cubic Hamiltonian as: 
\begin{equation}\label{effectivecubic}
\begin{array}{cccccccccc}
H^{(3)}
& \longrightarrow\quad
& \cubicdiag
& \quad + \quad
& \taddiag
& \quad + \quad
& \tadLdiag
& \quad + \quad
& \tadLLdiag
& \quad + \cdots.
\\[0.7mm]
&
&
&
&
\mathcal{I}^{(3)}_{0}
&
+
&
\mathcal{I}^{(3)}_{1}
&
+
&
\mathcal{I}^{(3)}_{2}
&
\end{array}
\end{equation}
Analogously, starting from the quartic interaction leads to the following series of diagrams: \begin{equation}\label{effectivequartic}
\begin{array}{cccccccccc}
H^{(4)}
& \longrightarrow\quad
& \quarticdiag
& \quad + \quad
& \tadquartic
& \quad + \quad
& \quarticonelong
& \quad + \quad
& \quartictwolong
& + \cdots.
\\[0.7mm]
&
&
%H^{(3)}
&
%+
&
\mathcal{I}^{(4)}_{0}
&
+
&
\mathcal{I}^{(4)}_{1}
&
+
&
\mathcal{I}^{(4)}_{2}
&
\end{array}
\end{equation}
For later convenience, we define the effective Hamiltonian induced by the inhomogeneous cutoff as
\begin{align}
    \label{eq:h_inh_def}
    H^{(n)}_{\text{inh}} \equiv \sum^\infty_{i=1} H^{(n)}_{\text{inh},i}, \qquad  H^{(n)}_{\text{inh},i} \equiv \int d^3x\, \mathcal I^{(n)}_i.
\end{align}
From these drawings, it is then clear that, at one loop, the inhomogeneous cutoff regularization affects the two-point function only through the third diagram ($\mathcal{I}_1^{(3)}$) in Eq.~\eqref{effectivecubic}.
These effective interactions are generated by the regulator independently of the correlator being computed. For the long-mode power spectrum, an external leg can contract with the $\zeta_L$ insertion provided its momentum belongs to the long-wavelength sector at the interaction time.

\subsection{Momentum-space interpretation}

The momentum-space structure makes explicit how the cutoff-induced insertions contribute to correlators. At leading order in long-wavelength gradients, the effective interaction obtained by contracting two fields within the cubic vertex can be written as \begin{equation}
\label{eq:h_eff_tad_s}
H^{(3)}_{\rm tad}(\tau) \equiv \int d^3 x \, \mathcal I^{(3)}_0(\tau) + H^{(3)}_{\rm inh}(\tau)
=
\sum_A \mathcal{T}_A^{(0)}(\tau)\,\zeta^A_{\mathbf{0}}(\tau)
+
\sum_A \mathcal{T}_A^{(1)}(\tau)\int\frac{d^3q}{(2\pi)^3}\,
\zeta^A_{\mathbf q}(\tau)\,\zeta_{L,-\mathbf q}
+\cdots,
\end{equation}
where $\int d^3 x \, \mathcal I^{(3)}_0$ is the tadpole contribution with homogeneous cutoff. 
The first term is the ordinary tadpole insertion: the homogeneous contraction carries zero momentum and forces the remaining leg to have $\mathbf q=0$. In the second term, the spatial dependence of the cutoff allows the contracted pair to exchange momentum with the long-wavelength background. This term therefore acts as an effective quadratic insertion for long-mode correlators. For example, contracting $\zeta^A_{\mathbf q}$ with an external mode of momentum $\mathbf p$ fixes $\mathbf q=-\mathbf p$, leaving the background factor $\zeta_{L,\mathbf p}$. Provided $\mathbf p$ belongs to the long-wavelength sector at the interaction time, this factor can be averaged with the second external mode, of momentum $\mathbf p'$. For a statistically homogeneous long-mode ensemble, this average supplies $\delta^{(3)}(\mathbf p+\mathbf p')$, yielding a diagonal contribution to the long-wavelength power spectrum.

\section{The tadpole and one-loop scalar power spectrum}

We first evaluate the tadpole using the inhomogeneous cutoff prescription. We then show how the dependence of the regulated tadpole on the long-wavelength background generates an additional contribution to the one-loop scalar power spectrum.

\label{sec:one_loop_scalar_power}
\subsection{The (inhomogeneous) tadpole}
Let us compute the conditional one-point function in configuration space: 
\begin{equation}
\begin{split}
\langle \zeta(\bx,\tau)\rangle_{\Lambda[\zetal]}
&=i\int^\tau d\tau'\int d^3y\,
\left\langle[\mathcal H^{(3)}(\by,\tau'),\zeta(\bx,\tau)]\right\rangle_{\Lambda[\zetal]}\\
&=i\int^\tau d\tau'\int d^3y\int\frac{d^3q}{(2\pi)^3}\,
\sum_A \mathcal T_A[\zeta_L(\by)](\tau')\,e^{-i\bq\cdot(\by-\bx)}
\left\langle[\zeta^A_{-\bq}(\tau'),\zeta_{\bq}(\tau)]\right\rangle'.
\end{split}
\label{eq:tadpole}
\end{equation}
As usual, the prime removes the momentum-conserving delta function from the Fourier-space expectation value. The corresponding tadpole contribution is obtained by expanding in $\zetal$ and averaging over the long-wavelength background. In the Gaussian approximation, odd powers average to zero, giving
\begin{equation}
\langle\zeta(\bx,\tau)\rangle
=\left\langle\langle\zeta(\bx,\tau)\rangle_{\Lambda[\zetal]}\right\rangle
\equiv\sum_{n=0}^{\infty}\langle\zeta(\bx,\tau)\rangle_{(2n)},
\end{equation}
where $\langle\zeta(\bx,\tau)\rangle_{(2n)}$ denotes the contribution obtained by averaging the term containing $2n$ long-mode fields.

At zeroth order in the long-wavelength background, $\mathcal T_A^{(0)}(\tau')$ is independent of $\by$. The spatial integral therefore produces $(2\pi)^3\delta^{(3)}(\bq)$, selecting the zero-momentum insertion and reproducing the usual homogeneous tadpole, in formulas:
\begin{equation}\label{firstordertadpole}
\langle \zeta(\bx,\tau)\rangle_{(0)} =  i  \int d\tau' \sum_A \mathcal{T}^{(0)}_A(\tau') \lim_{\bq\to \mathbf0}[\zeta^A_{-\bq}(\tau'),\zeta_{\bq}(\tau)]'.
\end{equation}
After averaging over the long-wavelength background, all nonvanishing terms beyond zeroth order contribute at higher loop order. For example, at quadratic order in $\zeta_L$, the background average contracts the two additional long-mode legs in the last diagram of Eq.~\eqref{effectivecubic}. This introduces a second loop, yielding a two-loop tadpole with zero external momentum:
\begin{equation}\label{secordertadpole}
\begin{split}
\langle \zeta(\bx,\tau)\rangle_{(2)} =  &\frac{i}{2}  \int d\tau'\int d^3 q\int d^3 y \sum_A \mathcal{T}^{(2)}_A(\tau')\langle \zeta_L^2(\by) \rangle\frac{e^{-i\bq \cdot \by}}{(2\pi)^3}e^{i \bq \cdot \bx} [\zeta^A_{-\bq}(\tau'),\zeta_{\bq}(\tau)]'\\[1mm]
=&\frac{i}{2} \int d\tau' \sum_A \mathcal{T}^{(2)}_A(\tau')\int d\ln k \frac{k^3}{2\pi^2}\langle \zeta_{L,\bk}(\tau')\zeta_{L,-\bk}(\tau')\rangle' \lim_{\bq \to \mathbf 0}[\zeta^A_{-\bq}(\tau'),\zeta_{\bq}(\tau)]'.
\end{split}
\end{equation}
This contribution still corresponds to a zero-mode shift and can therefore be absorbed into a tadpole counterterm. In particular, a counterterm of the form: 
\begin{equation}\label{tadcounter}
\mathcal{L}^{(1)}_{\rm tad} = c_A \zeta^A, \qquad  c_A(\tau)=\langle \mathcal{T}_A[\zeta_L(\bx,\tau)]\rangle 
\end{equation}
cancels all contributions $\langle\zeta(\bx,\tau)\rangle_{(2n)}$ to the one-point function. For $n=0$, $c_A(\tau) = \mathcal{T}_A^{(0)}(\tau)$ is the standard tadpole counterterm that cancels the one-loop tadpole diagram in Eq.~\eqref{firstordertadpole} and guarantees the cancellation of all one-particle-reducible (1-PR) contributions arising from contractions between two legs of the same cubic interaction. For $n=1$, $c_A(\tau) \supset (\mathcal{T}_A^{(2)}(\tau)/2)\int d\ln k \frac{k^3}{2\pi^2}\langle \zeta_{L,\bk}(\tau)\zeta_{L,-\bk}(\tau)\rangle'  $ and thus the counterterm in Eq.~\eqref{tadcounter} will cancel the two-loop contributions from Eq.~\eqref{secordertadpole}. The same pattern extends order by order, with the corresponding tadpole counterterm canceling the one-point contribution generated at each perturbative order.\footnote{
The inhomogeneous cutoff introduces a dependence on the long-wavelength background into the regulated field contractions. It does not introduce an independent spatial dependence into the Lagrangian coefficients $c_A$, which remain spatially homogeneous. The resulting effective insertions cannot therefore be absorbed into arbitrary patch-dependent choices of $c_A$.}

\subsection{Contribution to the one-loop superhorizon power spectrum}
Let us consider the one-loop two-point function of a mode which belongs to the long-wavelength sector during the relevant interaction times. 
We denote by $\tau_\star$ the time when the relevant interactions are turned on. The long modes are already outside the horizon at $\tau_\star$, which allows us to approximate $\zetal$ as constant. 
As mentioned extensively, the expansion of the regulated tadpole derived above produces an effective quadratic interaction denoted by $H^{(3)}_{\text{inh},1}$, whose coefficient $\mathcal{T}_A^{(1)}$ already contains a one-loop contraction.
It can therefore contribute to the two-point function through a single in-in insertion, 
\begin{equation}\label{twopointino}\begin{split}
\left\langle
\zeta_{\mathbf{p}}(\tau)
\zeta_{\mathbf{p}'}(\tau)
\right\rangle_{\rm inh}
&=
i\int^{\tau}_{\tau_\star}d\tau'\,
\left\langle
\left[
H^{(3)}_{\text{inh},1}(\tau'),
\zeta_{\mathbf{p}}(\tau)
\zeta_{\mathbf{p}'}(\tau)
\right]
\right\rangle\\[1mm]
&= i \int^{\tau}_{\tau_\star}d\tau'\,\sum_A \mathcal{T}_A^{(1)}(\tau')\int d^3 x \left\langle \left(\zeta_L(\bx)e^{-i\bp\cdot\bx}\zeta_{\bp'}[\zeta^A_{-\bp},\zeta_{\bp}]' + (\bp\leftrightarrow\bp')\right)\right\rangle\\[1mm]
&= (2\pi)^3\delta(\bp+\bp')i |\zeta_p|^2\times 2 \int^{\tau}_{\tau_\star}d\tau'\,\sum_A \mathcal{T}_A^{(1)}(\tau')[\zeta^A_{-\bp}(\tau'),\zeta_{\bp}(\tau)]',
\end{split}
\end{equation}
where $|\zeta_p|^2$ is the modulus square of the mode functions that correspond to the dimensionful tree-level power spectrum.
We recall that
\begin{equation}
\label{eq:cal_t_a_1}
\mathcal{T}_A^{(1)}(\tau')= \frac{\delta \mathcal{T}_A[\zeta_L]}{\delta\zeta_L}\Big|_{\zeta_L=0}=\frac{d}{d\ln k}\left\langle
\frac{\delta\mathcal{H}^{(3)}}
{\delta\zeta^A}
\right\rangle\Big|_{k=\Lambda(\tau')}=\left( \frac{k^3}{2\pi^2}\llangleb \frac{\delta \mathcal{H}^{(3)}}{\delta\zeta^A}\Big|_{\bk,-\bk}\rrangleb'\right)\Big|_{k=\Lambda(\tau')},
\end{equation}
where
\begin{equation}
\Lambda(\tau) \equiv \Lambda[0](\tau)= a(\tau)\Lambda_\uv
\end{equation}
is the homogeneous UV cutoff.
Note that, as explained below Eq.~\eqref{effective interaction}, $\zetal$ is introduced after the nested commutator is evaluated and therefore $\zetal$ cannot be inside the commutator in Eq.~(\ref{twopointino}).
One can equivalently define the perturbation induced by this effective interaction,
\begin{equation}
\zeta_{{\rm inh},\mathbf{p}}(\tau)
\equiv
i\int^{\tau}_{\tau_\star}d\tau'\,\sum_A\mathcal{T}_A^{(1)}(\tau')\int d^3 x\zeta_L(\bx)
\left[
\zeta^A(\bx,\tau'),
\zeta_{\mathbf{p}}(\tau)
\right].
\end{equation}
The correction to the two-point function can then be written as 
\begin{equation}
\left\langle
\zeta_{\mathbf{p}}
\zeta_{\mathbf{p}'}
\right\rangle_{\rm inh}
=
\left\langle
\zeta_{{\rm inh},\mathbf{p}}
\zeta_{\mathbf{p}'}
\right\rangle
+
\left\langle
\zeta_{\mathbf{p}}
\zeta_{{\rm inh},\mathbf{p}'}
\right\rangle.
\end{equation}
Thus one may regard $\zeta_{\rm inh}$ as a regulator-induced contribution to the third-order solution,
\begin{equation}
\zeta^{(3)}
=
\zeta_{\rm ordinary}^{(3)}
+
\zeta_{\rm inh}^{(3)}.
\end{equation}
The corresponding power-spectrum contribution is therefore naturally part of
\begin{equation}
P_{13}
=
\left\langle
\zeta^{(1)}
\zeta^{(3)}
\right\rangle
+
\left\langle
\zeta^{(3)}
\zeta^{(1)}
\right\rangle .
\end{equation}
In this sense, the inhomogeneous tadpole does not define a new loop topology beyond the usual $13$ sector. Rather, it supplies a regulator-induced contribution to the third-order long-mode response that is absent if the comoving cutoff is kept (artificially) homogeneous.

\subsection{Cancellation of UV boundary terms and superhorizon conservation - general proof for the scalar case}
\label{subsec:ge_pr_scalar} 

Superhorizon conservation requires the one-loop correction to become time-independent. Choosing $\tau_\star$ sufficiently late that $p/(a(\tau_\star)H(\tau_\star))\ll1$, we aim to show
\begin{equation}\label{difference}
\langle \Rh_{\bp}(\tz)\Rh_{\bp'}(\tz)\rangle'_{\mathrm{1-loop}}\simeq \langle \Rh_{\bp}(\tau_\star)\Rh_{\bp'}(\tau_\star)\rangle'_{\mathrm{1-loop}},
\end{equation}
where $\tz$ denotes the end of inflation. 
In the following, by the late-time one-loop contribution we mean
the difference between the left- and right-hand sides of
Eq.~\eqref{difference}. Equation~\eqref{difference} is understood to hold at leading order in the long-wavelength expansion.
Gradient-suppressed corrections and other terms that vanish at late
times are neglected. We show below how the inhomogeneous cutoff supplies the contribution needed to cancel the UV boundary term in this late-time evolution. 
In the previous section, we have shown that, in general, the contribution from the inhomogeneous regulator to the superhorizon two-point function is given by
\be\label{generictadpole}
\langle \Rh_{\bp}(\tz)\Rh_{\bp'}(\tz)\rangle'_{\mathrm{inh}} = 2|\zeta_p|^2 \frac{\delta\langle\zeta(\bx,\tau_f)\rangle_{\Lambda[\zeta_L]}}{\delta \zeta_L}\Big|_{\zeta_L= 0},
\ee
where
\be
\frac{\delta\langle\zeta(\bx,\tau_f)\rangle_{\Lambda[\zeta_L]}}{\delta \zeta_L}\Big|_{\zeta_L= 0}= i \int^{\tz}_{\tau_\star} d\tau' \sum_A[\Rh^A_{-\bp}(\tau'),\Rh_{\bp}(\tz)]' \left(\frac{k^3}{2\pi^2}\llangleb \frac{\delta \mathcal{H}^{(3)}(\tau')}{\delta\R^A}\Big|_{\bk,-\bk}\rrangleb'\right)_{k = \Lambda(\tau')}.
\ee
On the other hand, results in Refs.~\cite{Pimentel:2012tw,Kawaguchi:2024rsv,Fumagalli:2024jzz} show that, after properly combining cubic--cubic diagrams and quartic terms induced by the cubic Lagrangian and consistently
treating tadpoles, the standard late-time 
one-loop contribution
can be rewritten 
in terms of connected
three-point functions:\footnote{Unlike in dimensional regularization, grouping the loop
contributions into three-point functions requires additional care
with a time-dependent cutoff. See Appendix~\ref{app:cutoff-matching}
for details.}
\begin{align}
&\langle \Rh_{\bp}(\tz)\Rh_{\bp'}(\tz)\rangle'_{\mathrm{1-loop},\,\Lambda} \nonumber \\ 
&= i \int^{\tz}_{\tau_\star} d\tau' \sum_A [\Rh^A_{-\bp}(\tau'),\Rh_{\bp}]' \, 2\int^{\Lambda(\tau')} d\ln k   \,\left(\frac{k^3}{2\pi^2}\llangleb \frac{\delta \Hamt (\tau')}{\delta\R^A}\Big|_{\bk,-\bk}\Rh_{\bp}\rrangleb'\right),
\label{eq:1_loop}
\end{align}
where  
we have omitted the time argument $\tau_f$ on the right-hand side. The factor $2$ in front of $\int d\ln k$ arises because there are two external legs $\Rh_{\bp},\Rh_{\bp'}$ in the nested commutator determining the loop integral~\cite{Pimentel:2012tw,Fumagalli:2024jzz}.\footnote{In comparison with the formula appearing in Refs.~\cite{Pimentel:2012tw,Fumagalli:2024jzz}, we have simply replaced $\Hamt=-\Lt$ and use the notation $\langle \cdots \rangle'$ which is equivalent to $ \langle\langle \cdots \rangle\rangle $ in Ref.~\cite{Fumagalli:2024jzz}.}
The three-point function appearing in the late-time one-loop contribution $\llangleb \frac{\delta \Hamt}{\delta\R^A}\Big|_{\bk,-\bk}\Rh_{\bp}\rrangleb'$ should be regarded as the squeezed-limit bispectrum with $k \gg p$, which is why it was written in terms of only the two momenta, $\bk$ and $\bp$. 
In this limit three-point correlators satisfy the consistency relations:\footnote{Recall that, when the three-point function contains spatial derivatives, a consistency relation is not satisfied. However, the quartic Hamiltonian  provides the additional diagram contributions to make Eq.~\eqref{consistency} satisfied also in this case~\cite{Pimentel:2012tw,Fumagalli:2024jzz}.}
\be\label{consistency}
\frac{k^3}{2\pi^2}\llangleb \frac{\delta \Hamt}{\delta\R^A}\Big|_{\bk,-\bk}\Rh_{\bp}\rrangleb' = -|\zeta_p|^2 \frac{d}{d\ln k}\left(\frac{k^3}{2\pi^2}\llangleb \frac{\delta \Hamt}{\delta\R^A}\Big|_{\bk,-\bk} \rrangleb' \right).
\ee
Substituting this into Eq.~\eqref{eq:1_loop} and integrating over the comoving momenta,
\footnote{Formally, also this loop integration should be regularized with the inhomogeneous cutoff, but only the zeroth-order term in $\zeta_L$, i.e. $\Lambda[\zeta_L=0],$ will give a non-zero contribution at one-loop order once averaged over the stochastic realization of the regulator. For the one-loop bispectrum, the $\zetal$ expansion of $\Lambda[\zetal]$ provides an important contribution, as we will see in Eq.~(\ref{eq:b_33_inh}).} we obtain a contribution that is equal in magnitude and opposite in sign to Eq.~\eqref{generictadpole}. The opposite sign arises from the minus sign in the consistency relations above. Summing the two contributions thus provides an exact cancellation of the soft leading term, implying the constancy of $\zeta$ on superhorizon scales at one-loop order within cutoff regularization:
\be
\langle \Rh_{\bp}(\tz)\Rh_{\bp'}(\tz)\rangle'_{\mathrm{1-loop},\,\Lambda[0]} + \langle \Rh_{\bp}(\tz)\Rh_{\bp'}(\tz)\rangle'_{\rm inh}\simeq 0.
\ee
We thus prove that the divergence from the UV boundary term that arises in standard cutoff regularization is due to the violation of the separate universe picture caused by choosing a homogeneous UV cutoff.

\section{Tensor modes}\label{tensorsec}
A superhorizon tensor mode can be incorporated into the local background geometry within the same separate universe picture as a scalar long mode. While $\zeta_L$ produces an isotropic rescaling of spatial distances, the long-wavelength tensor modes, denoted by $\gamma_L^{ij}$, induce an anisotropic deformation. Accordingly, a fixed physical UV cutoff corresponds, in the original comoving coordinates, to an anisotropically deformed cutoff surface.
Following the same steps as in Sec.~\ref{inoscalar}, we introduce the inhomogeneous cutoff in the presence of tensor modes. We then show how the resulting cutoff surface regulates the amputated tadpole insertion and contributes to correlation functions. We emphasize from the outset that, as we will show explicitly, the vanishing of the ordinary tensor tadpole does not by itself imply that the regulated tadpole source gives no contribution. We show this explicitly in the one-loop tensor power spectrum for modes that are already superhorizon at the interaction time, thereby establishing, analogously to the scalar case, the crucial role of the inhomogeneous cutoff in preserving the conservation of the superhorizon tensor power spectrum at one loop.
\subsection{Inhomogeneous cutoff in the presence of long-wavelength tensor modes}
\begin{figure}
        \centering \includegraphics[width=0.8\columnwidth]{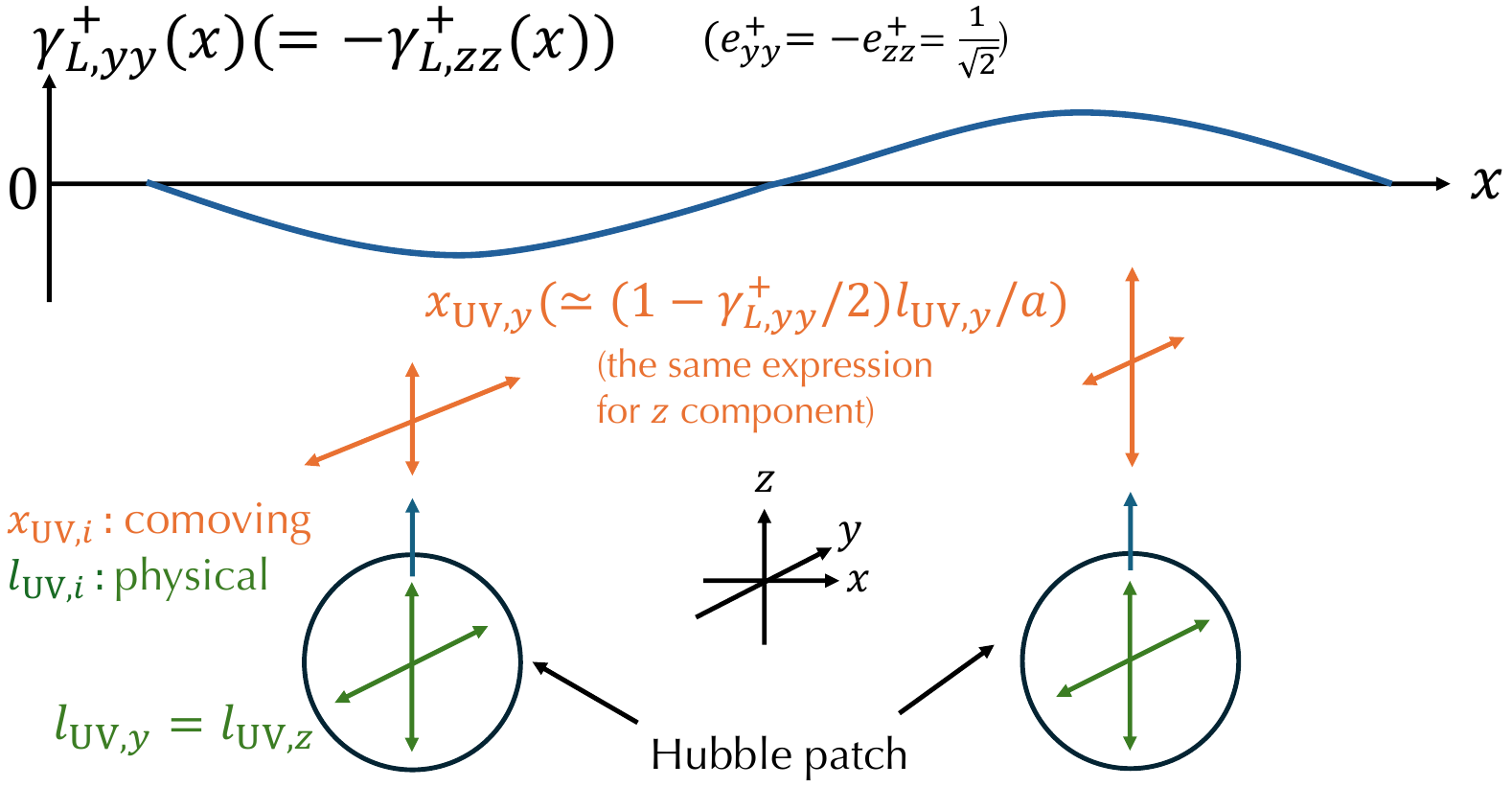}
        \caption{        
        Tensor perturbation version of Fig.~\ref{fig:l_uv_x_uv}.
        We show the situation where there is a plus mode of tensor perturbations with the $x$-axis as the longitudinal direction and $e^+_{yy} = -e^+_{zz} = 1/\sqrt{2}$.
        The lower region shows the comoving length, $x_{\uv,i}$, and the physical length, $l_{\uv,i}$, on the $yz$ plane ($i \in \{y,z\}$).
        Note that the $z$-axis of the upper region denotes the value of $\gamma^+_{L,yy}$, not the $z$-component.
    }
        \label{fig:l_uv_x_uv_tensor}
\end{figure}

As before, the fundamental physical prescription is to define the UV regulator using the local background spatial metric. For a general long-wavelength scalar and tensor background, the metric reads
\begin{equation}\label{metrictensor}
ds^2 \simeq -dt^2 + h^{\rm bg}_{ij}dx^idx^j,\qquad 
h^{\rm bg}_{ij}
=
a^2 e^{2\zeta_L(\bx)}
\left(e^{\gamma_L(\bx)}\right)_{ij},
\end{equation}
where $\gamma_{ij}$ denotes the transverse-traceless tensor perturbation
\begin{equation}
\partial_i\gamma_{ij}=0,
\qquad
\gamma_{ii}=0,
\end{equation}
and, as before, the subscript $L$ indicates the frozen perturbation outside the horizon.
A constant long tensor is locally (in each Hubble patch) removable by an anisotropic spatial redefinition.
The physical momentum corresponding to the local background metric is
\begin{equation}\label{condition}
k_{\rm phys}^2
=
h_{\rm bg}^{ij}k_i k_j
=
\frac{e^{-2\zeta_L}}{a^2}
\left(e^{-\gamma_L}\right)^{ij}
k_i k_j.
\end{equation}
The same physical UV cutoff in each Hubble patch is therefore imposed through
\begin{equation}
h_{\rm bg}^{ij}k_i k_j
<
\Lambda_{\rm UV}^2.
\end{equation}
As found previously, for a purely scalar long mode, i.e. $\gamma_L=0$, this reduces to a spherical comoving cutoff surface whose radius is different in each patch:
\begin{equation}
\Lambda[\zeta_L]=
a(t)\Lambda_{\rm UV} e^{\zeta_L},
\end{equation}
where, as before, $\Lambda[\ldots]$ denotes the cutoff on the magnitude of the comoving momentum.
With a tensor long mode included, inverting Eq.~\eqref{condition} gives
\begin{equation}
\Lambda[\zeta_L,\gamma_L,\hat{\bk}]=a(t) \Lambda_{\rm UV} e^{\zeta_L}\left(\hat{k}_i\hat{k}_j (e^{-\gamma_L})^{ij}\right)^{-1/2}.
\end{equation}
To first order in the tensor modes,
$\left(e^{-\gamma_L}\right)^{ij}
=
\delta^{ij}
-
\gamma_L^{ij}
+\mathcal O(\gamma_L^2)$, and the inhomogeneous cutoff becomes
\begin{equation}
\Lambda[\zeta_L,\gamma_L,\hat{\bk}]= 
a\Lambda_{\rm UV}e^{\zeta_L}
\left(
1+\frac{1}{2}
\gamma_L^{ij}
\hat{k}_i\hat{k}_j
+\cdots
\right).
\end{equation}
Thus, while a scalar long mode changes the radius of the cutoff sphere, a tensor long mode deforms it into an ellipsoid that differs in each Hubble patch. The corresponding comoving cutoff surface is therefore anisotropic. 
Figure~\ref{fig:l_uv_x_uv_tensor} shows the relation between the comoving and physical scales in the presence of the large-scale tensor perturbations, $\gamma_{L}$.

\subsection{Relation to anisotropic spatial rescaling symmetry}

The tensor counterpart of the scalar dilatation is associated with the symmetric traceless component $S_{ij}$ introduced previously in Sec.~\ref{dilatationsymmetry}, Eq.~\eqref{residualgauge}. Its leading generator and the corresponding transformation of the tensor perturbation are
\begin{equation}
\xi^i=S^i{}_j x^j,\qquad \delta_S\gamma_{ij}=2S_{ij}+ \mathcal O(S\gamma),
\end{equation}
where $S_{ij}=S_{ji}$, $S_{ii}=0$, and the coefficients are constant in space and time~\cite{Hinterbichler:2012nm,Hinterbichler:2013dpa}. The constant shift generates a homogeneous tensor perturbation, just as the scalar dilatation generates a homogeneous curvature perturbation.
Exponentiating the spatial generator gives a finite active anisotropic rescaling. Within each Hubble patch, we neglect the spatial variation of the background fields, so the transformed tensor background and comoving Fourier labels satisfy
\begin{equation}
e^{\widetilde\gamma_L}=e^S e^{\gamma_L}e^S,\qquad \widetilde{\mathbf k}=e^S\mathbf k.
\end{equation}
Here tildes denote the transformed quantities, with the spatial coordinates held fixed. Since $\det e^S=1$, the anisotropic rescaling changes the shape of the local spatial metric without changing its volume element. The constant scalar background $\zeta_L$, which determines this volume factor, is therefore unchanged. The Euclidean magnitude $\widetilde k=|\widetilde{\mathbf k}|$ and direction of the transformed comoving momentum are
\begin{equation}
\widetilde k=|e^S\hat{\mathbf k}|\,k,\qquad
\widehat{\widetilde{\mathbf k}}=\frac{e^S\hat{\mathbf k}}{|e^S\hat{\mathbf k}|}.
\end{equation}
Compatibility of the regulator with this symmetry requires
\begin{equation}
k<\Lambda[\zeta_L,\gamma_L,\hat{\mathbf k}]
\quad\Longleftrightarrow\quad
|e^S\hat{\mathbf k}|\,k
<
\Lambda[\zeta_L,\tilde{\gamma}_{L},\hat{\tilde{\mathbf k}}].
\end{equation}
The two integration boundaries must therefore obey
\begin{equation}
\Lambda[\zeta_L,\tilde{\gamma}_{L},\hat{\tilde{\mathbf k}}]
=
|e^S\hat{\mathbf k}|\,\Lambda[\zeta_L,\gamma_L,\hat{\mathbf k}].
\end{equation}
Choosing $S=-\gamma_L/2$, so that $\tilde{\gamma}_{L}=0$, the functional relation leads to
\begin{equation}
\Lambda[\zeta_L,\gamma_L,\hat{\mathbf k}]
=
a e^{\zeta_L}\Lambda_{\rm UV}
\left[\hat k_i(e^{-\gamma_L})^{ij}\hat k_j\right]^{-1/2}.
\end{equation}
Here the tensor dependence follows from the symmetry requirement on the integration domain, given the isotropic reference cutoff.
The directional dependence of the inhomogeneous cutoff thus implements the residual anisotropic spatial symmetry at leading order in long-wavelength gradients, analogously to the scalar dependence required by spatial dilatations. A tensor-independent spherical boundary would not transform consistently under this rescaling.

\subsection{Regulated tadpole source}
A one-loop tadpole leaving one tensor leg uncontracted produces a rank-two object that we denote by $\mathcal{T}_{ij}$.
Specifically, let us consider a cubic interaction with at least one tensor leg. This interaction can be written as $ H^{(3)}\supset \int d^3x\,\gamma_{ij}\,\mathcal O_{ij}$. 
Contracting the two fields inside $\mathcal O_{ij}$ leads to the amputated tadpole insertion, the object we aim to regularize with the inhomogeneous cutoff described in the previous section: 
\begin{equation}\label{tadpoletensor}
\mathcal{T}_{ij}[\gamma_L(\bx)] \equiv
\left\langle
\frac{\delta\mathcal{H}^{(3)}}
{\delta\gamma_{ij}}
\right\rangle_{\Lambda[\gamma_L,\hat{\bk}]}.
\end{equation}
To lighten the notation, we have omitted the subscript $A$ of $\mathcal T_{ij}$ labeling the uncontracted leg, as well as the regulator dependence on $\zeta_L$, which is irrelevant for the discussion in this section. Note that, after performing the angular integration in
Eq.~\eqref{tadpoletensor}, the position dependence of
$\mathcal{T}_{ij}$ enters only through $\gamma_L(\bx)$.
The dependence of $\Lambda$ on $\hat{\bk}$ implies that
the angular integration must be performed differently
in each patch, according to the local tensor background.
In a homogeneous and isotropic background, the amputated tadpole has the general form implied by rotational invariance,
\begin{equation}
\mathcal{T}_{ij}= \int d^3 k\,T_{ij}(\bk),\qquad T_{ij}(\mathbf{k})
=
A(k)\delta_{ij}
+
B(k)\hat{k}_i \hat{k}_j,
\end{equation}
where we have also omitted the time dependence.
In Fourier space, the effective Hamiltonian interaction is 
\begin{equation}
\label{eq:eff_hamil_tensor}
H_{\rm tad}^{\gamma}
=
\sum_\lambda \int \frac{d^3 q}{(2\pi)^3}\,
\gamma_{\mathbf q}^\lambda\,
e_{ij}^\lambda(\hat{\mathbf q})\,
\int d^3 x e^{i\bq\cdot\bx} \,\mathcal T_{ij}[\gamma_L(\bx)],
\end{equation}
where we have decomposed the tensor field as
\begin{align}\label{tensorexpansion}
    \gamma_{ij}(\mathbf x,\tau) = \int \frac{d^3 q}{(2\pi)^3} e^{i \mathbf q \cdot \mathbf x} \gamma_{ij,\mathbf q}(\tau) =\sum_{\lambda} \int \frac{d^3 q}{(2\pi)^3} e^{i \mathbf q \cdot \mathbf x} e^\lambda_{ij}(\hat {\bq}) \gamma^\lambda_{\mathbf q}(\tau),
\end{align}
$\lambda$ denotes the polarization, and we normalize the polarization tensors as 
\begin{align}
q^i e^\lambda_{ij}(\hat {\bq}) = 0, \quad e^{\lambda }_{ii}(\hat {\bq})=0, \quad e^{\lambda \, ij}(\hat {\bq}) e^{\lambda'*}_{ij}(\hat {\bq}) = \delta^{\lambda \lambda'}.
\end{align}
The Fourier mode of the tadpole source for the helicity state 
$\lambda$ is therefore given by 
\begin{equation}\label{Tsource}
\mathcal T_{\bq}^{\gamma,\lambda}=e_{ij}^\lambda(\hat{\mathbf q})\,
\int d^3 x e^{-i\bq\cdot\bx} \,\mathcal T_{ij}[\gamma_L(\bx)].
\end{equation}
With this, we can express the effective Hamiltonian interaction as 
\begin{align}
    \label{eq:h_tad_gamma}
    H_{\rm tad}^{\gamma}
=
\sum_\lambda \int \frac{d^3 q}{(2\pi)^3}\,
\gamma_{\mathbf q}^\lambda \mathcal T^{\gamma,\lambda}_{-\bq},
\end{align}
where this corresponds to the tensor version of Eq.~(\ref{eq:h_eff_tad_s}).

At zeroth order in $\gamma_L$, i.e. taking the homogeneous cutoff, the integration over the angular part inside the amputated tadpole is trivial and $\mathcal{T}_{ij}\propto \delta_{ij}$, which gives a zero contribution once contracted with the traceless polarization tensor. In other words,
without the dependence of $\mathcal T_{ij}$ on $\gamma_L$, the same statement follows directly from isotropy: the only available homogeneous rank-two tensor is proportional to $\delta_{ij}$, which is removed by the traceless projection. Thus the ordinary tensor tadpole vanishes,
\begin{equation}
\mathcal{T}^{\lambda\,(0)}_{\gamma}=0.
\end{equation}
This is the usual argument ensuring that the graviton tadpole vanishes, $\langle \gamma^\lambda_{\bq}\rangle=0$.
The relevant question for the inhomogeneous-cutoff construction is instead whether the response of the regulated tensor tadpole to a long-wavelength tensor background can be nonzero.
Let us consider the tensor amputated tadpole expanded around the homogeneous background as
\begin{equation}\label{expansion}
\mathcal{T}_{ij}[\gamma_L]
=
\mathcal{T}^{(0)}_{ij}
+
\mathcal{T}^{(1)\,kl}_{ij}\gamma^L_{kl}
+\cdots.
\end{equation}
Although the zeroth-order contribution vanishes after transverse-traceless projection, there is no symmetry reason forcing the first-order response to vanish. Indeed, the long tensor itself provides the spin-two structure required to construct a nonzero tensor response. Schematically, angular integrals of the form
\begin{equation}
\gamma_L^{kl}
\int d\Omega_{\hat{k}}\,
\hat{k}_i\hat{k}_j
\hat{k}_k\hat{k}_l\,
T(k)
\end{equation}
contain 
\begin{equation}
\int d\Omega_{\hat{k}}\,
\hat{k}_i\hat{k}_j
\hat{k}_k\hat{k}_l
= \frac{4\pi}{15}
(\delta_{ij}\delta_{kl}
+
\delta_{ik}\delta_{jl}
+
\delta_{il}\delta_{jk}),
\label{eq:angle_int}
\end{equation}
and, after contraction with the traceless tensor $\gamma_L^{kl}$, the first term vanishes while the remaining terms generate a contribution proportional to $
\gamma^L_{ij}$. Inserting Eq.~\eqref{expansion} in Eq.~\eqref{Tsource} thus leads to $\mathcal{T}_\gamma^{\lambda}\supset  e^\lambda_{ij}(\bq) \gamma_{L,ij} (-\bq)$.
Therefore, one may have
\begin{equation}
\mathcal{T}^{\lambda (0)}_{\gamma}=0,
\qquad
\mathcal{T}^{\lambda (1)}_{\gamma}[\gamma_L]\neq0.
\end{equation}
This response generates an effective quadratic interaction of the schematic form
\begin{equation}
    \label{eq:eff_h_tad2}
H_{\rm tad}^{\gamma(2)}
= \int d^3 x\,
\gamma_{ij}\,
\mathcal{T}_{ij}^{(1) kl} \gamma^L_{kl}
\propto
\int d^3 x\,
\gamma_{ij}\gamma_L^{ij}.
\end{equation}
The conceptual structure is therefore analogous to the scalar case, with an important difference. For scalars, the ordinary tadpole is generally nonzero and must first be canceled by the standard tadpole counterterm, while the $\zeta_L$-dependent response survives as an effective quadratic insertion. For tensors, isotropy and transverse-tracelessness force the ordinary tadpole to vanish, but the response of the regulated tadpole to a long tensor background can in principle be nonzero.

This suggests that a total-derivative boundary term appearing in the long-wavelength tensor power spectrum may have a regulator interpretation analogous to the scalar case, with the derivative of the regulator arising from the anisotropic deformation of the cutoff surface under the long tensor background. 
\subsection{Cancellation of UV boundary terms and superhorizon conservation - general proof for the tensor case}
We here see how the anisotropic cutoff contributions cancel the one-loop power spectrum of the superhorizon tensor perturbations, which shows the conservation of superhorizon tensor perturbations at one-loop level. 
This subsection is the tensor counterpart of Sec.~\ref{subsec:ge_pr_scalar}.
In the soft limit, the standard contributions involving two cubic vertices and one quartic vertex combine into a total derivative in loop momentum~\cite{Ema:2025ftj,Fang:2025hid}. This structure underlies the cancellation of the scale-invariant correction to superhorizon tensor power sourced by enhanced small-scale scalar fluctuations, previously claimed in Refs.~\cite{Ota:2022hvh,Ota:2022xni}. With a homogeneous momentum cutoff, however, the total derivative leaves a UV boundary term. We show that the additional contribution generated by the tensor dependence of the inhomogeneous cutoff cancels this term, providing a realization of the soft tensor cancellation within cutoff regularization.

We here show the concrete calculation. 
We consider the tensor perturbations induced by scalar perturbations during inflation. 
To focus on the inhomogeneous cutoff induced by $\gamma_L$, we neglect $\zeta_L$ in the following.
As in Ref.~\cite{Ema:2025ftj},
we use the quite generic scalar kinetic action,\footnote{Our notation is related to that of Ref.~\cite{Ema:2025ftj} by $h_{ij}=\gamma_{ij}$, with $\chi$ unchanged. For simplicity, we specialize their Eq.~(3.2) to $d=4$, $f(\tau)=a(\tau)$, and $c_s(\tau)=1$.}
\begin{equation}
 S_\chi=-\frac12\int d^4x\,\sqrt{-g}\, \left(g^{\mu\nu}\partial_\mu\chi\,\partial_\nu\chi +2 V(\chi)\right). 
\end{equation}
Using the metric specified in Eq.~\eqref{metrictensor} with $\zeta_L = 0$ gives $\sqrt{-g}=a^4$, and the interaction Hamiltonian relevant for scalar-tensor interaction only arises from the spatial-gradient action:
\begin{equation} 
S_{\chi,\mathrm{grad}} =-\frac12\int d\tau\,d^3 x\,a^2 (e^{-\gamma})^{ij}\partial_i\chi\,\partial_j\chi. 
\end{equation}
Expanding to second order in the tensor perturbation provides the cubic and quartic
interaction Hamiltonians relevant for the one-loop tensor power spectrum sourced by the scalar field $\chi$.  
Together, the two vertices are
\begin{equation}\label{interactions}
H_{\mathrm{int}}^{(3)}+H_{\mathrm{int}}^{(4)} =\int d^3 x\,a^2(\tau) \left[-\frac12\gamma^{ij} +\frac14\gamma^i{}_k\gamma^{kj}\right] \partial_i\chi\,\partial_j\chi.
\end{equation}
Each helicity mode is quantized in the standard way: $\gamma_{\bq}^{\lambda}(\tau)= \gamma_{q}(\tau)a^{\lambda}_{\bq}+\gamma^*_{q}(\tau)a^{\lambda\dagger}_{-\bq}$, where $[a^{\lambda}_{\bq},a^{\dagger\lambda'}_{\bq'}]=(2\pi)^3\delta^{\lambda\lambda'}\delta(\bq-\bq')$ and the tensor power spectrum at tree level is given by
\begin{equation}
\langle \gamma_{\bq}^\lambda(\tau)\gamma_{\bq'}^{\lambda'}(\tau) \rangle = (2\pi)^3\delta^{\lambda\lambda'}\delta(\bq+\bq')|\gamma_q(\tau)|^2.
\end{equation}
$\chi_k$ is given by the following relations:
\begin{align}
    \chi(\mathbf x) &= \int \frac{d^3 k}{(2\pi)^3} e^{i \mathbf k \cdot \mathbf x}\chi_{\mathbf k}, \\
    \expval{\chi_{\mathbf k} \chi_{\mathbf k'}} &= (2\pi)^3 \delta(\mathbf k + \mathbf k') |\chi_k|^2.
\end{align}
The cubic interaction, which is the one relevant for the regulated tadpole, corresponds to the equal-time quadratic composite source 
\begin{align}
    \mathcal O_{ij} = -\frac{a^2(\tau)}{2} \partial_i \chi \partial_j \chi.
\end{align}
To proceed, let us calculate explicitly the regulated two-point contraction, more precisely Eq.~\eqref{Tsource}, which is the tadpole source for the helicity state 
$\lambda$: 
\begin{align}\label{Tsource2}
\mathcal T_{\gamma,-\bq}^\lambda&=e_{ij}^\lambda(\hat \bq)\,
\int d^3 x e^{i\bq\cdot\bx} \,\langle \mathcal{O}_{ij}\rangle_{\Lambda[\gamma_L(\bx),\hat{\bk}]}\\[1mm]
 &= -e^{\lambda}_{ij}(\hat\bq)\int d^3 x e^{i\bq \cdot\bx} \frac{a^2}{2} \int d \Omega_{\hat k} \int^{
\Lambda
\left(
1+\frac{1}{2}
\gamma_L^{mn}
\hat{k}_m\hat{k}_n
+\cdots
\right)} \frac{k^2 d k}{(2\pi)^3} \hat k_i \hat k_j k^2 |\chi_k|^2,
\end{align}
where we have omitted the time argument and note again that $\Lambda = \Lambda[0] = a\Lambda_{\rm UV}$ is the homogeneous cutoff.
At zeroth order in the long-wavelength perturbation:
\begin{equation}
    \mathcal{T}^{\lambda\,(0)}_{\gamma,-\bq}  
    = -(2\pi)^3\delta(\bq)e^\lambda_{ij}(\hat \bq)\frac{a^2}{2}   \int^{\Lambda} \frac{k^2 d k}{(2\pi)^3} \int d \Omega_{\hat k} \hat k_i \hat k_j k^2 |\chi_k|^2 = 0.
\end{equation}
The equation above gives zero as expected, namely the graviton has no tadpole, because the integration over the angles is proportional to $\delta_{ij}$ which is then zero once contracted with any $e^\lambda_{ij}$.
Expanding the cutoff at first order in the tensor mode gives
\begin{equation}
    \mathcal{T}^{\lambda\,(1)}_{\gamma,-\bq}
    = -\frac{a^2}{2}  \Lambda^5 |\chi_{\Lambda}|^2 e^{\lambda}_{ij}(\hat{\bq})\frac{\gamma^{mn}_{L,-\bq}}{2 (2\pi)^3}\int d \Omega_{\hat k} \hat k_i \hat k_j \hat k_m \hat k_n = -\frac{a^2}{60 \pi^2}  \Lambda^5 |\chi_\Lambda|^2 \gamma^{\lambda}_{L,-\bq}, 
\end{equation}
where we have used Eq.~(\ref{eq:angle_int}) and neglected $\mathcal O(\gamma_L^2)$ to focus on the one-loop contributions.
From Eqs.~(\ref{eq:h_tad_gamma}) and (\ref{eq:eff_h_tad2}), the induced interaction Hamiltonian is given by 
\begin{align}
    H_{\rm tad}^{\gamma(2)}(\tau) &= -\int d^3 x \,\gamma_{ij}(\mathbf x,\tau) \gamma^{ij}_L(\mathbf x) \frac{a^2(\tau)}{60 \pi^2}  \Lambda^5(\tau) |\chi_{\Lambda}(\tau)|^2 \nonumber\\[1mm]
     &= -\sum_{\lambda}\int \frac{d^3 q}{(2\pi)^3}\, \gamma^{\lambda}_{\bq}\gamma^{\lambda}_{L,-\bq}\frac{a^2(\tau)}{60 \pi^2}  \Lambda^5(\tau) |\chi_{\Lambda}(\tau)|^2. 
\end{align}
This effective Hamiltonian leads to the following new contribution, 
\begin{align}
\langle
\gamma^\lambda_{\mathbf{p}}(\tau)
\gamma^{\lambda'}_{\mathbf{p}'}(\tau)
\rangle'_{\rm inh}
&= -i \delta^{\lambda \lambda'}\int^\tau_{\tau_\star} d \tau' [\gamma_{-\mathbf p}(\tau'), \gamma_{\mathbf p}(\tau)]' |\gamma_p|^2 \frac{a^2(\tau')}{30 \pi^2}  \Lambda^5(\tau') |\chi_{\Lambda}(\tau')|^2,
\label{eq:inh_tensor}
\end{align}
where we have expanded the tensor perturbation as 
in Eq.~\eqref{tensorexpansion}.
We note that, in Eq.~(\ref{eq:inh_tensor}), we used the fact that $|\gamma_p(\tau)|^2$ is independent of $\tau$ because we consider the limit of $|p\tau| \to 0$. 

Next, as found in Refs.~\cite{Ema:2025ftj,Fang:2025hid}, the one-loop result for the leading soft contribution from the cubic and quartic Hamiltonian interactions~\eqref{interactions} reduces to a total derivative term of the form:\footnote{We have multiplied Eq.~(3.8) in Ref.~\cite{Fang:2025hid} by 2 because we are interested in the total contribution, which can be expressed as $\expval{\gamma_{\mathbf p}^{(1)} \gamma_{\mathbf p'}^{(3)}} + \expval{\gamma_{\mathbf p}^{(3)} \gamma_{\mathbf p'}^{(1)}}$ with the notation in Ref.~\cite{Fang:2025hid}.}
\begin{equation}
\langle\gamma^\lambda_{\mathbf p}(\tau) \gamma^{\lambda'}_{\mathbf p'}(\tau)\rangle'_{(3)+(4)}  = \frac{i\,\delta^{\lambda\lambda'}|\gamma_p|^2}{30\pi^2} \int^\tau_{\tau_\star} d\tau'\,a^2(\tau')\, [\gamma_{-\mathbf p}(\tau'),\gamma_{\mathbf p}(\tau)]'\int dk\, \frac{d}{dk}\left( k^5|\chi_k(\tau')|^2\right).
\end{equation}
Integrating the total derivative leads to\footnote{The IR boundary term is negligible unless $|\chi_k|^2 \propto k^{n}$ with $n \leq -5$. Note that the scale-invariant power spectrum corresponds to $|\chi_k|^2 \propto k^{-3}$.} a UV boundary term which is equal in magnitude and opposite in sign to the contribution arising from the inhomogeneous cutoff just computed in Eq.~\eqref{eq:inh_tensor}. 

The UV boundary term is not retained in Refs.~\cite{Ema:2025ftj,Fang:2025hid}.
In particular, Ref.~\cite{Ema:2025ftj} sets aside the momentum boundary terms, noting that
the UV contribution should be treated with a symmetry-preserving
regularization, such as dimensional regularization, and that their
analysis concerns enhanced fluctuations at finite loop momenta.
In our finite-cutoff prescription, the UV boundary is instead
explicit and cancels the contribution induced by the inhomogeneous
cutoff. 

\section{Squeezed one-loop bispectrum}
\label{sec:cq-squeezed-bispectrum}
In this and in the next section, we discuss the one-loop bispectrum. See Refs.~\cite{Arkani-Hamed:2015bza,Chen:2016uwp,Chen:2016hrz} for early works on the one-loop bispectrum in the context of cosmological collider physics.

The dependence of the UV cutoff on the long-wavelength background also generates contributions to higher-point functions. For instance, contracting two fields at a quartic vertex and expanding the cutoff dependence of the contraction to first order in $\zetal$ produces a contribution to the one-loop squeezed bispectrum, represented by $\mathcal{I}^{(4)}_{1}$ in Eq.~\eqref{effectivequartic}. The same prescription applies to loops connecting two different interaction vertices.

Throughout this section, we consider the squeezed bispectrum
\begin{equation}
B(p,k)\equiv \langle \zeta_{\bp}\zeta_{\bk}\zeta_{\bk'}\rangle',
\qquad p\ll k,k'.
\label{eq:b_squeezed}
\end{equation}
We discuss the bispectrum with a general configuration in the next section.
When the soft leg is an adiabatic long mode, the leading squeezed bispectrum satisfies the consistency relation~\cite{Maldacena:2002vr,Creminelli:2004yq,Senatore:2012wy}\footnote{As elsewhere in the paper, we retain only the constant adiabatic
component of the external soft mode.}
\be\label{eq:d_k_def}
B(p,k)\simeq -P(p)D_k P(k),\qquad D_k\equiv 3+k\frac{\partial}{\partial k}.
\ee
At tree level, this relation follows from the residual spatial dilatation that locally removes the constant long mode. At loop level, the same argument must also account for the transformation of the regulator. In particular, incorrectly holding the comoving integration boundary fixed as the long background varies can leave an additional UV boundary term.

In this section, we aim to show that within cutoff regularization and using our inhomogeneous cutoff prescription, the one-loop consistency relation below is satisfied\footnote{For a related example of regulator-induced symmetry breaking in the one-loop tensor power spectrum, see Ref.~\cite{Ballesteros:2024cef}.}
\begin{equation}
 B_\Lambda^{(1)}(p,k)\simeq
 -P^{(0)}(p)D_kP_\Lambda^{(1)}(k)-P_\Lambda^{(1)}(p)D_kP^{(0)}(k).
 \label{eq:bw-one-loop-consistency}
\end{equation}
The superscripts denote the loop order, so that $P=P^{(0)}+P_\Lambda^{(1)}+\cdots$ and $B=B^{(0)}+B_\Lambda^{(1)}+\cdots$. Here the subscript $\Lambda$ indicates dependence on the reference cutoff $\Lambda=\Lambda[0]$ about which we expand the inhomogeneous cutoff. The one-loop quantities in Eq.~\eqref{eq:bw-one-loop-consistency} include the additional contributions generated by this expansion.
The first term describes the long-mode dependence of the one-loop hard
power spectrum. The second term describes the one-loop correction to the soft propagator
connecting the external long mode to the tree-level hard two-point response.

To establish this relation in general, it is convenient to first compute
the squeezed bispectrum using the background-wave method, reviewed in
Appendix~\ref{app:background-wave}. We therefore begin by recalling how
the method relates the tree-level consistency relation to dilatation
symmetry, and then show how the inhomogeneous cutoff preserves this
relation for regulated loops.
As a concrete example, we next introduce a
derivative-interaction model, explain its symmetry completion, and
verify the result by ordinary in-in perturbation theory, including
explicit massless de Sitter mode functions. The details
of the in-in calculation are in Appendix~\ref{app:cq-ordinary-inin}.
Note that, unlike in the previous sections (and the next section), we do not specify the time when the interactions are switched on, which means that the interactions can be on even before the external (non-loop) modes exit the horizon.

Before proceeding, we make an important remark on the scope of our result. Our prescription incorporates the dependence of the UV cutoff on the long-wavelength geometry, generating the additional contributions required by the consistency relation. We impose the tadpole condition, including the nonlinear completion of its counterterm, but do not carry out the full UV renormalization of the correlators.
Our calculation therefore establishes the consistency relation at finite cutoff, without asserting that the regulated correlators remain finite when the cutoff is removed. We leave for future work the explicit construction and inclusion of the additional counterterms needed to render the correlators UV finite while preserving the consistency relation.

\subsection{Dilatation symmetry and the regulated consistency relation: general argument}
\label{subsec:cq-general-proof}

A brute-force in-in proof of the one-loop consistency relation, including the additional diagrams generated by the inhomogeneous cutoff, risks missing the forest for the trees. Conversely, from the perspective of dilatation symmetry, the reason why our cutoff prescription preserves the relation is almost trivial: the consistency relation is a consequence of dilatation symmetry, which is preserved by our regulator. The relation therefore continues to hold at loop order.

As already stressed in Sec.~\ref{inoscalar}, in comoving gauge, the curvature perturbation enjoys a residual spatial dilatation symmetry, realized nonlinearly as a rescaling of its argument accompanied by a constant shift. A constant adiabatic long mode is therefore locally equivalent to a spatial rescaling.
The symmetry also fixes how spatial derivatives are accompanied by
powers of the undifferentiated field. For example, invariant kinetic
and gradient terms take the form
\begin{equation}
S\supset\int d\tau\,d^3x\,a^2\epsilon
\left[e^{3\zeta}\zeta'^2-e^\zeta(\partial_i\zeta)^2\right],
\end{equation}
where $\epsilon \equiv -\dot H/H^2$ with $H$ being the Hubble parameter and the dot denoting its time derivative.
These terms exhibit the coupling of a constant adiabatic long mode to the quadratic short-mode action, which is the coupling relevant for the leading tree-level squeezed bispectrum.
Let us now use the background-wave method 
\cite{Creminelli:2004yq,Ganc:2010ff,Senatore:2012wy}, reviewed in
Appendix~\ref{app:background-wave}, to derive this bispectrum. The idea
is to reduce the three-point calculation to that of the conditioned
power spectrum: the short-mode power spectrum evaluated at fixed
$\zetal$, treating the adiabatic long mode as a constant classical
background within the local patch. The leading squeezed bispectrum
follows by correlating the term linear in $\zetal$ with the external soft mode.
Splitting the field as $\zeta=\zetal+\short$, we take the part of the action above that is quadratic in $\short$
as the free action, keeping its full dependence on the constant
background $\zetal$. The short interaction-picture field then takes
the form
\begin{equation}
{\zeta}_S(\mathbf{x},\tau)
=\int\frac{d^3k}{(2\pi)^3}
\left[
{a}_{\bk}e^{i\mathbf{k}\cdot\mathbf{x}}u_{k,\zetal}(\tau)
+{a}_{\bk}^{\dagger}e^{-i\mathbf{k}\cdot\mathbf{x}}
u^*_{k,\zetal}(\tau)
\right],
\label{standard}
\end{equation}
where
\begin{equation}
u_{k,\zetal}(\tau)
=e^{-3\zetal/2}u_{ke^{-\zetal}}(\tau).
\label{eq:bw-mode-transformation0}
\end{equation}
Here $u_{k,\zetal}$ denotes the normalized short-mode function in the
presence of the constant long background $\zetal$, whereas
$u_{ke^{-\zetal}}$ is the usual mode function in the absence of the
long background, evaluated at momentum $ke^{-\zetal}$. See
Appendix~\ref{app:background-wave} for the derivation. We denote the
corresponding conditioned short-mode power spectrum by $P(k\mid\zetal)$.
At tree level, the relation between the mode functions in Eq.~\eqref{eq:bw-mode-transformation0} immediately gives
\begin{equation}
P^{(0)}(k\mid\zetal)
=e^{-3\zetal}P^{(0)}(ke^{-\zetal}).
\label{treelevelrelation}
\end{equation}
The background-wave method then obtains the leading squeezed bispectrum
by correlating the term linear in the background with an external long
mode:
\begin{align}
B^{(0)}(p,k)
&\simeq
\left\langle\zeta_{\mathbf p}
\left\langle\zeta_{\mathbf k_1}\zeta_{\mathbf k_2}
\right\rangle_{\zetal}\right\rangle'
\nonumber\\
&\simeq P^{(0)}(p)
\left.\partial_{\zetal}P^{(0)}(k\mid\zetal)\right|_{\zetal=0}
=-P^{(0)}(p)D_kP^{(0)}(k).
\label{eq:cq-tree-background-bispectrum}
\end{align}
Equation~\eqref{treelevelrelation} is thus sufficient to derive the
consistency relation at tree level.

At loop order, the question is whether the regulated integrals preserve
the same dilatation symmetry. We will show that, with the inhomogeneous
cutoff, the conditioned one-loop hard power spectrum obeys the same
scaling law as at tree level,
\begin{equation}
P_{\Lambda[\zetal]}^{(1)}(k\mid\zetal)
=e^{-3\zetal}P_\Lambda^{(1)}(ke^{-\zetal}),
\qquad
\left.\partial_{\zetal}
P_{\Lambda[\zetal]}^{(1)}(k\mid\zetal)\right|_{\zetal=0}
=-D_kP_\Lambda^{(1)}(k).
\label{eq:cq-general-covariant-scaling}
\end{equation}
Inserting this response into the one-loop analogue of
Eq.~\eqref{eq:cq-tree-background-bispectrum} gives the first term on
the right-hand side of Eq.~\eqref{eq:bw-one-loop-consistency}.
The second term comes from the one-loop correction to the soft power
spectrum multiplying the tree-level response of the hard power
spectrum. Together, these terms give the one-loop consistency relation.
We establish Eq.~\eqref{eq:cq-general-covariant-scaling} explicitly
for the two one-loop topologies that contribute to the power spectrum,
the quartic seagull and the cubic--cubic diagram. For each topology,
the relation follows without evaluating the time integrals.

Before continuing, let us consider the conditioned one-loop power
spectrum with the comoving cutoff held fixed as $\zetal$ varies.
This comparison identifies the additional contribution from the
inhomogeneous cutoff that we should recover in the explicit in-in
calculation. Applying Eq.~\eqref{eq:cq-general-covariant-scaling}
with this choice gives
\begin{equation}
P_\Lambda^{(1)}(k\mid\zetal)
=e^{-3\zetal}P_{e^{-\zetal}\Lambda}^{(1)}(ke^{-\zetal}).
\label{eq:cq-general-hom-scaling}
\end{equation}
Here $e^{-\zetal}\Lambda$ denotes a common rescaling of the comoving
cutoff boundary at every vertex time at which it is imposed.

For any regulated kernel $F_\Lambda$, we define its response to a
common rescaling of the cutoff boundaries by
\begin{equation}
\mathscr R_\Lambda F_\Lambda
\equiv
\left.\frac{\partial}{\partial\alpha}
F_{e^\alpha\Lambda}\right|_{\alpha=0},
\label{eq:cq-regulator-derivative}
\end{equation}
with the mode functions, vertex times, and external momenta held fixed.
For a single sharp boundary, for example,
\begin{equation}
\mathscr R_\Lambda\int dk\,\Theta(\cutoff-k)K(k)
=\int dk\,\cutoff\,\delta(\cutoff-k)K(k).
\label{eq:cq-boundary-derivative}
\end{equation}
Differentiating Eq.~\eqref{eq:cq-general-hom-scaling} gives
\begin{equation}
\left.\partial_{\zetal}P_\Lambda^{(1)}(k\mid\zetal)
\right|_{\zetal=0}
=-(D_k+\mathscr R_\Lambda)P_\Lambda^{(1)}(k).
\end{equation}
The extra boundary term $-\mathscr R_\Lambda P_\Lambda^{(1)}(k)$
is the violation of the dilatation relation for the conditioned hard
power spectrum regulated with a homogeneous cutoff.
To isolate the contribution that cancels this term, we denote by
$P_{\Lambda[\zetal]}^{(1)}(k)$ the power spectrum in which only the
cutoff depends on $\zetal$, with the fields and vertices evaluated
at zero background. The fully conditioned spectrum
$P_{\Lambda[\zetal]}^{(1)}(k\mid\zetal)$ also includes the background
dependence of the fields and vertices. Schematically,
\begin{equation}
P_{\Lambda[\zetal]}^{(1)}(k)-P_\Lambda^{(1)}(k)
=\vcenter{\hbox{\quarticPScutoffboundary}}
+\vcenter{\hbox{\cubicPScutoffboundary}}
=\zetal\,\mathscr R_\Lambda P_\Lambda^{(1)}(k)
+\mathcal O(\zetal^2).
\end{equation}
The crossed diagrams denote the full change in the cutoff boundary
for each loop topology. Correlating their terms linear in $\zetal$
with the external soft mode gives
\begin{align}
 B_{\mathrm{inh}}^{(1)}(p,k)
 &\simeq\left\langle\zeta_{\mathbf p}
    \left\langle\zeta_{\mathbf k_1}\zeta_{\mathbf k_2}
    \right\rangle_{\Lambda[\zetal]}^{(1)}\right\rangle'
 \nonumber\\
 &\simeq P^{(0)}(p)\left.\partial_{\zetal}
    P_{\Lambda[\zetal]}^{(1)}(k)\right|_0
 =P^{(0)}(p)\mathscr R_\Lambda P_\Lambda^{(1)}(k)
 =\vcenter{\hbox{\quarticBScutoffboundary}}+\vcenter{\hbox{\cubicBcutoffboundary}} .
\end{align}
This contribution cancels the extra UV boundary term in the squeezed
bispectrum computed with the homogeneous cutoff.

\subsection{How the loop diagrams transform}
\label{subsec:cq-background-interpretation}

We now prove Eq.~\eqref{eq:cq-general-covariant-scaling} for the two one-loop topologies entering the one-loop power spectrum, namely the seagull diagram and the cubic-cubic one-loop diagram.

We denote by $H^{(n)}_{\zetal}$ the $n$-short-field\footnote{The ``short-field'' here includes both the short external fields, $\zeta_S$, and the fields contributing to the loop integral.} interaction Hamiltonian
in the conditioned interaction picture. 
Recall that we are performing a conditioned in-in calculation. Here, $\zeta_{\bk}$ denotes the
short-mode interaction-picture field evolved with the quadratic
Hamiltonian at fixed $\zetal$. Its dependence on $\zetal$ is implicit
in the field notation and appears explicitly in the mode functions
$u_{k,\zetal}$.

\medskip
\noindent\textbf{The seagull diagram.}
We include the background dependence of the cutoff from the outset: 
\begin{equation}
 \begin{aligned}
 P_{\Lambda[\zetal],4}^{(1)}(k\mid\zetal)
 &=i\int_{-\infty}^{\tau_f}d\tau\,
 \left\langle
 [H^{(4)}_{\zetal}(\tau),\zeta_{\bk}\zeta_{-\bk}]
 \right\rangle_{\Lambda[\zetal]}'\\[1mm]
 &=\vcenter{\hbox{\quarticPSconditionedboundary}}
   +\vcenter{\hbox{\quarticPScutoffboundary}}.
 \end{aligned}
 \label{eq:cq-conditioned-seagull-complete}
\end{equation}
The gray drawing contains the conditioned fields and vertex at
fixed homogeneous boundary. The crossed drawing represents the
full additional contribution from replacing that boundary by
$\Lambda[\zetal]$. This decomposition holds to all orders in the background. For reference, the fixed-boundary drawing expands schematically as
\begin{equation}
 \begin{aligned}
 P_{\Lambda,4}^{(1)}(k\mid\zetal)=\vcenter{\hbox{\quarticPSconditionedboundary}}=\vcenter{\hbox{\quarticPSboundary}}
+\vcenter{\hbox{\quarticPSvertexbackground}}
+\vcenter{\hbox{\quarticPSloopbackground}}\quad+\left[
   \vcenter{\hbox{\quarticPShardbackground}}
   + \rm{permutation}
   \right]+\mathcal O(\zetal^2).
 \end{aligned}
 \label{eq:cq-conditioned-diagrams}
\end{equation}
Consider a quartic interaction with $N$ spatial derivatives. Dilatation symmetry fixes the dependence on the constant long-wavelength
background $\zetal$, so the interaction-picture Hamiltonian contains
terms of the form
\begin{equation}
H^{(4)}_{\zetal}(\tau)
\supset g_4(\tau)e^{(3-N)\zetal}
\int d^3x\,\mathcal O^{(N)}(\short).
\end{equation}
The factor $e^{3\zetal}$ comes from the spatial volume element, while
the $N$ spatial derivatives give the factor $e^{-N\zetal}$.
After angular integration, a representative Wick contraction
contributes
\begin{align}
P_{\Lambda[\zetal],4}^{(1)}(k\mid\zetal)
\supset 2\,\operatorname{Im} \Big[ \int_{-\infty}^{\tau_f}d\tau\,
&\kappa\,g_4(\tau)e^{(3-N)\zetal}\,
k^n\Sigma_{\Lambda[\zetal]}^{(m)}(\tau\mid\zetal)
\nonumber\\
&\times C_{k,\zetal}^{0c}(\tau_f,\tau)
C_{k,\zetal}^{0d}(\tau_f,\tau) \Big],
\label{eq:cq-seagull-factorized}
\end{align}
where we define $C^{ab}_{k,\zetal}$ and $\Sigma^{(m)}_{\Lambda[\zetal]}$ in Eqs.~(\ref{eq:c_ab_def}) and (\ref{eq:cq-general-seagull-contraction}), respectively.
Here $\kappa$ collects the numerical factors from the Wick
contractions and angular integration. We have defined the conditioned
Wightman contraction by\footnote{These contractions are the unordered two-point functions from which the time-ordered and retarded propagators entering the in-in calculation are constructed.}
\begin{equation}
    \label{eq:c_ab_def}
C_{k,\zetal}^{ab}(\tau_1,\tau_2)
\equiv u_{k,\zetal}^{(a)}(\tau_1)
       u_{k,\zetal}^{(b)*}(\tau_2),
\end{equation}
where $a$ and $b$ count time derivatives at the two endpoints.
Since the dilatation leaves the time arguments unchanged,
Eq.~\eqref{eq:bw-mode-transformation0} gives
\begin{equation}
C_{k,\zetal}^{ab}(\tau_1,\tau_2)
=e^{-3\zetal}C_{ke^{-\zetal}}^{ab}(\tau_1,\tau_2),
\label{eq:cq-conditioned-modes}
\end{equation}
where $C_k^{ab}$ denotes the contraction at zero background.
In Eq.~\eqref{eq:cq-seagull-factorized}, the two external fields are
undifferentiated, while the vertex legs to which they contract carry
$c$ and $d$ time derivatives. This gives the two contractions
$C_{k,\zetal}^{0c}$ and $C_{k,\zetal}^{0d}$. These two vertex legs
also carry $n$ spatial derivatives in total, giving the factor $k^n$
after angular integration. The remaining two vertex legs carry
$2m$ spatial derivatives, so $N=n+2m$. Their loop contraction is
\begin{equation}
\Sigma_{\Lambda[\zetal]}^{(m)}(\tau\mid\zetal)
\equiv
\int^{e^{\zetal}\cutoff(\tau)}
\frac{d^3q}{(2\pi)^3}\,
q^{2m}C_{q,\zetal}^{ab}(\tau,\tau),
\label{eq:cq-general-seagull-contraction}
\end{equation}
where the time-derivative orders $a,b$ on the two loop legs are
implicit in the notation for $\Sigma^{(m)}$.
Performing the rescaling $\mathbf Q=e^{-\zetal}\mathbf q$, the upper limit becomes
$\cutoff(\tau)$. The loop measure contributes $e^{3\zetal}$ and the
loop Wightman contraction contributes $e^{-3\zetal}$, leaving the
factor $e^{2m\zetal}$ from $q^{2m}$. Thus
\begin{equation}
\Sigma_{\Lambda[\zetal]}^{(m)}(\tau\mid\zetal)
=e^{2m\zetal}\Sigma_\Lambda^{(m)}(\tau),
\label{eq:cq-seagull-contraction-scaling}
\end{equation}
where $\Sigma_\Lambda^{(m)}$ is evaluated at zero background. Thus the contraction without spatial derivatives is invariant,
whereas the gradient contraction retains its momentum weight.
Writing the power spectrum in terms of $K=ke^{-\zetal}$, the factor $k^n$ contributes $e^{n\zetal}$,
and the two external Wightman contractions together contribute
$e^{-6\zetal}$. The net factor in Eq.~(\ref{eq:cq-seagull-factorized}) is therefore
\begin{equation}
e^{(3-N)\zetal}\,e^{n\zetal}\,e^{2m\zetal}\,e^{-6\zetal}
=e^{-3\zetal},
\qquad N=n+2m.
\label{eq:factor_product}
\end{equation}
The remaining integrand is the zero-background integrand evaluated at
external momentum $K$. Since the same factor holds for each
contraction, the full seagull contribution satisfies
\begin{equation}
P_{\Lambda[\zetal],4}^{(1)}(k\mid\zetal)
=e^{-3\zetal}P_{\Lambda,4}^{(1)}(ke^{-\zetal}).
\end{equation}

Finally, the cancellation supplied by the cutoff can be seen by separating
    the two linear contributions in $\Sigma_{\Lambda[\zetal]}$: 
\begin{equation}
 \begin{aligned}
 \Sigma_{\Lambda[\zetal]}^{(m)}(\tau\mid\zetal)
 =\Sigma_\Lambda^{(m)}(\tau)
   +\zetal(2m-\mathscr R_\Lambda)
      \Sigma_\Lambda^{(m)}(\tau)+\zetal\mathscr R_\Lambda
      \Sigma_\Lambda^{(m)}(\tau)+\mathcal O(\zetal^2).
 \end{aligned}
 \label{eq:cq-seagull-general-cutoff-cancellation}
\end{equation}
The second term comes from conditioning the propagator at fixed boundary. The third is the linear term of the full cutoff
contribution. Their boundary terms cancel for every $m$.
For a seagull contribution containing $\Sigma^{(0)}$, the inhomogeneous cutoff addition removes
the dependence on $\zetal$ of the bubble:
\begin{equation}
 \vcenter{\hbox{\quarticPSloopbackground}}
 +\underset{\text{linear in }\zetal}{
    \vcenter{\hbox{\quarticPScutoffboundary}}}
 =0.
 \label{eq:cq-seagull-loop-cutoff-cancellation}
\end{equation}
The remaining diagram factors are held at their zero-background
values in this equation. The two terms are respectively
$-\zetal\mathscr R_\Lambda P_{\Lambda,4}^{(1)}$ and
$+\zetal\mathscr R_\Lambda P_{\Lambda,4}^{(1)}$ for this contribution.

\medskip
\noindent\textbf{The cubic--cubic diagram.} We consider now the cubic-cubic contribution where we include the background dependence of the cutoff from the outset,
\begin{equation}
\begin{aligned}
P_{\Lambda[\zetal],33}^{(1)}(k\mid\zetal)
&=-\int_{-\infty}^{\tau_f}d\tau_1
     \int_{-\infty}^{\tau_1}d\tau_2
\left\langle
[H_{3,\zetal}(\tau_2),
[H_{3,\zetal}(\tau_1),\zeta_{\bk}(\tz)\zeta_{-\bk}(\tz)]]
\right\rangle_{\Lambda[\zetal]}'\\[1mm]
&=\vcenter{\hbox{\cubicPSconditionedboundary}}
 +\vcenter{\hbox{\cubicBcutoffboundaryPower}}.
\end{aligned}
\label{eq:cq-conditioned-33-complete}
\end{equation}
The crossed diagram represents the full
change of the two-line cutoff window, with all other factors still
evaluated at fixed $\zetal$. The fixed-boundary diagram expands as the one in Eq.~\eqref{eq:cq-conditioned-diagrams}.

Consider two cubic vertices with $N_1$ and $N_2$ spatial derivatives.
Dilatation symmetry fixes their dependence on the constant
long-wavelength background $\zetal$, so the interaction-picture
Hamiltonian contains terms of the form
\begin{equation}
H_{3,\zetal}(\tau_i)
\supset g_i(\tau_i)e^{(3-N_i)\zetal}
\int d^3x\,\mathcal O_3^{(N_i)}(\short),
\qquad i=1,2.
\label{cubicgenericscaling}
\end{equation}
Similarly to the quartic vertex, the factor $e^{3\zetal}$ comes from the
spatial volume element, and the $N_i$ spatial derivatives give
$e^{-N_i\zetal}$.
For the scaling argument, it is enough to write a representative cubic--cubic
contribution as
\begin{align}
P_{\Lambda[\zetal],33}^{(1)}(k\mid\zetal)
\sim
\int_{-\infty}^{\tau_f}d\tau_1
\int_{-\infty}^{\tau_1}d\tau_2\,
&g_1(\tau_1)g_2(\tau_2)e^{(6-N_1-N_2)\zetal}
\nonumber\\
&\times C_{k,\zetal}^{0c}(\tau_f,\tau_1)
C_{k,\zetal}^{0d}(\tau_f,\tau_2)
\mathcal K_{\Lambda[\zetal]}(k,\tau_1,\tau_2\mid\zetal).
\label{eq:cq-cubic-cubic-factorized}
\end{align}
Here $c$ and $d$ count the time derivatives on the vertex legs
contracted with the external fields. The sum over Wick contractions
and the complex conjugations required by the nested commutator are
implicit. The kernel $\mathcal K$ contains the two internal Wightman
contractions and all spatial-momentum factors from the vertices,
\begin{equation}
\begin{aligned}
\mathcal K_{\Lambda[\zetal]}(k,\tau_1,\tau_2\mid\zetal)
=\int \frac{d^3q}{(2\pi)^3}\,
& W_{\Lambda[\zetal]}(\mathbf k,\mathbf q,\tau_1,\tau_2)\,
\mathcal M(\mathbf k,\mathbf q)\\
&\times C_{q,\zetal}^{a_1a_2}(\tau_1,\tau_2)
 C_{|\mathbf k-\mathbf q|,\zetal}^{b_1b_2}(\tau_1,\tau_2).
\end{aligned}
\label{eq:cq-bg-two-line-kernel}
\end{equation}
The indices $a_i,b_i$ count the time derivatives on the two internal
legs at vertex $i$. The factor $\mathcal M$ collects the spatial momenta from both
vertices. Each spatial derivative contributes one power of momentum,
so $\mathcal M$ contains ${N_1+N_2}$ powers of momentum in total. For a two-time loop, the equal-time cutoff prescription must be
extended to the two-line kernel~\cite{Senatore:2009cf}. For example,
the earlier-time prescription bounds both internal momenta by
$\Lambda_{12}\equiv\cutoff(\min\{\tau_1,\tau_2\})$, so that
\begin{equation}
W_\Lambda(\mathbf k,\mathbf q,\tau_1,\tau_2)
=\Theta(\Lambda_{12}-q)\,
 \Theta(\Lambda_{12}-|\mathbf k-\mathbf q|).
\label{windowW}
\end{equation}
The inhomogeneous cutoff replaces both boundaries by
$e^{\zetal}\Lambda_{12}$. Writing
$\mathbf Q=e^{-\zetal}\mathbf q$ and
$\mathbf K=e^{-\zetal}\mathbf k$, we therefore have
\begin{equation}
W_{\Lambda[\zetal]}(\mathbf k,\mathbf q,\tau_1,\tau_2)
=W_\Lambda(\mathbf K,\mathbf Q,\tau_1,\tau_2).
\end{equation}
This transformation law is the only property of the window needed
for the scaling argument.
Under this rescaling, the loop measure contributes $e^{3\zetal}$,
the two internal Wightman contractions together contribute
$e^{-6\zetal}$, and $\mathcal M$ contributes $e^{(N_1+N_2)\zetal}$. The kernel thus transforms as
\begin{equation}
\mathcal K_{\Lambda[\zetal]}(k,\tau_1,\tau_2\mid\zetal)
=e^{(N_1+N_2-3)\zetal}
 \mathcal K_\Lambda(ke^{-\zetal},\tau_1,\tau_2).
\label{eq:cq-bg-bubble-scaling}
\end{equation}
Note that even in the absence of spatial derivatives ($N_1+N_2 = 0$), the bubble kernel $\mathcal K$ regulated
with the inhomogeneous cutoff generally retains a dependence on $\zetal$
through both the overall factor $e^{-3\zetal}$ and the rescaled external
momentum $ke^{-\zetal}$, whereas the corresponding seagull contraction without the spatial derivatives, 
$\Sigma^{(0)}$, is independent of $\zetal$.
The two external Wightman contractions in
Eq.~\eqref{eq:cq-cubic-cubic-factorized} supply a further factor
$e^{-6\zetal}$. Combining this with the two vertex coefficients
and the kernel in Eq.~(\ref{eq:cq-cubic-cubic-factorized}) gives
\begin{equation}
e^{(6-N_1-N_2)\zetal}\,
e^{(N_1+N_2-3)\zetal}\,
e^{-6\zetal}
=e^{-3\zetal}.
\end{equation}
The remaining integrand is the zero-background integrand evaluated at
external momentum $ke^{-\zetal}$. Since the same factor holds for
each contraction, the full cubic--cubic contribution satisfies
\begin{equation}
P_{\Lambda[\zetal],33}^{(1)}(k\mid\zetal)
=e^{-3\zetal}P_{\Lambda,33}^{(1)}(ke^{-\zetal}).
\end{equation}

\subsection{Cutoff insertions involving superhorizon hard modes}
\label{subsec:cq-additional-cutoff-insertions}
We highlight here an additional family of diagrams that arises only when the inhomogeneous cutoff is introduced. At finite cutoff, this family can modify both the power spectrum and the squeezed bispectrum. We show that the corresponding contributions satisfy the same consistency relation.

The label ``short'' in the squeezed configuration refers to the hierarchy
$p\ll k$ between the external momenta. It does not imply that the
$k$-mode is subhorizon. At sufficiently late times, the hierarchy can be
\begin{equation}
p\ll k\ll aH\ll q,
\qquad q\sim\Lambda(\tau)=a(\tau)\Lambda_{\rm UV}.
\end{equation}
Both external modes are then on superhorizon scales and approximately constant
within a local Hubble patch. The $k$-mode can therefore contribute to
the geometry seen by the UV loop, even though it is the hard
external mode of the squeezed correlator. 
Therefore, when a $\zeta_S$ mode belongs to the adiabatic superhorizon background sector at the relevant interaction time, it should also enter the cutoff: $\Lambda[\zetal+\short]=e^{\zetal+\short}\Lambda$. 
We then expand this dependence on $\short$ perturbatively and contract the resulting insertions with the external fields as usual. This generates the additional contributions we investigate here, i.e. diagrams contributing to the one-loop squeezed
bispectrum where one of the short external legs is generated by expanding the inhomogeneous cutoff.

In the in-in description, one contribution comes from expanding the cutoff in a loop connecting two cubic vertices. The cutoff supplies one hard external leg, while the soft leg and the other hard leg attach to the ordinary vertices. Another contribution comes from expanding the cutoff of a cubic tadpole to second order, supplying one soft and one hard external leg. 
In the background-wave calculation, we condition the theory of $\zeta_S$ on the external soft mode $\zeta_L$.\footnote{Note that the free theory is conditioned on the fixed
long-wavelength background $\zetal$. No additional conditioning
of the free theory on $\zeta_S$ is required. The dependence of the
cutoff on $\zeta_S$ is treated perturbatively, generating insertions
that are contracted with the external fields in the same conditioned
theory.}
Both previous diagrams can then be seen as generated from \textit{the conditioned inhomogeneous tadpole}. After the first cutoff expansion supplies a hard leg, expanding the conditioned loop or external propagator or each vertex in the soft background reproduces the cubic–cubic contribution. 
Expanding the cutoff once more instead gives the cross term involving one soft and one hard background field.
Diagrammatically, the two expansions giving rise to this loop contribution can be represented as\footnote{Here we retain the grey dot marking the conditioned propagator
even after expanding in the long mode, to distinguish insertions
generated by the propagator expansion from those generated by
the regulator expansion. 
}
\begin{equation}
  \icTadpoleConditioned\qquad \longrightarrow\qquad 
  \icTadpoleHardInsertion
 \qquad   \longrightarrow\qquad 
  \icTadpoleSoftFromLoop
  +
  \icTadpoleSoftFromCutoff.
  \label{eq:diagram_cut_cond}
\end{equation}
Note that, although we do not show here the diagrams with a leg from the background conditioning attached to a Wightman line or a vertex, the following discussion also includes such contributions.
Equivalently, in equation form:
\begin{equation}
\langle \langle\zeta_{\bk_1}\rangle_{\Lambda[\zetal+\short],\,\zeta_{L} }\zeta_{\bk_2}\zeta_{\bp}\rangle\simeq \langle \langle\zeta_{\bk_1} \zeta_{\bk_2}\rangle_{\rm inh, \,\Lambda[\zetal],\,\zeta_{L}}\zeta_{\bp}\rangle,
\end{equation}
where $\langle\zeta_{\bk_1} \zeta_{\bk_2}\rangle_{\rm inh, \,\Lambda[\zetal],\,\zeta_{L}}$ is the contribution to the one-loop power spectrum of the hard modes coming from the inhomogeneous cutoff expanded at first order in the short modes. We then keep $\Lambda[\zetal]$ and $\zetal$ in the subscript to mark the two different possible expansions in $\zetal$ from the cutoff and from the conditioned propagator respectively.

Let us now make this argument more explicit and determine how this
contribution scales with the long-wavelength background. Contracting two UV fields at a cubic vertex leaves one short
field factor $\short^A$ uncontracted. Here $A$ specifies the derivatives on
that factor. Suppressing the vertex coefficient, the resulting
conditional insertion has the schematic form
\begin{equation}
\short^A\,\Sigma_{\Lambda[\zetal+\short]}^{(m)}(\tau\mid\zetal).
\label{eq:zeta_s_a_sigma}
\end{equation}
The argument after the vertical bar specifies the background entering
the free mode functions. Expanding only the cutoff dependence in
$\short$, with $\zetal$ held fixed, gives
\begin{equation}
\begin{aligned}
\Sigma_{\Lambda[\zetal+\short]}^{(m)}(\tau\mid\zetal)
={}&\Sigma_{\Lambda[\zetal]}^{(m)}(\tau\mid\zetal)
 +\short\,\mathscr R_\Lambda
  \Sigma_{\Lambda[\zetal]}^{(m)}(\tau\mid\zetal)
 +\mathcal O(\short^2),
\end{aligned}
\label{eq:cq-extra-tadpole-short-expansion}
\end{equation}
where the operator $
\mathscr R_\Lambda F_\Lambda
$ defined in Eq.~\eqref{eq:cq-regulator-derivative}
varies the cutoff boundaries with the fields, vertex times, and external
momenta held fixed.

The first term in Eq.~\eqref{eq:cq-extra-tadpole-short-expansion}
is a tadpole insertion with one surviving short field. Multiplying
the second term by $\short^A$ gives two short-field factors,
$\short^A \short$. It thus contributes to the conditioned hard power
spectrum. We denote this contribution by
$P_{\Lambda[\zetal],3,\mathrm{inh}}^{(1)}(k\mid\zetal)$, and its
zero-background value by $P_{\Lambda,3,\mathrm{inh}}^{(1)}(k)$.
It is one loop because the only loop comes from the original contracted pair.
Expanding its dependence on $\zetal$ subsequently produces the
$\zetal \short^A \short$ terms relevant to the squeezed bispectrum.
This entire family contains a background-dependent cutoff insertion
and is absent if that regulator dependence is switched off.

Let us explicitly see the transformation of the additional diagrams.
Consider a cubic interaction with $N$ spatial derivatives. At fixed
$\zetal$, its Hamiltonian contains a term of the form
\begin{equation}
H^{(3)}_{\zetal}(\tau)
\supset g_3(\tau)e^{(3-N)\zetal}
\int d^3x\,\mathcal O_3^{(N)}.
\end{equation}
For a representative nonvanishing contraction, let $2m$ spatial
derivatives act on the UV pair and $n$ on the surviving short
leg, so $N=2m+n$. Its loop contraction is $\Sigma_{\Lambda[\zetal]}^{(m)}(\tau\mid\zetal)
$ defined in Eq.~\eqref{eq:cq-general-seagull-contraction}.
The additional power contribution consequently has the schematic form
\begin{align}
P_{\Lambda[\zetal],3,\mathrm{inh}}^{(1)}(k\mid\zetal)
\sim\int_{-\infty}^{\tau_f}d\tau\,
&g_3(\tau)e^{(3-N)\zetal}k^n
\mathscr R_\Lambda\Sigma_{\Lambda[\zetal]}^{(m)}(\tau\mid\zetal)
 C_{k,\zetal}^{0c}(\tau_f,\tau)
C_{k,\zetal}^{00}(\tau_f,\tau).
\label{eq:cq-extra-power-factorized}
\end{align}
The
conditioned Wightman contractions satisfy Eq.~\eqref{eq:cq-conditioned-modes}.
Here $c$ counts time derivatives on the surviving vertex leg, which corresponds to $\zeta^A_S$ in Eq.~\eqref{eq:zeta_s_a_sigma}. The two
external contractions attach to that leg and to the cutoff insertion,
respectively. The symbol $\sim$ suppresses numerical factors,
commutator combinations, complex conjugations, and permutations.
Each term in the full in-in expression has the same scaling.
Changing the loop variable to $\mathbf Q=e^{-\zetal}\mathbf q$ gives 
\begin{equation}
\Sigma_{\Lambda[\zetal]}^{(m)}(\tau\mid\zetal)
=e^{2m\zetal}\Sigma_\Lambda^{(m)}(\tau),
\qquad
\mathscr R_\Lambda\Sigma_{\Lambda[\zetal]}^{(m)}(\tau\mid\zetal)
=e^{2m\zetal}\mathscr R_\Lambda\Sigma_\Lambda^{(m)}(\tau).
\end{equation}
The cutoff derivative preserves the scaling weight because it acts
only on the reference boundary. Writing $k=e^{\zetal}K$, the factors in
Eq.~\eqref{eq:cq-extra-power-factorized} combine as
\begin{equation}
e^{(3-N)\zetal}\,e^{n\zetal}\,e^{2m\zetal}\,e^{-6\zetal}
=e^{-3\zetal},
\qquad N=n+2m.
\end{equation}
The remaining integrand is evaluated at $K$. Hence
\begin{equation}
P_{\Lambda[\zetal],3,\mathrm{inh}}^{(1)}(k\mid\zetal)
=e^{-3\zetal}P_{\Lambda,3,\mathrm{inh}}^{(1)}(ke^{-\zetal}).
\label{eq:cq-extra-power-scaling}
\end{equation}
Differentiating this equation and correlating
with the external soft mode gives the consistency condition for this part of the squeezed bispectrum.

\subsection{An explicit example: Squeezed bispectrum from ordinary in-in diagrams}
\label{subsec:cq-power}
We now verify the general argument in an explicit model. The action and Hamiltonians used for the power spectrum are written in terms of the unsplit field $\zeta$. Consider the seed action 
\begin{equation}
 S_\text{seed}[\zeta]=\int d\tau\,d^3x\left\{a^2\epsilon[\zeta'^2-(\partial\zeta)^2]-\frac{g_4}{24}\zeta'^4\right\}. \label{eq:cq-action}
\end{equation}
The seed specifies the quadratic theory and a selected quartic interaction. A simple completion respecting the dilatation symmetry is
\begin{equation}
 S[\zeta]=\int d\tau\,d^3x\left\{a^2\epsilon[e^{3\zeta}\zeta'^2-e^\zeta(\partial\zeta)^2]-\frac{g_4}{24}e^{3\zeta}\zeta'^4\right\}. \label{eq:cq-completed-action}
\end{equation}
The exponentials supply the spatial-volume weights explained above and fix the additional cubic, quartic, and quintic vertices needed for computing the one-loop bispectrum.\footnote{This is a simple dilatation-symmetry-preserving completion of the retained operators. Dilatation symmetry alone does not uniquely specify a full inflationary action.}
We present the diagrams and results below.  
See Appendix~\ref{app:cq-ordinary-inin} for the details.

\subsubsection*{Contribution proportional to $g_4$}
The one-loop correction to the power spectrum can be represented as
\begin{equation}
 P_{\Lambda,4}^{(1,g_4)}(k)=i\int_{-\infty}^{\tau_f}d\tau\,\langle[H^{(4,g_4)}(\tau),\zeta_{\mathbf k}(\tau_f)\zeta_{-\mathbf k}(\tau_f)]\rangle'=\vcenter{\hbox{\quarticPSboundary}}.
 \end{equation}
We denote here $P_4(k)\equiv P_{\Lambda,4}^{(1,g_4)}(k)$ for simplicity. Let $B_5$ denote the quintic vertex attachment and $B_{34}^{\rm loop}$ and $B_{34}^{\rm hard}$ the cubic--quartic attachments to the internal pair and external hard lines, respectively. The ordinary diagrams give 
\begin{align}
 B_5&=\vcenter{\hbox{\quarticBSvertexboundary}}=-9P^{(0)}(p)P_4(k),\qquad B_{34}^{\rm loop}=\vcenter{\hbox{\quarticBSloopboundary}}=P^{(0)}(p)(6-\mathscr R_\Lambda)P_4(k),\nonumber\\[1mm] B_{34}^{\rm hard}&=\vcenter{\hbox{\quarticBShardboundary}}=-P^{(0)}(p)k\partial_kP_4(k). \label{eq:cq-main-g4-diagrams}
\end{align}
Thus their sum is $-P^{(0)}(p)(D_k+\mathscr R_\Lambda)P_4(k)$. The additional contribution follows by expanding the cutoff in the unconditioned contraction:
\begin{equation}
 \Sigma_{\Lambda[\zetal]}(\tau)=\int^{e^{\zetal}\cutoff(\tau)}\frac{d^3q}{(2\pi)^3}|u'_q(\tau)|^2=\Sigma_\cutoff(\tau)+\zetal\mathscr R_\Lambda\Sigma_\cutoff(\tau)+\mathcal O(\zetal^2). \label{eq:cq-cutoff-expansion}
\end{equation}
Correlating the term linear in $\zetal$ with the external soft
mode gives the additional contribution
\begin{equation}
 B_{4,\mathrm{cut}}^{(1,g_4)}
 =\vcenter{\hbox{\quarticBScutoffboundary}}
 =P^{(0)}(p)\mathscr R_\Lambda P_4(k).
 \label{eq:cq-cutoff-bispectrum}
\end{equation}
Here the crossed drawing denotes the term linear in the soft field, contracted with the external soft mode.
Adding this contribution cancels the boundary term left by
the ordinary diagrams, giving
\begin{equation}
 B_{4,\mathrm{total}}^{(1,g_4)}=B_5+B_{34}^{\rm loop}+B_{34}^{\rm hard}+B_{4,\mathrm{cut}}^{(1,g_4)}=-P^{(0)}(p)D_kP_4(k). \label{eq:cq-g4-completed-bispectrum}
\end{equation}
Using the de Sitter mode functions, the coefficients in units of $P^{(0)}(p)P_4(k)$ are
\begin{equation}
 \begin{array}{c@{\qquad}c@{\qquad}c@{\qquad}c} \vcenter{\hbox{\quarticBSvertexboundary}}&\vcenter{\hbox{\quarticBSloopboundary}}&\vcenter{\hbox{\quarticBShardboundary}}&\vcenter{\hbox{\quarticBScutoffboundary}}\\[1mm] -9&+2&+3&+4 \end{array}. \label{eq:cq-completed-g4-ds}
\end{equation}
See Appendix \ref{app:cq-ordinary-inin} for the explicit calculation with the de Sitter mode functions. 
The external-line drawing includes both hard attachments. The four displayed coefficients sum to zero, $-9+2+3+4=0$: completing the action fixes the ordinary result $-4$, and the cutoff supplies the remaining $+4$. The leading squeezed coefficient vanishes, while the power correction remains nonzero.\footnote{Note that the separate coefficients differ from those in the background-wave
decomposition above. Here we use the mode functions not conditioned on $\zetal$, whereas the earlier discussion uses derivatives
of the conditioned fields. For derivative interactions, the relation
between these operators redistributes contributions among the vertex,
loop, and external-line attachments, as explained in
Appendix~\ref{app:cq-line-insertions}.
Consequently, $B_{34}^{\rm loop}$ and $B_{4,\mathrm{cut}}^{(1,g_4)}$
do not cancel separately in this organization.
With the de Sitter modes, the respective vertex, loop, external-line,
and cutoff coefficients in the canonical and conditioned-field
organizations give the same total:
\begin{equation}
 -9+2+3+4=3-4-3+4=0.
\end{equation}
}

\subsubsection*{Contributions independent of $g_4$}
The completed quadratic seed gives a quartic seagull and a cubic--cubic power spectrum correction 
\begin{equation}
 P_\Lambda^{(1,g_4^0)}(k)=\vcenter{\hbox{\quarticPSboundary}}+\vcenter{\hbox{\cubicPSboundary}}=P_{\Lambda,4}^{(1,g_4^0)}(k)+P_{\Lambda,33}^{(1)}(k). \label{eq:cq-power-zero-decomposition}
\end{equation}
Their explicit integrals are given in Appendix~\ref{app:cq-ordinary-inin}. Below, we write $P_4=P_{\Lambda,4}^{(1,g_4^0)}=P_{4,K}+P_{4,\partial}$ and $P_{33}=P_{\Lambda,33}^{(1)}$. The subscripts $K$ and $\partial$ refer to the kinetic and gradient quartic vertices, associated with the $\zeta^2 (\zeta')^2$ and $\zeta^2 (\partial \zeta)^2$ terms in the Hamiltonian, respectively.

\medskip
\noindent\textbf{Quartic seagull.}
The ordinary vertex and line attachments give
\begin{align}
 B_{4,\mathrm{vertex}}^{(1,g_4^0)}&=\vcenter{\hbox{\quarticBSvertexboundary}}=P^{(0)}(p)(-3P_{4,K}+P_{4,\partial}),\nonumber\\[1mm] B_{4,\mathrm{lines}}^{(1,g_4^0)}&=\vcenter{\hbox{\quarticBSloopboundary}}+\vcenter{\hbox{\quarticBShardboundary}}\nonumber\\ &=-P^{(0)}(p)[(k\partial_k+\mathscr R_\Lambda)P_{4,K}+(4+k\partial_k+\mathscr R_\Lambda)P_{4,\partial}]. \label{eq:cq-main-zero-attachments}
\end{align}
Their sum is $-P^{(0)}(p)(D_k+\mathscr R_\Lambda)P_4$. The inhomogeneous cutoff adds
\begin{equation}
 \label{eq:b_4_cut}
B_{4,\mathrm{cut}}^{(1,g_4^0)}=\vcenter{\hbox{\quarticBScutoffboundary}}=P^{(0)}(p)\mathscr R_\Lambda P_4,\end{equation}
 which thus implies
 \begin{equation}
 B_{4,\mathrm{total}}^{(1,g_4^0)}=-P^{(0)}(p)D_kP_4. \label{eq:cq-cutoff-quartic-zero}
\end{equation}

\medskip
\noindent\textbf{Cubic--cubic loop.}
Decompose $P_{33}$ into terms $P_{33}^{[n_K,n_\partial]}$ with $n_K$ kinetic and $n_\partial$ gradient cubic vertices, $n_K+n_\partial=2$. The two families of ordinary attachments are (permutations understood):
\begin{align}
 B_{33,\mathrm{vertex}}^{(1)}&=\vcenter{\hbox{\cubicBvertexboundary}}=P^{(0)}(p)\sum_{n_K+n_\partial=2}(-3n_K+n_\partial)P_{33}^{[n_K,n_\partial]},\nonumber\\[1mm] B_{33,\mathrm{lines}}^{(1)}&=\vcenter{\hbox{\cubicBloopboundary}}+\vcenter{\hbox{\cubicBhardboundary}}\nonumber\\ &=P^{(0)}(p)\sum_{n_K+n_\partial=2}[6n_K-9+2n_\partial-k\partial_k-\mathscr R_\Lambda]P_{33}^{[n_K,n_\partial]}. \label{eq:cq-main-33-diagrams}
\end{align}
Using \(n_K+n_\partial=2\), the constant parts of the vertex and line coefficients sum to \(-3\) in every kinetic/gradient combination, giving $B_{33,\mathrm{vertex}}^{(1)} + B_{33,\mathrm{lines}}^{(1)}=-P^{(0)}(p)(D_k+\mathscr R_\Lambda)P_{33}$. The contribution from the inhomogeneous cutoff provides 
\begin{equation}
 B_{33,\mathrm{inh}}^{(1)}=\vcenter{\hbox{\cubicBcutoffboundary}}=P^{(0)}(p)\mathscr R_\Lambda P_{33},\label{eq:cq-cutoff-cubic-cubic}
 \end{equation}
therefore
\begin{equation}
 B_{33,\mathrm{total}}^{(1)}=-P^{(0)}(p)D_kP_{33}.\label{eq:cq-completed-cubic-cubic}
\end{equation}
The hard-loop families therefore supply $-P^{(0)}(p)D_kP_\Lambda^{(1)}(k)$. The soft-propagator correction supplies the second term of Eq.~\eqref{eq:bw-one-loop-consistency}, $-P_\Lambda^{(1)}(p)D_kP^{(0)}(k)$. Its value is not needed here. In the de Sitter approximation, we find $D_kP^{(0)}(k)=0$. A brief additional check of the undifferentiated canonical soft source appears at the end of Appendix~\ref{app:cq-ordinary-inin}.

\section{Late-time constancy of the one-loop bispectrum}
\label{sec:bispectrum}

In this section, we discuss the one-loop corrections to the bispectrum of superhorizon curvature perturbations. 
With a simplifying assumption, we see how the inhomogeneous cutoff contributions cancel some parts of the one-loop bispectrum for a general configuration, which can be different from the squeezed configuration, discussed in the previous section.
Similarly to what we have seen in Sec.~\ref{sec:one_loop_scalar_power}, this cancellation is associated with the late-time constancy of the bispectrum.  
Specifically, we consider 
\begin{align}
    B(p_1,p_2,p_3) \equiv \expval{Q_3(\bp_1, \bp_2, \bp_3)}', \qquad Q_3(\bp_1, \bp_2, \bp_3) \equiv \zeta_{\bp_1} \zeta_{\bp_2} \zeta_{\bp_3},
\end{align}
where we do not impose the squeezed configuration for $\bp_i$ throughout this section. 

\begin{figure}[t]
  \centering
  \begin{tikzpicture}[
    x=1cm, y=1cm,
    external/.style={draw=black, line width=0.7pt},
    internal/.style={draw=black, line width=0.7pt},
    boundary/.style={draw=black, line width=0.7pt},
    line cap=round,
    line join=round
  ]

    % ===================== 1-PI =====================
    \node[anchor=west] at (-1.25,0.35) {1-PI};

    % Three cubic vertices: three attachments to one loop.
    \begin{scope}
      \coordinate (O) at (0,-1.05);
      \coordinate (A) at
        ({0.48*cos(140)},{-1.05+0.48*sin(140)});
      \coordinate (B) at (0,-0.57);
      \coordinate (C) at
        ({0.48*cos(40)},{-1.05+0.48*sin(40)});

      \draw[boundary] (-0.8,0) -- (0.8,0);
      \draw[external] (-0.70,0) -- (A);
      \draw[external] (0,0) -- (B);
      \draw[external] (0.70,0) -- (C);
      \draw[internal] (O) circle[radius=0.48];
      \foreach \v in {A,B,C}
  \fill (\v) circle[radius=1.1pt];
    \end{scope}

    % Cubic + quartic: two legs meet at one loop vertex.
    \begin{scope}[xshift=3cm]
      \coordinate (O) at (0,-1.05);
      \coordinate (A) at
        ({0.48*cos(125)},{-1.05+0.48*sin(125)});
      \coordinate (B) at
        ({0.48*cos(55)},{-1.05+0.48*sin(55)});

      \draw[boundary] (-1.1,0) -- (0.9,0);
      \draw[external] (-0.95,0) -- (A) -- (0.10,0);
      \draw[external] (0.75,0) -- (B);
      \draw[internal] (O) circle[radius=0.48];
      \foreach \v in {A,B}
  \fill (\v) circle[radius=1.1pt];
    \end{scope}

    % Quintic: all three legs meet at the tadpole vertex.
    \begin{scope}[xshift=6cm]
      \coordinate (A) at (0,-0.57);

      \draw[boundary] (-0.9,0) -- (0.9,0);
      \draw[external] (-0.80,0) -- (A) -- (0.80,0);
      \draw[external] (0,0) -- (A);
      \draw[internal] (0,-1.05) circle[radius=0.48];
      \fill (A) circle[radius=1.1pt];
    \end{scope}

    % ===================== 1-PR =====================
    \node[anchor=west] at (-1.25,-2.25) {1-PR};
    % Bubble inserted in the right external branch.
    \begin{scope}[shift={(0.7,-2.6)}]
      \coordinate (V) at (0,-1.65);
      \fill (V) circle[radius=1.1pt];

      \draw[boundary] (-1.,0) -- (1.,0);
      \draw[external] (-0.80,0) -- (V) -- (0,0);

      % Local y-axis follows the inclined right branch.
      \begin{scope}[shift={(V)},rotate=-25]
        \draw[external] (0,0) -- (0,0.70);
        \draw[external] (0,1.30) -- (0,1.8206);
        \draw[internal] (0,1)
          ellipse[x radius=0.32,y radius=0.30];
          \fill (0,0.70) circle[radius=1.1pt];
\fill (0,1.30) circle[radius=1.1pt];
      \end{scope}
    \end{scope}
    % Tadpole attached at one vertex of the right branch.
    \begin{scope}[shift={(5.3,-2.6)}]
      \coordinate (V) at (0,-1.65);
      \fill (V) circle[radius=1.1pt];

      \draw[boundary] (-1.,0) -- (1.,0);
      \draw[external] (-0.80,0) -- (V) -- (0,0);

      \begin{scope}[shift={(V)},rotate=-25]
        \draw[external] (0,0) -- (0,1.8206);
        \draw[internal] (0.34,0.95) circle[radius=0.34];
        \fill (0,0.95) circle[radius=1.1pt];
      \end{scope}
    \end{scope}
  \end{tikzpicture}
  \caption{The 1-PI and 1-PR diagrams for the one-loop bispectrum.}
  \label{fig:1pi_1pr}
\end{figure}
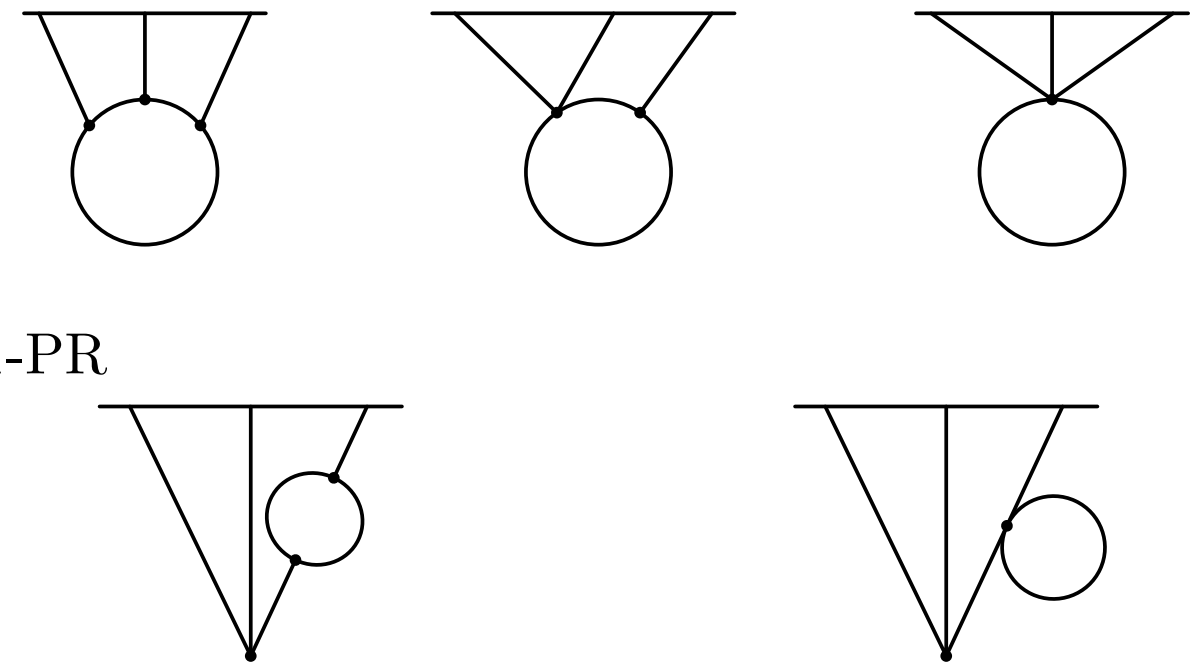

In this section, we restrict the scope of our calculation and make one simplifying assumption:

\begin{itemize}
\item
We focus on a subset of the one-loop contributions.
Figure~\ref{fig:1pi_1pr} shows the one-particle-irreducible (1-PI) and
1-PR diagrams contributing to the one-loop bispectrum.
For the 1-PR diagrams, we expect the inhomogeneous cutoff to play the same role
as it does in the two-point function, as discussed in
Sec.~\ref{sec:one_loop_scalar_power}.
We therefore focus primarily on the 1-PI diagrams shown in
Fig.~\ref{fig:1pi_1pr}.

\item
We assume that the one-loop bispectrum has a structure similar to that of
the one-loop power spectrum. In particular, we assume that some contributions
to the one-loop bispectrum satisfy the consistency relation, as do some
contributions to the one-loop power spectrum.
We then examine how the terms associated with the consistency relation
cancel against the inhomogeneous cutoff contributions.
A complete calculation of the one-loop bispectrum is beyond the scope of
this paper.
\end{itemize}
Throughout this section, we use the shorthand notations for the interactions:
\begin{align}
    h = - \mathcal L_3, \ l = -\mathcal L_4, \ m = - \mathcal L_5,
\end{align}
where $\mathcal L_n$ denotes the $n$-th order in $\zeta$ and can include time or spatial derivatives of $\zeta$.
We express the interaction Hamiltonian as 
\begin{align}
    \mathcal H^{(3)} = h, \ \mathcal H^{(4)} = l + \mathcal H^{(4)}_\ind, \ \mathcal H^{(5)} = m + \mathcal H^{(5)}_\ind,
\end{align}
where $H_\ind$ is the induced Hamiltonian associated with $\zeta'$ in $h$ or $l$.
See Appendix~\ref{app:induced_hamiltonian} for the explicit expressions of $\mathcal H_\ind$.

\subsection{Contributions with the first commutator involving the cubic interaction}
First, we focus on the one-loop contributions for which the first commutator involves the cubic interaction,\footnote{Here, ``first'' refers to the first commutator taken with the external operators. The cubic Hamiltonian in this commutator has the latest time argument.} which contribute to both the left and the middle 1-PI diagrams in Fig.~\ref{fig:1pi_1pr}.
There are two contributions in this category. One comes with three cubic interactions, and the other comes with a cubic and a quartic interaction.
The former is given by 
\begin{align}
    B_{333} &= -i\int^{\tau_f}_{\tau_\star} d \tau_1 \int^{\tau_1}_{-\infty} d\tau_2 \int^{\tau_2}_{-\infty} d\tau_3  \expval{[H^{(3)}(\tau_3),[H^{(3)}(\tau_2),[H^{(3)}(\tau_1),Q_3]]]}' \nonumber \\ 
    &= -i\int^{\tau_f}_{\tau_\star} d \tau_1 \int^{\tau_1}_{-\infty} d\tau_2 \int^{\tau_2}_{-\infty} d\tau_3  \nonumber \\ 
    &\qquad \times \sum_{\text{cyc(1,2,3)}} \sum_A [\zeta^A_{-\bp_1}(\tau_1),\zeta_{\bp_1}]' \expval{[H^{(3)}(\tau_3),[H^{(3)}(\tau_2),h_{,A}|_{\tau_1,\bp_1}\zeta_{\bp_2} \zeta_{\bp_3}]]}',
\end{align}
where the time argument $\tau_f$ has been omitted, $\sum_A$ denotes the sum over $\zeta^A \in \{\zeta,\zeta',\partial_i \zeta, \cdots\}$, $h_{,A} = \delta h/\delta \zeta^A$, and $\sum_{\text{cyc}(1,2,3)}$ denotes the sum over the cyclic permutations of the momenta, $(\bp_1,\bp_2,\bp_3), (\bp_2,\bp_3,\bp_1)$, and $(\bp_3,\bp_1,\bp_2)$.
Note that $X|_{\tau,\bp_i} \equiv \int d^3 x e^{-i \bp_i \cdot \bx} X(\bx,\tau)$ with $X$ being an arbitrary product of fields.
Since we are interested in the conservation of the bispectrum of superhorizon curvature perturbations, the lower bound of the $\tau_1$ integral ($\tau_\star$) is set to be the time when all the external momenta are outside the horizon ($|p_i \tau_\star| \ll 1$). 
We split $B_{333}$ into two parts as $B_{333} = B_{333,a} + B_{333,b}$ with 
\begin{align}
    B_\text{333,b} &= -i\int^{\tau_f}_{\tau_\star} d \tau_1 \int^{\tau_1}_{-\infty} d\tau_2 \int^{\tau_2}_{-\infty} d\tau_3 \sum_{\sigma \in S_3}\sum_{A,B} [\zeta^A_{-\bp_{\sigma(1)}}(\tau_1),\zeta_{\bp_{\sigma(1)}}]' 
     \langle h_{,AB}|_{\tau_1,\bp_{\sigma(2)}}\zeta_{-\bp_{\sigma(2)}} \rangle' \nonumber \\ 
     &\qquad \qquad \qquad \qquad \qquad 
     \times \expval{[H^{(3)}(\tau_3),[H^{(3)}(\tau_2), \zeta^B_{\bp_{\sigma(1)} + \bp_{\sigma(2)}}(\tau_1) \zeta_{\bp_{\sigma(3)}}]]}',
\end{align}
where $S_3$ denotes all six permutations of $(1,2,3)$.
Note that $\sum_{\sigma \in S_3}$ is different from $\sum_{\text{cyc}(1,2,3)}$, which only includes three permutations, $(1,2,3), (2,3,1)$, and $(3,1,2)$. 
$B_\text{333,b}$ only consists of the left 1-PR diagram in Fig.~\ref{fig:1pi_1pr}, while $B_\text{333,a}$ consists of both the left 1-PI and the left 1-PR diagrams in Fig.~\ref{fig:1pi_1pr}.

The other contribution for which the first commutator involves the cubic interaction comes with a cubic and a quartic interaction:
\begin{align}
    B_{43} &= -\int^{\tau_f}_{\tau_\star} d \tau_1 \int^{\tau_1}_{-\infty} d\tau_2 \expval{[H^{(4)}(\tau_2),[H^{(3)}(\tau_1),Q_3]]}' \nonumber \\ 
    &= B_{43,\text{1PI}} + B_{43,\text{1PR}}, 
\end{align}
where the 1-PR part is given by 
\begin{align}
    B_{43,\text{1PR}} &= -\int^{\tau_f}_{\tau_\star} d \tau_1 \int^{\tau_1}_{-\infty} d\tau_2 \sum_{\sigma \in S_3}\sum_{A,B}[\zeta^A_{-\bp_{\sigma(1)}}(\tau_1),\zeta_{\bp_{\sigma(1)}}]' \nonumber \\ 
    &\qquad \times 
     \langle h_{,AB}|_{\tau_1,-\bp_{\sigma(2)}} \zeta_{\bp_{\sigma(2)}} \rangle'\expval{[H^{(4)}(\tau_2), \zeta^B_{\bp_{\sigma(1)}+\bp_{\sigma(2)}}(\tau_1) \zeta_{\bp_{\sigma(3)}}]}'.
\end{align}
If $h$ includes $\zeta'$, $\mathcal H$ includes the induced contributions associated with $\delta h/\delta \zeta'$, denoted by $\mathcal H_\ind$.
Combining $B_{333,\text{1PI}}, B_{43,\text{1PI}}$ and some parts of the induced Hamiltonian contributions, we obtain 
\begin{align}
    \label{eq:q_3_c}
    B_{3,a} &\equiv B_{333,a} + B_{43,\text{1PI}} + B_{\ind,X} \nonumber \\ 
    &=  i\int^{\tau_f}_{\tau_\star} d \tau_1   \sum_{\text{cyc(1,2,3)}} \sum_A [\zeta^A_{-\bp_1}(\tau_1),\zeta_{\bp_1}]' \expval{h_{,A}|_{\tau_1,\bp_1}\zeta_{\bp_2} \zeta_{\bp_3}}_{c}',
\end{align}
where the subscript $c$ denotes the connected part of the four-point function.  
The explicit expression of $B_{\ind,X}$ can be seen in Eq.~(\ref{eq:q_3_c_app}) in  Appendix~\ref{app:one_loop_bispectrum}. 
If $h_{,A}$ includes $\zeta'$, $\expval{h_{,A}|_{\tau_1,\bp_1}\zeta_{\bp_2} \zeta_{\bp_3}}_{c}$ includes the so-called contact terms~\cite{Pimentel:2012tw,Fumagalli:2024jzz}, which are taken into account through $B_{\ind,X}$. See Appendix~\ref{app:one_loop_bispectrum} for the details.

To proceed, we express $h_{,A}$ as 
\begin{align}
    h_{,A} = \sum_{B,C} a_{ABC} \zeta^B \zeta^C,
\end{align}
where $\zeta^B, \zeta^C \in \{\zeta, \zeta', \partial_i \zeta, \cdots\}$ and $a_{ABC}$ is a coefficient.
With this, we can express $\expval{\cdots}_{c}$ in Eq.~(\ref{eq:q_3_c}) as
\begin{align}
    \expval{h_{,A}|_{\tau_1,\bp_1}\zeta_{\bp_2} \zeta_{\bp_3}}'_{c} = \sum_{B,C} a_{ABC}(\tau_1) \int \frac{d^3k}{(2\pi)^3} \expval{\zeta^B_{\bk}(\tau_1) \zeta^C_{\bp_1-\bk}(\tau_1)\zeta_{\bp_2} \zeta_{\bp_3}}'_{c}.
\end{align}
We focus on the case with a large hierarchy between the loop momentum $k$ and the external momenta $p_i$ ($k \gg p_i$).
This means that the loop modes can be on subhorizon scales in our analysis.
Note that, when both the external and the loop modes are on superhorizon scales, the constancy of the superhorizon curvature perturbations was probed at the nonlinear level in Ref.~\cite{Maldacena:2002vr}. 
If $\zeta^B, \zeta^C \in \{\zeta, \zeta'\}$, the double-soft consistency relation reads~\cite{Joyce:2014aqa}\footnote{
Strictly speaking, we have generalized the double-soft consistency relation in Ref.~\cite{Joyce:2014aqa} so that $\zeta^B$ and/or $\zeta^C$ can be $\zeta'$.
This is justified because the time derivative and the dilatation transformation commute.
}
\begin{align}
    &\frac{\bigl\langle \zeta^B_{\bk}(\tau_1) \zeta^C_{\bp_1-\bk}(\tau_1) \zeta_{\bp_2} \zeta_{\bp_3}\bigr\rangle'_{c}}{P(p_2) P(p_3)} \nonumber \\ 
    &= -\frac{\bigl\langle \zeta_{\bp_1} \zeta_{\bp_2} \zeta_{\bp_3} \bigr\rangle'}{P(p_2)P(p_3)} D_k \bigl\langle \zeta^B_{\bk}(\tau_1) \zeta^C_{-\bk}(\tau_1)  
       \bigr\rangle' \nonumber \\ 
       &\qquad + D^2_k \bigl\langle \zeta^B_{\bk}(\tau_1) \zeta^C_{-\bk}(\tau_1) 
       \bigr\rangle' + \mathcal O(p/k) \quad (\text{if } \zeta^B, \zeta^C \in \{\zeta, \zeta'\}),
       \label{eq:double_cons_rel}
\end{align}
where $p = \max[p_1,p_2,p_3]$ and the operator $D_k$ is defined in Eq.~(\ref{eq:d_k_def}).
The first term in Eq.~(\ref{eq:double_cons_rel}) corresponds to the 1-PR diagram in $B_{333,a}$.
Although we can in principle extract the contribution of the left 1-PI diagram in Fig.~\ref{fig:1pi_1pr} by removing the first term, we here keep it for a general discussion.
Using Eq.~(\ref{eq:double_cons_rel}), we obtain the following relation:
\begin{align}
    &\int \frac{d^3 k}{(2\pi)^3} \bigl\langle \zeta^B_{\bk}(\tau_1) \zeta^C_{\bp_1-\bk}(\tau_1) \zeta_{\bp_2} \zeta_{\bp_3} 
       \bigr\rangle'_{c} \nonumber \\
       &= \int \frac{d^3 k}{(2\pi)^3} \left[ -\bigl\langle \zeta_{\bp_1} \zeta_{\bp_2} \zeta_{\bp_3} \bigr\rangle' D_k \bigl\langle \zeta^B_{\bk}(\tau_1) \zeta^C_{-\bk}(\tau_1)  
       \bigr\rangle'  + P(p_2) P(p_3) D_k^2 \bigl\langle \zeta^B_{\bk}(\tau_1) \zeta^C_{-\bk}(\tau_1) 
       \bigr\rangle' \right] \nonumber \\ 
       &= - \bigl\langle \zeta_{\bp_1} \zeta_{\bp_2} \zeta_{\bp_3} \bigr\rangle' \mathcal P^{BC}(\Lambda,\tau_1) \nonumber \\ 
       &\qquad + P(p_2) P(p_3) \frac{d}{d \ln \Lambda} \mathcal P^{BC}(\Lambda,\tau_1) \quad (\text{if } \zeta^B, \zeta^C \in \{\zeta, \zeta'\}),
\end{align}
where we have assumed that the loop momentum always satisfies $k \gg p_i$ and 
\begin{align}
\mathcal P^{BC}(\Lambda,\tau) = \frac{k^3}{2\pi^2}\langle \zeta^B_{\bk}(\tau) \zeta^C_{-\bk}(\tau) 
       \rangle' \Big|_{k = \Lambda(\tau)}.
\label{eq:calp_bc}
\end{align}
We here remove our assumption on $\zeta^B$ and $\zeta^C$ by allowing them to be spatial derivatives of $\zeta$. 
In this case, the double-soft consistency relation, Eq.~(\ref{eq:double_cons_rel}), is not satisfied because the spatial derivative and the dilatation transformation do not commute. 
As in the power spectrum case~\cite{Pimentel:2012tw,Fumagalli:2024jzz}, to keep the right-hand side of Eq.~(\ref{eq:double_cons_rel}) even in that case, we need to additionally take into account the terms associated with a residual diffeomorphism invariance. 
We here simply assume that those terms make the consistency relation hold even if $\zeta^B$ and $\zeta^C$ include spatial derivatives:
\begin{align}
    \label{eq:cubic_one_loop_bi}
    &B_{3,a} + B_{3,\text{diff}} = i
  \int_{\tau_\star}^{\tau_f}d\tau_1 \sum_{\text{cyc}(1,2,3)} \sum_{A,B,C} a_{ABC}(\tau_1)\,\nonumber \\ 
&  \times    
    [\zeta^A_{-\bp_1}(\tau_1), \zeta_{\bp_1}]' \left( - \bigl\langle \zeta_{\bp_1} \zeta_{\bp_2} \zeta_{\bp_3} \bigr\rangle' \mathcal P^{BC}(\Lambda,\tau_1) + P(p_2) P(p_3) \frac{d \, \mathcal P^{BC}(\Lambda,\tau_1)}{d \ln \Lambda} \right),
\end{align}
where we have neglected $p/\Lambda$ and $B_{3,\text{diff}}$ is the additional contribution associated with a residual diffeomorphism invariance.
We leave the explicit calculation of $B_{3,\text{diff}}$ for future work. 
Note again that the first term in Eq.~\eqref{eq:cubic_one_loop_bi} corresponds to the 1-PR part, which is included in $B_{333,a}$, while the second term corresponds to the 1-PI part. 

Next, let us see how the UV boundary terms in Eq.~\eqref{eq:cubic_one_loop_bi} are canceled by the inhomogeneous cutoff contributions.
Three contributions are associated with the UV boundary terms in Eq.~\eqref{eq:cubic_one_loop_bi}. 
The first contribution comes from $H^{(3)}_{\inh,1}$, 
\begin{align}
     B_\text{3,inh,1} &=  B_\text{3,inh,1a} + B_\text{3,inh,1b}, \\ 
    B_\text{3,inh,1a} &= -
  \int_{\tau_\star}^{\tau_f}d\tau_1 \int_{-\infty }^{\tau_1}d\tau_2 \bigl\langle[H^{(3)}(\tau_2),
       [H^{(3)}_{\inh,1}(\tau_1),Q_3]]\bigr\rangle', \\
    B_\text{3,inh,1b} &= -
  \int_{\tau_\star}^{\tau_f}d\tau_1 \int_{-\infty }^{\tau_1}d\tau_2 \bigl\langle[H^{(3)}_{\inh,1}(\tau_2),
       [H^{(3)}(\tau_1),Q_3]]\bigr\rangle',        
\end{align}
where $H^{(n)}_{\inh,i}$ is given by Eq.~\eqref{eq:h_inh_def}.
We find\footnote{One of the $\zeta$ factors in $\bigl\langle \zeta_{\bp_1} \zeta_{\bp_2} \zeta_{\bp_3} \bigr\rangle$ in Eq.~(\ref{eq:tad_1a}) is from $\zeta_L$ in $\mathcal 
I^{(3)}_1$. Although $\bigl\langle \zeta_{\bp_1} \zeta_{\bp_2} \zeta_{\bp_3} \bigr\rangle$ is constant in $\tau_\star < \tau < \tau_f$, its constant value is determined from the past history of that mode including when it was inside the horizon.
}
\begin{align}
    \label{eq:tad_1a}
    B_\text{3,inh,1a} &= i
  \int_{\tau_\star}^{\tau_f}d\tau_1 \sum_{\text{cyc}(1,2,3)}\sum_{A,
   B,C}a_{ABC}(\tau_1)[\zeta^A_{-\bp_1}(\tau_1), \zeta_{\bp_1}]' \bigl\langle \zeta_{\bp_1} \zeta_{\bp_2} \zeta_{\bp_3} \bigr\rangle' \mathcal P^{BC}(\Lambda,\tau_1).
\end{align}
This cancels the first term in the second line of Eq.~(\ref{eq:cubic_one_loop_bi}), which is the 1-PR part.
The contribution of $B_\text{3,inh,1b}$ also contributes to the cancellation in the other 1-PR parts. 

The second contribution comes from $H^{(3)}_{\inh,2}$:
\begin{align}
    \label{eq:tad_2}
  B_\text{3,inh,2} &= i
  \int_{\tau_\star}^{\tau_f}d\tau_1 
  \bigl\langle[ H^{(3)}_{\inh,2}(\tau_1),Q_3]\bigr\rangle' \nonumber \\ 
  &= i
  \int_{\tau_\star}^{\tau_f}d\tau_1 
   \sum_{\text{cyc}(1,2,3)}\sum_{A,B,C} a_{ABC}(\tau_1)[\zeta^A_{-\bp_1}(\tau_1), \zeta_{\bp_1}]' P(p_2) P(p_3) \frac{d \, \mathcal P^{BC}(\Lambda,\tau_1)}{d \ln \Lambda}.
\end{align}
Note that this has the same sign as the second term in the second line of Eq.~(\ref{eq:cubic_one_loop_bi}) and they do not cancel. 

The third contribution is 
\begin{align}
    B_\text{33,inh} &= -
  \int_{\tau_\star}^{\tau_f}d\tau_1 \int_{-\infty }^{\tau_1}d\tau_2 \bigl\langle[H^{(3)}(\tau_2),
       [H^{(3)}(\tau_1),Q_3]]\bigr\rangle'_{\Lambda[\zetal]}.
       \label{eq:b_33_inh}
\end{align}
The subscript $\Lambda[\zetal]$ denotes the contribution obtained by replacing the loop regulator by its variation linear in the long-wavelength mode. After performing the hard-mode contractions, the factor of $\zeta_L$ supplied by this variation is included in the remaining long-mode contractions.
This contribution actually corresponds to the term linear in \(\zetal\) obtained by replacing \(\Lambda\) with \(\Lambda[\zetal]\) in Eq.~(\ref{eq:1_loop}) and expanding \(\Lambda[\zetal]\) in powers of \(\zetal\).
Note that, in the squeezed limit, this corresponds to Eq.~(\ref{eq:cq-cutoff-cubic-cubic}). 
We can express $B_\text{33,inh}$ as
\begin{align}
    B_\text{33,inh} &= \sum_{\text{cyc}(1,2,3)}\expval{\zeta_{\bp_1} \expval{\zeta_{\bp_2} \zeta_{\bp_3}}_{33,\Lambda[\zetal]}}'.
    \label{eq:til_b_33_cut}
\end{align}
$\expval{\zeta_{\bp} \zeta_{\bp'}}_{33,\Lambda[\zetal]}$ with $\bp + \bp' \neq 0$ is given by 
\begin{align}
    \expval{\zeta_{\bp} \zeta_{\bp'}}_{33,\Lambda[\zetal]}     
    &= -i \int^{\tau_f}_{\tau_\star}d\tau' \sum_{A,B,C} a_{ABC}(\tau')\nonumber \\
    &\quad \times ([\Rh^A_{-\bp}(\tau'),\Rh_{\bp}]' P(p') + (\bp \leftrightarrow \bp')) \frac{d \, \mathcal P^{BC}(\Lambda ,\tau')}{d \ln \Lambda} \zeta_{\bp+\bp'},
\end{align}
where $\zeta_{\bp+\bp'}$ originates from the linear expansion of $\zetal$ in $\Lambda[\zetal]$.
Using this, we obtain 
\begin{align}
    B_\text{33,inh} &= -2i \int^{\tau_f}_{\tau_\star}d\tau' \sum_{\text{cyc}(1,2,3)}\sum_{A,B,C}a_{ABC}(\tau')[\Rh^A_{-\bp_1}(\tau'),\Rh_{\bp_1}(\tau_f)]' P(p_2) P(p_3) \frac{d \, \mathcal P^{BC}(\Lambda ,\tau')}{d \ln \Lambda}.
\end{align}
This cancels Eq.~(\ref{eq:tad_2}) and the second term in the second line of Eq.~(\ref{eq:cubic_one_loop_bi}), which corresponds to the 1-PI part.
Through the above discussion, we have seen the cancellation of all the contributions in Eq.~(\ref{eq:cubic_one_loop_bi}).

\subsection{Contributions with the first commutator involving the quartic interaction}

Next, we focus on the 1-PI contributions for which the first commutator involves the non-induced quartic Hamiltonian:
\begin{align}
    \label{eq:q_34}
    B_{34,\text{non-ind}} &= -\int^{\tau_f}_{\tau_\star} d \tau_1 \int^{\tau_1}_{-\infty} d\tau_2 \int d^3 x \expval{[H^{(3)}(\tau_2),[l(\bx,\tau_1),Q_3]]}'.
\end{align}
We here focus on the non-induced quartic interaction Hamiltonian because the induced quartic Hamiltonian was already used to reproduce the contact terms in Eq.~(\ref{eq:q_3_c}). See Appendix~\ref{app:one_loop_bispectrum} for the details.
We split this as $B_{34,\text{non-ind}} = B_{34,\text{1PI}} + B_{34,\text{1PR}}$ with 
\begin{align}
    \label{eq:q_34_c}
    B_{34,\text{1PI}} &= -\int^{\tau_f}_{\tau_\star} d \tau_1 \int^{\tau_1}_{-\infty} d\tau_2\sum_{\sigma \in S_3}\sum_{A,B} \nonumber \\ 
    &\qquad \times  [\zeta^A_{-\bp_{\sigma(1)}}(\tau_1),\zeta_{\bp_{\sigma(1)}}]'
     \langle \zeta^B_{-\bp_{\sigma(2)}}(\tau_1) \zeta_{\bp_{\sigma(2)}} \rangle'\expval{[H^{(3)}(\tau_2), l_{,AB}|_{\tau_1,\bp_{\sigma(1)}+\bp_{\sigma(2)}} \zeta_{\bp_{\sigma(3)}}]}', \\
    B_{34,\text{1PR}} &= -\int^{\tau_f}_{\tau_\star} d \tau_1 \int^{\tau_1}_{-\infty} d\tau_2 \sum_{\text{cyc}(1,2,3)}\sum_{A,B}\nonumber \\ 
    &\qquad \times [\zeta^A_{-\bp_1}(\tau_1),\zeta_{\bp_1}]'
     \int \frac{d^3 k}{(2\pi)^3}\langle l_{,AB}|_{\tau_1,-\bk,\bk}\rangle'
     \expval{[H^{(3)}(\tau_2), \zeta^B_{\bp_1}(\tau_1) \zeta_{\bp_2} \zeta_{\bp_3}]}',     
\end{align}
where $\int d^3 k/(2\pi)^3 \langle l_{,AB}|_{\tau_1,-\bk,\bk}\rangle' = \expval{l_{,AB}(\bx,\tau_1)}$.
The 1-PI part, $B_{34,\text{1PI}}$, corresponds to the middle 1-PI diagram in Fig.~\ref{fig:1pi_1pr}.
With some part of the contributions from the induced Hamiltonian, we obtain
\begin{align}
    \label{eq:q_3_4c}
    B_{4,\text{1PI}} &\equiv B_{34,\text{1PI}} + B_{\ind,Y} \nonumber \\ 
    &=  i\int^{\tau_f}_{\tau_\star} d \tau_1 \sum_{\sigma \in S_3}\sum_{A,B} \nonumber \\ 
    &\qquad \times  [\zeta^A_{-\bp_{\sigma(1)}}(\tau_1),\zeta_{\bp_{\sigma(1)}}]'
     \langle \zeta^B_{-\bp_{\sigma(2)}}(\tau_1) \zeta_{\bp_{\sigma(2)}} \rangle'\expval{ l_{,AB}|_{\tau_1,\bp_{\sigma(1)}+\bp_{\sigma(2)}} \zeta_{\bp_{\sigma(3)}}}', 
\end{align}
where $B_{\ind,Y}$ reproduces the contact terms in $\expval{ l_{,AB}|_{\tau_1,\bp_{\sigma(1)}+\bp_{\sigma(2)}} \zeta_{\bp_{\sigma(3)}}}$.
The explicit expression of $B_{\ind,Y}$ is given in Eq.~(\ref{eq:q_4_app}) in Appendix~\ref{app:one_loop_bispectrum}.

To proceed, we express $l_{,AB}$ as 
\begin{align}
    l_{,AB} = \sum_{C,D} b_{ABCD} \zeta^C \zeta^D. 
\end{align}
Then, we can reexpress Eq.~(\ref{eq:q_3_4c}) as 
\begin{align}
        B_{4,\text{1PI}}
    &=  i\int^{\tau_f}_{\tau_\star} d \tau_1 \sum_{\sigma \in S_3}\sum_{A,B,C,D}b_{ABCD}(\tau_1) \nonumber \\ 
    &\qquad \times [\zeta^A_{-\bp_{\sigma(1)}}(\tau_1),\zeta_{\bp_{\sigma(1)}}]'
     \langle \zeta^B_{-\bp_{\sigma(2)}}(\tau_1) \zeta_{\bp_{\sigma(2)}} \rangle' \int \frac{d^3 k}{(2\pi)^3} \expval{ \zeta^C_{\bk}(\tau_1) \zeta^D_{\bp_{\sigma(1)}+\bp_{\sigma(2)}-\bk} (\tau_1) \zeta_{\bp_{\sigma(3)}}}'.
\end{align}
Similarly to the previous subsection, we assume a large hierarchy between the loop momentum and the external momenta, which allows us to use the consistency relation: 
\begin{align}
    \label{eq:quart_bi_a}
        B_{4,\text{1PI}} + B_{4,\text{diff}}
    &=  -i\int^{\tau_f}_{\tau_\star} d \tau_1 \sum_{\sigma \in S_3}\sum_{A,B,C,D}b_{ABCD}(\tau_1) [\zeta^A_{-\bp_{\sigma(1)}}(\tau_1),\zeta_{\bp_{\sigma(1)}}]'\nonumber \\ 
    &\qquad \times 
     \langle \zeta^B_{-\bp_{\sigma(2)}}(\tau_1) \zeta_{\bp_{\sigma(2)}} \rangle' P(p_{\sigma(3)})\mathcal P^{CD}(\Lambda,\tau_1),
\end{align}
where we have added $B_{4,\text{diff}}$, which originates from a residual diffeomorphism invariance, to make the consistency relation hold even if $\zeta^C$ and/or $\zeta^D$ include spatial derivatives. 
We leave the derivation of the explicit $B_{4,\text{diff}}$ for future work.

We here consider another contribution induced by the inhomogeneous cutoff,
\begin{align}
    \label{eq:bi_inh_4}
    B_\text{4,inh,1} &= i
  \int_{\tau_\star}^{\tau_f}d\tau_1 \bigl\langle
       [H^{(4)}_{\inh,1}(\tau_1),Q_3]\bigr\rangle' \nonumber \\
       &= \sum_{\text{cyc}(1,2,3)}\expval{\zeta_{\bp_1} \expval{\zeta_{\bp_2} \zeta_{\bp_3}}_{4,\Lambda[\zetal]}}',
\end{align}
where 
\begin{align}
    \expval{\zeta_{\bp} \zeta_{\bp'}}_{4,\Lambda[\zetal]} 
    &= i \int^{\tau_f}_{\tau_\star}d\tau' \sum_{A,B,C,D}b_{ABCD}(\tau')  \nonumber \\ 
    &\qquad \times ([\Rh^A_{-\bp}(\tau'),\Rh_{\bp}]'\langle \zeta^B_{-\bp'}(\tau') \zeta_{\bp'} \rangle' + (\bp \leftrightarrow \bp')) \mathcal P^{CD}(\Lambda,\tau')\zeta_{\bp + \bp'}.
\end{align}
Eq.~(\ref{eq:bi_inh_4}) cancels Eq.~(\ref{eq:quart_bi_a}).
Note that this contribution corresponds to Eq.~(\ref{eq:b_4_cut}) in the squeezed limit. 

Before closing this section, we mention the right 1-PI diagram in Fig.~\ref{fig:1pi_1pr}, which comes from $\mathcal H^{(5)}$. 
In particular, the genuine part of $\mathcal H^{(5)}$, which is irrelevant to either the induced Hamiltonian or the part associated with a residual diffeomorphism invariance, can provide an independent contribution.
Given that the genuine part of $\mathcal H^{(4)}$ does not provide a UV boundary term for the one-loop power spectrum~\cite{Pimentel:2012tw,Fumagalli:2024jzz}, we similarly expect that the genuine part of $\mathcal H^{(5)}$ does not provide the UV boundary terms.
We leave the detailed analysis of this contribution for future work.

\section{Conclusions}

In an inhomogeneous spacetime, a cutoff fixed in physical units defines an inhomogeneous boundary in comoving momentum space. In this work,
we have implemented this geometric principle in the cutoff regularization of
cosmological loop correlators. At leading order in long-wavelength gradients,
the local spatial metric determines both the radius and the shape of the
comoving integration domain. Its dependence on scalar and tensor backgrounds
generates additional contributions when regulated contractions are expanded
in the long modes. These contributions admit a diagrammatic interpretation
as insertions on the corresponding loop contractions and implement the
response of the regulator under residual spatial dilatations and
volume-preserving anisotropic rescalings.

At one loop, we showed that the curvature-background response of the regulated
tadpole cancels the UV boundary term in the late-time evolution of
the superhorizon curvature power spectrum. The analogous tensor cancellation
requires the anisotropic deformation of the cutoff surface. In particular,
the vanishing of the ordinary tensor tadpole is compatible with a nonzero
response of its regulated source to a long tensor background. Together, these
calculations provide a geometric explanation and an explicit cutoff
realization of the expected conservation of the power spectra on superhorizon scales. 

For the squeezed scalar bispectrum, we established the consistency relation
at finite cutoff and verified the interplay between the ordinary diagrams
and the regulator contributions in an explicit derivative-interaction
example. 
A practical consequence is that the cutoff response of a
lower-point loop amplitude determines part of the higher-point calculation
with an additional soft leg. 
Under the stated assumption about certain one-loop contributions, our analysis of the bispectrum with all
external modes outside the horizon identifies analogous cancellations. 
Completing the spatial-derivative contributions and the remaining quintic
sector would establish the full extent of this cancellation.
Our results concern regulated correlators and their symmetry
relations. A complete UV renormalization additionally requires
counterterms and renormalization conditions consistent with these
relations.

Possible applications include one-loop squeezed bispectra in models with primordial features and mixed tensor--scalar--scalar correlators, providing settings with nontrivial scale and angular dependence~\cite{Braglia:2026fle,Ghosh:2023agt}. A complementary direction is to combine the prescription with a complete UV renormalization procedure.
Matching to other symmetry-preserving schemes, including dimensional regularization, would then provide a systematic check of the interplay between ordinary loop diagrams, regulator-induced contributions, and counterterms~\cite{Bhowmick:2024kld,Ballesteros:2025nhz}.
One-loop bispectra with massive fields offer a setting for comparing the physical cutoff with Pauli--Villars and dimensional regularization~\cite{Wang:2021qez,Xianyu:2022jwk,Borinsky:2026roj}. A comparison with the same initial state and renormalization conditions could assess how practical the prescription proposed in this paper is when the mode functions must be obtained numerically.
Finally, the distinction between physical and fiducial cutoffs plays an explicit role in recent studies connecting volume renormalization to stochastic and eternal inflation~\cite{Cotler:2026lna,Seo:2026ajv}. Related work connects the renormalization of field products to stochastic evolution~\cite{Cohen:2021fzf,Green:2024fsz}. Establishing how our regulator-induced contributions relate to the renormalization of geometric composite operators and the associated stochastic description remains a direction for future investigation.
\acknowledgments 
The authors thank Matteo Braglia, Carlos Duaso Pueyo and Marko Simonović for useful discussion.
The authors thank the organizers and the attendees of the ``Looping in the Primordial Universe'' workshop at CERN for fruitful discussions.
The work of JF was carried out as part of research activities funded by the Scuola Normale Superiore, Pisa, Italy.
The authors used OpenAI ChatGPT as an interactive assistant for manuscript preparation. 
All scientific judgments, derivations, calculations, interpretation, and responsibility for the manuscript remain solely with the author.

\appendix
\section{Initial-state matching in cutoff regularization}
\label{app:cutoff-matching}
This appendix explains the initial-state prescription underlying the three-point representation used in Eq.~\eqref{eq:1_loop}. 
The identification of the inner time integral in Eq.~\eqref{eq:1_loop} with a tree-level three-point function requires some care in cutoff regularization. In the cubic--cubic contribution, the homogeneous physical cutoff restricts the loop momentum at the earlier vertex time, $\tau_2<\tau_1$, according to $k<a(\tau_2)\Lambda_\uv$~\cite{Senatore:2009cf}. The momentum integration domain therefore depends on the time being integrated. We cannot simply perform the $\tau_2$ integral over the entire past at fixed $k$ and subsequently impose the cutoff at $\tau_1$.

To expose the issue, define the cutoff-entry time $\tau_k$ by $
k=a(\tau_k)\Lambda_\uv.$
The mode $k$ is included in the regulated description only after $\tau_k$, when its physical momentum falls below $\Lambda_\uv$. Exchanging the integrations gives
\begin{equation}
\int_{-\infty}^{\tau_1}d\tau_2\int^{a(\tau_2)\Lambda_\uv}d\ln k\,F(k,\tau_1,\tau_2)
=
\int^{a(\tau_1)\Lambda_\uv}d\ln k\int_{\tau_k}^{\tau_1}d\tau_2\,F(k,\tau_1,\tau_2),
\end{equation}
where $F$ denotes the corresponding integrand and adiabatic convergence in the past is understood.\footnote{Both internal momenta, $\bk$ and $\bp-\bk$, must lie below the cutoff. The inner time integral therefore starts at the later of their two cutoff-entry times. In the leading soft limit $p/k\to0$, these times coincide, allowing us to denote the lower boundary by $\tau_k$.}
The inner integral thus starts at $\tau_k$, rather than in the adiabatic past. By itself, it does not reproduce the usual Bunch--Davies three-point function.
This observation also makes explicit an additional ingredient of the regulated description: the state of the modes when they enter. Restricting the integration domain does not specify that state. In particular, an interacting state at $\tau_k$ can already contain a three-point correlation, which must be included alongside the correlation generated by subsequent evolution.
For compactness, denote the three-field operator entering the loop source by
\begin{equation}
\mathcal O_{A,\bk}(\tau_1)\equiv
\left.\frac{\delta\Hamt}{\delta\R^A}(\tau_1)\right|_{\bk,-\bk}\Rh_{\bp}.
\end{equation}
The momentum integral of this three-field correlator defines a source $\mathcal S_{A,\Lambda}(p,\tau_1)$ for the remaining outer time integration. Schematically, its contribution to the late-time cubic--cubic correction is
\begin{equation}
\Delta P_{\Lambda,33}^{(1)}(p)\sim
\sum_A\int_{\tau_\star}^{\tz}d\tau_1\,
\mathcal K_{A,p}(\tz,\tau_1)\mathcal S_{A,\Lambda}(p,\tau_1),
\end{equation}
where $\mathcal K_{A,p}$ represents the remaining external retarded contraction, with the derivative and numerical factors appearing in Eq.~\eqref{eq:1_loop}.

To evaluate this source, write the initial state as
\begin{equation}
|\Psi(\tau_k)\rangle=|0\rangle+|\delta\Psi(\tau_k)\rangle+\cdots,
\end{equation}
where $|0 \rangle$ is the free vacuum and $|\delta\Psi(\tau_k)\rangle$ denotes the correction linear in the cubic interaction. Since the Gaussian expectation of $\mathcal O_{A,\bk}$ vanishes, expanding the finite-time in-in expression to the required order gives
\begin{align}
\mathcal S_{A,\Lambda}(p,\tau_1)
={}&\int^{a(\tau_1)\Lambda_\uv}d\ln k\,\frac{k^3}{2\pi^2}
\Bigg[
\bra{0}\mathcal O_{A,\bk}(\tau_1)\ket{\delta\Psi(\tau_k)}'
+\bra{\delta\Psi(\tau_k)}\mathcal O_{A,\bk}(\tau_1)\ket{0}'
\nonumber\\
&\hspace{20mm}
+i\int_{\tau_k}^{\tau_1}d\tau_2\,
\bra{0}[H^{(3)}(\tau_2),\mathcal O_{A,\bk}(\tau_1)]\ket{0}'
\Bigg].
\label{eq:int_bi}
\end{align}
The first two terms describe the initial three-point correlation propagated freely to the operator times. The last term describes the correlation generated after cutoff entry. All three terms have the same perturbative order. Corrections to the state inside the last term contribute only at higher order.

To recover the standard Bunch--Davies correlator, we adopt an explicit matching prescription: the state at entry is matched to the interacting state prepared adiabatically from the same free Bunch--Davies vacuum,
\begin{equation}
|\Psi(\tau_k)\rangle=U_I(\tau_k,-\infty)|0\rangle,\qquad
U_I(\tau,\tau')=T\exp\left[-i\int_{\tau'}^\tau d\eta\,H_{\rm int}(\eta)\right].
\end{equation}
The past limit includes the usual adiabatic prescription, with conjugate contours understood. All momentum configurations use the same state preparation, evaluated at their corresponding matching times. Its perturbative non-Gaussian correlations encode the interaction history carried into the regulated description at entry.
This matching is thus consistent with spatial dilatation symmetry. 

The recovery of the usual three-point function is now explicit. At the required order, the cubic part of the state correction is
\begin{equation}
|\delta\Psi(\tau_k)\rangle=-i\int_{-\infty}^{\tau_k}d\tau_2\,H^{(3)}(\tau_2)|0\rangle.
\end{equation}
Adding this term to the evolution from $\tau_k$ to $\tau_1$ in Eq.~\eqref{eq:int_bi} gives
\begin{equation}
\mathcal S_{A,\Lambda}(p,\tau_1)=
\int^{a(\tau_1)\Lambda_\uv}d\ln k\,\frac{k^3}{2\pi^2}
i\int_{-\infty}^{\tau_1}d\tau_2\,
\bra{0}[H^{(3)}(\tau_2),\mathcal O_{A,\bk}(\tau_1)]\ket{0}'.
\end{equation}
The inner correlator is therefore the usual tree-level Bunch--Davies three-point function, with the remaining momentum integral regulated at $\tau_1$. This recovers the representation used in Eq.~\eqref{eq:1_loop} and Refs.~\cite{Pimentel:2012tw,Fumagalli:2024jzz} for the cutoff prescription supplemented by the stated initial-state matching.

Single-vertex bulk contributions  have no analogous exchange-of-integrations issue and are regulated at their vertex time. In main-text calculations, any replacement of a truncated interaction history by a standard Bunch--Davies correlator is understood with the same matching prescription, retaining the initial correlations required at the perturbative order considered.
\section{The background-wave method and soft consistency relations}
\label{app:background-wave}
In this appendix, we review, perhaps somewhat pedantically, the background-wave method for computing squeezed correlation functions and relate it to our inhomogeneous-cutoff prescription. 
In particular, we show how this
prescription arises naturally when a fixed local physical cutoff is
expressed in the original coordinates in the presence of a
long-wavelength background.  

Consider a squeezed correlator of the form
\begin{equation}
    \langle \zeta_{\bp}\zeta_{\bk}\zeta_{\bk'}\cdots\rangle,
    \qquad p\ll k,k',\ldots,
\end{equation}
where $p=|\bp|$, and similarly for the hard momenta. A direct calculation
would evaluate the correlator using the interaction Hamiltonian to the
required perturbative order and then take the soft limit. The background-wave method instead computes the leading contribution in the squeezed limit by treating the constant adiabatic long mode as a
background for the quantum short modes
\cite{Creminelli:2004yq,Ganc:2010ff,Senatore:2012wy}.
The essential observation is that a constant adiabatic long mode can
be absorbed locally into a rescaling of the spatial coordinates.
In these local coordinates, the short-scale correlators coincide with
those evaluated in the absence of the long mode. Re-expressing them in terms of the original coordinates yields correlators conditioned on the long-wavelength mode and therefore explicitly dependent on the long-wavelength perturbation $\zetal$. Correlators with one or more soft external legs can then be obtained by expanding the conditional correlator to the corresponding order in $\zetal$ and correlating the resulting terms with the external long-wavelength modes. 
For instance, the squeezed three-point function is computed by averaging the two-point function conditioned on the long-wavelength background with one long external mode:
\begin{equation}
\langle\zeta_{\mathbf q}\zeta_{\mathbf k_1}\zeta_{\mathbf k_2}\rangle= \langle\zeta_{\mathbf q}\langle\zeta_{\mathbf k_1}\zeta_{\mathbf k_2}\rangle_{\zetal}\rangle.
\end{equation}
We provide here
the procedure explicitly, keeping track of the coordinate transformation and
Fourier normalization. 

\subsection{The conditional action and local coordinates}
\label{appsub:bw-action}
 
We split the comoving curvature perturbation as
$\zeta=\zetal+\short$, where $\zetal$ denotes the long-wavelength
sector and $\short$ the quantum short-wavelength fluctuations.
We assume that the long modes are frozen throughout the period
relevant to the evolution of the short-mode correlators.
At leading order in long-mode gradients, the spatial metric is
$a^2(\tau)e^{2\zetal}\delta_{ij}$. The quadratic action for $\short$ in the
constant background $\zetal$ can be written as
\begin{equation}
 S_2[\zetal,\short]=\int d\tau\,d^3x\,\epsilon a^2
        \left[e^{3\zetal}\short'^2-e^{\zetal}(\partial_i \short)^2\right].
 \label{eq:bw-conditional-action}
\end{equation}
The factors in Eq.~\eqref{eq:bw-conditional-action}
follow from the spatial volume element and inverse spatial metric. 
One can first remove the background by the local coordinate transformation: 
\begin{equation}
 \widetilde{\mathbf x}=e^{\zetal}\mathbf x,
 \qquad
 \widetilde {\short}(\widetilde{\mathbf x},\tau)
       =\short(e^{-\zetal}\widetilde{\mathbf x},\tau).
 \label{eq:bw-local-coordinates}
\end{equation}
The full curvature perturbation transforms as
$\widetilde\zeta(\widetilde{\mathbf x},\tau)
=\zeta(e^{-\zetal}\widetilde{\mathbf x},\tau)-\zetal
=\widetilde {\short}(\widetilde{\mathbf x},\tau)$, so the constant background
has been removed. Since $\zetal$ is time-independent,
substituting in the quadratic action for $\short$ gives 
\begin{equation}
 S_2[\zetal,\short]=\int d\tau\,d^3\widetilde x\,\epsilon a^2
       \left[\widetilde {\short}'^2-(\partial_{\widetilde x^i}\widetilde {\short})^2\right]
       =S_2[0,\widetilde {\short}].
 \label{eq:bw-local-action}
\end{equation}
Thus the local short-mode problem has the same quadratic action as at
$\zetal=0$. 
 More generally, denote the full conditional short-mode action
by $S_{\rm short}[\zetal,\zeta_S]$, including its interactions. Residual spatial
diffeomorphism invariance implies
\begin{equation}
 S_{\rm short}[\zetal,\short]=S_{\rm short}[0,\widetilde {\short}],
 \label{eq:bw-full-action}
\end{equation}
where the right-hand side is evaluated in the local coordinates.\footnote{This identity holds provided that the dependence of the action on
the constant long background is retained consistently. In
particular, the factors arising from the spatial volume element
and spatial derivatives generate couplings between the long
background and the short fields that must be included.}
One may therefore quantize the short modes directly in the local
coordinates, using the ordinary action at $\zetal=0$, and then return to
the original momentum variables. 
The conjugate momenta are densities with respect to their respective
coordinate volumes. For the quadratic theory,
\begin{equation}
 \Pi_s(\mathbf x)=2\epsilon a^2e^{3\zetal}\short'(\mathbf x)=e^{3\zetal}\widetilde\Pi_s(\widetilde{\mathbf x}),
 \qquad
 \widetilde\Pi_s(\widetilde{\mathbf x})=2 \epsilon a^2\widetilde {\short}'(\widetilde{\mathbf x}).
 \label{eq:bw-momentum-density}
\end{equation}
This transformation preserves the canonical commutator because
$\delta^{(3)}(\widetilde{\mathbf x}-\widetilde{\mathbf y})
=e^{-3\zetal}\delta^{(3)}(\mathbf x-\mathbf y)$.
\subsection{Mode functions and Fourier modes in the two coordinate systems}
To obtain the transformation of the normalized free mode function, we start
from the usual quantum expansion in local coordinates, which is a solution of the quadratic action with $\zetal =0$:
\begin{equation}
 {\widetilde \zeta}(\widetilde{\mathbf x},\tau)
  =\int\frac{d^3\widetilde {k}}{(2\pi)^3}
 \left[{\widetilde{a}}_{\widetilde{\bk}}\widetilde{f}_{\widetilde{\bk}}(\widetilde{\bx},\tau)
 +{\widetilde{a}}_{\widetilde{\bk}}^{\dagger}\widetilde{f}_{\widetilde{\bk}}^{*}(\widetilde{\bx},\tau)\right],
 \qquad
 [{\widetilde{a}}_{\widetilde{\bk}},{\widetilde{a}}_{\widetilde{\bk}'}^{\dagger}]
 =(2\pi)^3\delta^{(3)}(\widetilde{\bk}-\widetilde{\bk}'),
 \label{eq:bw-local-mode-expansion}
\end{equation}
where in the homogeneous background (local coordinates) the solution of the linear mode equation is
\begin{equation}
\widetilde{f}_{\widetilde{\bk}}(\widetilde{\bx},\tau)=e^{i\widetilde{\bk}\cdot\widetilde{\bx}}u_{\widetilde{k}}(\tau),
\end{equation}
and $u_{\widetilde{k}}(\tau)$ is the standard time-dependent mode function.
The Fourier phase fixes the relation between momentum labels: $
\widetilde{\mathbf k}=e^{-\zetal}\mathbf k$,
$\widetilde{\mathbf k}\cdot\widetilde{\mathbf x}
       =\mathbf k\cdot\mathbf x$,
$d^3\widetilde k=e^{-3\zetal}d^3k$.
With $\zeta_{\mathbf k}=\int d^3x\,e^{-i\mathbf k\cdot\mathbf x}\zeta(\mathbf x)$,
the complete Fourier coefficient of the field therefore obeys
\begin{equation}
\zeta_{\mathbf k}=e^{-3\zetal}
{\widetilde \zeta}_{e^{-\zetal}\mathbf k}.
 \label{eq:bw-fourier-field}
\end{equation}
Changing integration variables to $\mathbf k$ gives
\begin{equation}
{\widetilde \zeta}(\widetilde{\mathbf x},\tau)
  = \zeta(\mathbf x,\tau)
 =\int\frac{d^3k}{(2\pi)^3}
       \left[e^{-3\zetal}\widetilde a_{e^{-\zetal}\mathbf k}
       u_{ke^{-\zetal}}(\tau)e^{i\mathbf k\cdot\mathbf x}
       +\mathrm{h.c.}\right].
 \label{eq:bw-global-mode-expansion}
\end{equation}
The operators in this expression are still normalized with respect to
$\widetilde{\mathbf k}$. Their commutator, expressed in the original labels,
is $
 [\widetilde a_{e^{-\zetal}\mathbf k},
  \widetilde a^\dagger_{e^{-\zetal}\mathbf k'}]
 =e^{3\zetal}(2\pi)^3\delta^{(3)}(\mathbf k-\mathbf k')$.
Consequently, defining \begin{equation}
a_{\mathbf k}
=e^{-3\zetal/2}\widetilde a_{e^{-\zetal}\mathbf k}\end{equation} restores the standard
commutator in the original momentum coordinates. Equation~\eqref{eq:bw-global-mode-expansion} then takes the usual form:
\begin{equation}
 \zeta(\mathbf x,\tau)
  =\int\frac{d^3 {k}}{(2\pi)^3}
 \left[{a}_{\bk}f_{\bk,\zetal}(\bx,\tau)
 +{a}_{\bk}^{\dagger}f_{\bk,\zetal}^{*}(\bx,\tau)\right],
 \label{standard_app}
 \end{equation}
with
\begin{equation}
 \begin{aligned}
 f_{\mathbf k,\zetal}(\mathbf x,\tau)
 =e^{-3\zetal/2}
   \widetilde f_{\mathbf k e^{-\zetal}}(e^{\zetal}\mathbf x,\tau)
 \equiv e^{i\mathbf k\cdot\mathbf x}u_{k,\zetal},\qquad
 u_{k,\zetal}(\tau)=e^{-3\zetal/2}u_{ke^{-\zetal}},
 \end{aligned}
 \label{eq:bw-mode-transformation}
\end{equation}
where $u_{k,\zetal}$ denotes the normalized short-mode function
in the presence of the constant long background $\zetal$, expressed
in the original momentum variable $k$.
$u_{ke^{-\zetal}}$ is the usual mode function in the absence of the
long background, evaluated at the local rescaled momentum
$\widetilde k=ke^{-\zetal}$.
For a spatially varying long background, note that $f_{\mathbf k}(\mathbf x,\tau)$ generally cannot be written as a single plane wave times $u_k(\tau)$.

\subsection{Power spectra and squeezed correlators with the background wave method}
\label{appsub:bw-soft-relation}

At a fixed observation time, the local-coordinate spectrum is given by
\begin{equation}
\langle \label{homogeneousp}{\widetilde{\zeta}}_{\widetilde{\bk}} {\widetilde{\zeta}}_{\widetilde{\bk}'}\rangle=(2\pi)^3\delta(\widetilde{\bk}+\widetilde{\bk}')P(\tilde{k}),
\end{equation}
where $P(\widetilde k)$ is the standard dimensionful power spectrum which has no explicit
$\zetal$-dependence. At tree level $P(\tilde k)=|u_{\tilde k}|^2$. We now want to relate this to the original coordinates. 

Let us start by computing the conditioned two-point correlator of two short fields in position space. We omit subscript $S$ on the short modes and the time dependencies, and write:
\begin{equation}
\begin{split}
\langle\zeta(\bx_1)\zeta(\bx_2)\rangle_{\zetal} 
\simeq 
\langle{\widetilde \zeta}(\widetilde{\bx}_1){\widetilde{\zeta}}(\widetilde{\bx}_2)\rangle= 
\int \frac{d^3 \widetilde{p}}{(2\pi)^3} e^{i\widetilde{\bp}\cdot\mathbf{\widetilde{x}}}P(\widetilde{p}),
\label{correlatorcond}
\end{split}
\end{equation}
where we defined $\widetilde{\bx} =\widetilde{\bx}_1-\widetilde{\bx}_2$ and we used Eq.~\eqref{homogeneousp}. The homogeneous correlator computed in local coordinates is standard and only depends on the difference between the two spatial coordinates $\widetilde{\bx}$. We define the conditional power spectrum $P(p|\zetal)$ as 
\begin{equation}\label{eq:defPcond}
\langle\zeta(\bx_1)\zeta(\bx_2)\rangle_{\zetal} \equiv \int\frac{d^3 p}{(2\pi)^3} e^{i\bp \cdot \bx} P(p|\zetal),
\end{equation}
where $\bx =\bx_1-\bx_2$.
Equating Eq.~\eqref{correlatorcond} with Eq.~\eqref{eq:defPcond} leads to
\begin{equation}
 P(k|\zetal)\simeq e^{-3\zetal}P(ke^{-\zetal}),\qquad
 \left.\partial_{\zetal} P(k|\zetal)\right|_0
       \simeq -(3+k\partial_k)P(k).
 \label{eq:bw-power-response}
\end{equation}
For a dimensionless spectrum
$\mathcal P(k)=k^3P(k)/(2\pi^2)$, the same transformation is simply
$\mathcal P(k|\zetal)=\mathcal P(ke^{-\zetal})$.
In the previous equation, $\zetal$ can be considered evaluated at the midpoint $\mathbf{X}=(\bx_1+\bx_2)/2$. Another reference point in a neighborhood where the long mode is approximately constant will be equivalent at leading order in the long-mode gradient.

To compute the conditional hard two-point function in momentum space, we Fourier transform Eq.~\eqref{correlatorcond}:
\begin{equation}
\begin{split}
\langle \zeta_{\bk_1}\zeta_{\bk_2}\rangle_{\zetal}=&\int d^3x_1 d^3x_2 e^{-i \bk_1 \cdot \bx_1}e^{-i \bk_2 \cdot\bx_2} \langle\zeta(\bx_1)\zeta(\bx_2)\rangle_{\zetal}\\[1mm]
=& \int d^3 X d^3x e^{-i \bK\cdot\mathbf{X}}e^{-i \bk\cdot\bx}\int \frac{d^3p}{(2\pi)^3} e^{i\bp\cdot\mathbf{x}}P(p|\zetal)= \int d^3 X e^{-i \bK\cdot\mathbf{X}} P(k|\zetal(\mathbf{X})).
\end{split}
\end{equation}
In the first step, we have performed the change of variables, $\mathbf X=(\bx_1+\bx_2)/2$, $\mathbf{x}= \bx_1-\bx_2$, and defined $\bk =(\bk_1-\bk_2)/2$, $\mathbf{K}= \bk_1+\bk_2$.
Expanding $P(k|\zetal)$ around $\zetal=0$, and using Eq.~\eqref{eq:bw-power-response}, we obtain 
\begin{equation}
\langle \zeta_{\bk_1}\zeta_{\bk_2}\rangle_{\zetal}= (2\pi)^3\delta(\bk_1+\bk_2)P(k)+ \zeta_{L,\bk_1+\bk_2}R(k)+\mathcal O(\zetal^2),\qquad R(k)=\left.\partial_{\zetal} P(k|\zetal)\right|_{0},
 \label{eq:bw-offdiagonal-response}
\end{equation}
where $P(k|\zetal=0)= P(k)$ and $k=|\mathbf k_1-\mathbf k_2|/2$. The homogeneous term remains momentum-diagonal, whereas the term linear in the long background is proportional to its Fourier coefficient $\zeta_{L,\mathbf k_1+\mathbf k_2}$.
We can now compute a three-point squeezed correlator such as
\begin{equation}
 \langle\zeta_{\mathbf q}\zeta_{\mathbf k_1}
               \zeta_{\mathbf k_2}\rangle_c
 =(2\pi)^3\delta^{(3)}(\mathbf q+\mathbf k_1+\mathbf k_2)
              B(q,k),\qquad q\ll k_1,k_2.
\end{equation}
Using the background wave method amounts to writing 
\begin{equation}
\langle\zeta_{\mathbf q}\zeta_{\mathbf k_1}\zeta_{\mathbf k_2}\rangle_c= \langle\zeta_{\mathbf q}\langle\zeta_{\mathbf k_1}\zeta_{\mathbf k_2}\rangle_{\zetal}\rangle_c.
\label{squeezedbackgroundwave}
\end{equation}
The conditioned two-point function of the hard modes contains a series of $\zetal$ legs.  Averaging Eq.~\eqref{eq:bw-offdiagonal-response} with the external long mode yields
\begin{equation}
 \langle\zeta_{\mathbf q}\zeta_{\mathbf k_1}
               \zeta_{\mathbf k_2}\rangle_c
 \simeq(2\pi)^3\delta^{(3)}(\mathbf q+\mathbf k_1+\mathbf k_2)
               P(q)R(k).
 \label{eq:bw-long-average}
\end{equation}
Here $\simeq$ retains the leading adiabatic squeezed contribution as
$q/k\to0$. The short fields remain quantum at fixed background; only the
long growing-mode amplitude is treated as a classical variable at this
order in the soft expansion \cite{Ganc:2010ff}.
Combining with Eq.~\eqref{eq:bw-power-response} gives the scalar soft
consistency relation
\begin{equation}
 B(q,k)
 \simeq 
 -P(q)(3+k\partial_k)P(k).
 \label{eq:bw-soft-bispectrum}
\end{equation}
This connects the background-wave construction to the dilatation Ward
identity \cite{Maldacena:2002vr,Creminelli:2004yq,Goldberger:2013rsa}.
Note that we have never specified the loop order of the correlator in the previous derivation.\footnote{
The consistency relation holds beyond tree level for the full
renormalized correlators, provided the regularization and
renormalization preserve the dilatation symmetry
\cite{Goldberger:2013rsa}. There is a subtlety when deriving it
from a correlator conditioned on the entire long-wavelength
sector. Expanding this conditional correlator generates terms
with several long-mode insertions. To obtain a squeezed
bispectrum, one must include all contributions that leave one
long leg to be correlated with the external soft mode, allowing
the remaining long legs to contract among themselves. For
example, a term cubic in $\zetal$ contributes at one loop when
two of its long legs are contracted and the third is correlated
with the external mode. Such contributions are included in the
full correlators entering the consistency relation.}

The consistency relation~\eqref{eq:bw-soft-bispectrum} is not
restricted to tree level: it holds order by order in the loop
expansion, provided regularization and renormalization respect
the residual dilatation symmetry. We verify it explicitly at
one loop in the main text.
As an additional caveat, beyond tree level, extra care should be taken when using Eq.~\eqref{squeezedbackgroundwave} by correcting the long propagator that connects the external soft mode to the hard two-point response.

\subsection{Relation to the inhomogeneous cutoff}
\label{appsub:bw-cutoff}

The inhomogeneous cutoff introduced in the main text has a direct
interpretation in the background-wave description. Our prescription
regulates coincident short-field contractions with a momentum boundary
that takes the long-wavelength background into account, which is equivalent to imposing a homogeneous cutoff in local coordinates.

Split the curvature perturbation into long and short modes and consider a
frozen adiabatic long mode, neglecting its spatial gradients as in the background wave method. After rescaling the spatial coordinates, $\zetal$ disappears from
the short-mode action. To regularize loops of the local-coordinate field
$\widetilde{\zeta}$ at a fixed physical UV scale, it is then natural
to impose the homogeneous cutoff
$K(\tau)=a(\tau)\Lambda_{\rm UV}$ on the local momentum coordinate $\tilde{k}$.
Returning to the original coordinates gives
\begin{equation}
 \widetilde k<K(\tau)
 \quad\Longleftrightarrow\quad
 k<e^{\zetal}K(\tau).
 \label{eq:bw-cutoff-transformation}
\end{equation}
From the background-wave
perspective, the inhomogeneous cutoff prescription therefore follows simply from
imposing a homogeneous cutoff in the local coordinates.

For example, consider the free short-field contraction of two time
derivatives at the same spacetime point. With primes denoting conformal-time
derivatives, it is
\begin{align}
 \mathcal T_{\Lambda[\zetal]}(\tau|\zetal)=\int_{\widetilde k<K(\tau)}
       \frac{d^3\widetilde k}{(2\pi)^3}
       |u'_{\widetilde k}(\tau)|^2
 =\int_{k<e^{\zetal}K(\tau)}
       \frac{d^3k}{(2\pi)^3}
       e^{-3\zetal}|u'_{ke^{-\zetal}}(\tau)|^2.
 \label{eq:bw-invariant-contraction}
\end{align}
The second equality follows by changing the momentum variable.
In $\mathcal T_{\Lambda[\zetal]}(\tau\mid\zetal)$, the subscript
specifies the integration boundary, while the argument after
the vertical bar specifies the long-wavelength background
entering the short-mode functions. 

These two appearances of $\zetal$ in
Eq.~\eqref{eq:bw-invariant-contraction} have complementary roles.
The integrand  
is the short-field
contraction evaluated using the quadratic action in the long background.
Expanding it in $\zetal$ reproduces the effects of interaction terms
containing two short fields and one or more long fields. These effects
may therefore be incorporated either through the background-dependent
mode functions or perturbatively through explicit interaction insertions. The main-text calculation (at least for the power spectra) uses the latter organization: the free
contractions are built from the homogeneous mode functions, and the
long--short couplings are included through the interaction Hamiltonian.
The momentum boundary must also be expanded in $\zetal$. Holding the
homogeneous contraction fixed while expanding this boundary isolates the
additional regulator-induced contribution. This contribution is then one part
of the full conditional contraction.  
The distinction between the
background response of the integrand and the cutoff response is therefore
a perturbative decomposition of the same regulated quantity, which
transforms consistently under the residual spatial dilatation.
\section{Details of the in-in calculation for the one-loop squeezed bispectrum example}
\label{app:cq-ordinary-inin}

We give the power-spectrum integrals and the detailed ordinary in-in
calculation of the squeezed bispectrum for the illustrative example of Sec.~\ref{subsec:cq-power}. Fields and Hamiltonians use the unconditioned interaction
picture, with its subscript $I$ omitted as above. Write $Q_3=\zetal(\tau_f,\mathbf p)\short(\tau_f,\mathbf k_1)\short(\tau_f,\mathbf k_2)$, and let $H_{\rm int}$ be the interaction Hamiltonian. The calculation starts from the standard nested commutators,
\begin{align}
 \langle Q_3\rangle ={}&i\int^{\tau_f}d\tau_1\, \langle[H_{\rm int}(\tau_1),Q_3]\rangle -\int^{\tau_f}d\tau_1\int^{\tau_1}d\tau_2\, \langle[H_{\rm int}(\tau_2),[H_{\rm int}(\tau_1),Q_3]]\rangle \nonumber\\ &-i\int^{\tau_f}d\tau_1\int^{\tau_1}d\tau_2 \int^{\tau_2}d\tau_3\, \langle[H_{\rm int}(\tau_3),[H_{\rm int}(\tau_2),[H_{\rm int}(\tau_1),Q_3]]]\rangle +\cdots. \label{eq:cq-app-nested-inin}
\end{align}
We retain connected contractions and remove the overall momentum delta
function throughout.
Lower time limits carry the adiabatic prescription.

\subsection{Vertices, organization of the calculation, and one-loop spectrum integrals}
The starting action is given in Eq.~\eqref{eq:cq-action}.
Expanding the canonical Hamiltonian before using $\Pi=2\epsilon a^2\zeta'$
gives the additional quintic densities
\begin{equation}
 \mathcal H^{(5,g_4^0)} =\frac {\epsilon a^2}{6}[-27\zeta^3\zeta'^2 +\zeta^3(\partial_i\zeta)^2], \qquad \mathcal H^{(5,g_4)}=-\frac{9g_4}{24}\zeta\zeta'^4. \label{eq:cq-app-quintic}
\end{equation}
After $\zeta=\zetal+\short$, the vertices with one
undifferentiated long field are
\begin{align}
 H_{LSS} &=\epsilon a^2\int d^3x\,\zetal [-3\short'^2+(\partial_i\short)^2], \label{eq:cq-app-hlss}\\ H_{LSSS}^{(g_4^0)} &=\epsilon a^2\int d^3x\,\zetal [9\short\short'^2+\short(\partial_i\short)^2], \label{eq:cq-app-hlsss}\\ H_{LSSSS}^{(g_4^0)} &=\frac {\epsilon a^2}{2}\int d^3x\,\zetal [-27\short^2\short'^2 +\short^2(\partial_i\short)^2], \label{eq:cq-app-hlssss-zero}\\ H_{LSSSS}^{(g_4)} &=-\frac{9g_4}{24}\int d^3x\,\zetal\short'^4. \label{eq:cq-app-hlssss-g4}
\end{align}
The purely short vertices $H_{SSS}$ and $H_{SSSS}$ are the original
cubic and quartic Hamiltonians with $\zeta$ replaced by $\short$.
All products containing canonical momenta are understood with a
consistent Hermitian ordering.

For a quartic hard diagram, the ordinary squeezed contributions are a
single insertion of $H_{LSSSS}$ and a double insertion of $H_{SSSS}$
with $H_{LSS}$. For the cubic--cubic hard diagram, they are the double
insertion of $H_{LSSS}$ with $H_{SSS}$ and the triple insertion of two
$H_{SSS}$ vertices with $H_{LSS}$.

The $H_{LSS}$ insertions admit a useful simplification.
Once the soft field is replaced by its constant background value,
$H_{LSS}$ is quadratic in the short fields. We evaluate its
time-integrated insertion once and incorporate it into a corrected
short propagator, which can then be used on each internal or external
hard line. This reorganization is analogous to the mixed-propagator
method of Ref.~\cite{Chen:2017ryl}.
Direct
attachments through $H_{LSSS}$ or $H_{LSSSS}$ are not included in this
step and we add them separately.
We keep the field unsplit and set $Q_2=\zeta_{\mathbf k}(\tau_f)\zeta_{-\mathbf k}(\tau_f)$. 

The contribution to the power spectrum proportional to $g_4$ reads
\begin{equation}
 \begin{aligned} P_{\Lambda,4}^{(1,g_4)}(k)&=i\int_{-\infty}^{\tau_f}d\tau\,\langle[H^{(4,g_4)}(\tau),Q_2]\rangle'\\ &=\vcenter{\hbox{\quarticPSboundary}}=-g_4\int_{-\infty}^{\tau_f}d\tau\,\Sigma_0(\tau)\operatorname{Im}[G_k'(\tau)^2],\qquad G_k(\tau)\equiv u_k(\tau)u_k^*(\tau_f). \end{aligned} \label{eq:cq-power-integral}
\end{equation}
Here $\Sigma_0(\tau)=\int_{q<\cutoff(\tau)}d^3q\,|u'_q(\tau)|^2/(2\pi)^3$ is evaluated at the quartic-vertex time. In the leading de Sitter mode approximation, with $H$, $\epsilon$, and $g_4$ constant, we have
\begin{equation}
 u_k(\tau)=\frac{H}{2\sqrt \epsilon\,k^{3/2}}(1+ik\tau)e^{-ik\tau},\qquad u'_k(\tau)=\frac{H\sqrt k\,\tau}{2\sqrt \epsilon}e^{-ik\tau},\qquad P^{(0)}(k)=\frac{H^2}{4 \epsilon k^3}. \label{eq:cq-modes}
\end{equation}
Taking $\tau_f\to0^-$ yields
\begin{equation}
\Sigma_0(\tau)=\frac{a^2(\tau)\Lambda_{\rm UV}^4}{32\pi^2 \epsilon},\qquad P_{\Lambda,4}^{(1,g_4)}(k)=-\frac{g_4\Lambda_{\rm UV}^4}{256\pi^2\epsilon^2}P^{(0)}(k). \label{eq:cq-power-result}
\end{equation}
The correction is nonzero and retains the $k^{-3}$ dependence. 

For the seagull diagram independent of $g_4$, we define the remaining contractions at the same vertex time,
\begin{equation}
T_\Lambda=\int_{q<\cutoff}\frac{d^3q}{(2\pi)^3}|u_q|^2,\qquad V_\Lambda=\int_{q<\cutoff}\frac{d^3q}{(2\pi)^3}q^2|u_q|^2,\qquad M_\Lambda=\int_{q<\cutoff}\frac{d^3q}{(2\pi)^3}\operatorname{Re}(u_qu_q'^*). \label{eq:cq-zero-contractions}
\end{equation}
Contracting one pair in $\mathcal H^{(4,g_4^0)}$ gives the quadratic insertion
\begin{equation}
 \mathcal I_4^{(g_4^0)}=\frac {\epsilon a^2}{2}\int d^3x\left[9T_\Lambda\zeta'^2+(9\Sigma_0+V_\Lambda)\zeta^2+36M_\Lambda\frac{\zeta\zeta'+\zeta'\zeta}{2}+T_\Lambda(\partial\zeta)^2\right].
\end{equation}
The mixed contraction has four choices of one differentiated and one undifferentiated field. The first-order commutator consequently yields
\begin{equation}
 P_{\Lambda,4}^{(1,g_4^0)}(k)=\vcenter{\hbox{\quarticPSboundary}}=-2\int_{-\infty}^{\tau_f}d\tau\,\epsilon a^2\operatorname{Im}\left[9T_\Lambda G_k'^2+(9\Sigma_0+V_\Lambda+k^2T_\Lambda)G_k^2+36M_\Lambda G_kG_k'\right]. \label{eq:cq-power-quartic-zero}
\end{equation}
Since the cutoff moves with time, $M_\Lambda\ne\tfrac12\partial_\tau T_\Lambda$: the latter also differentiates the boundary. The separation $P_{\Lambda,4}^{(1,g_4^0)}=P_{4,K}+P_{4,\partial}$ used in the main text is specified in Eq.~\eqref{eq:cq-app-quartic-pieces}.

The other topology is the two-line loop joining two cubic vertices:
\begin{equation}
 P_{\Lambda,33}^{(1)}(k)=\vcenter{\hbox{\cubicPSboundary}}=-\int_{-\infty}^{\tau_f}d\tau_1\int_{-\infty}^{\tau_1}d\tau_2\,\langle[H^{(3)}(\tau_2),[H^{(3)}(\tau_1),Q_2]]\rangle_{\Lambda}'\big|. \label{eq:cq-power-cubic-cubic}
\end{equation}
Every Wick term contains two external contractions and two internal contractions, with all time and spatial derivatives assigned by the cubic vertices. The latter carry momenta $\mathbf q$ and $\mathbf k-\mathbf q$ and the complete window $W_\Lambda$ defined in Eq.~\eqref{windowW}. Both restrictions are retained in all momentum integrations by parts. One-point subgraphs are removed by the tadpole condition.

\subsection{Cubic soft insertion on a short line}
\label{app:cq-line-insertions}

For the purpose of this insertion, replace the contracted long field by
a constant bookkeeping parameter $\zetal$, and write $H_{LSS}=\zetal J$,
with ($A\equiv \epsilon a^2$ from now on)
\begin{equation}
 J(\tau)=\int d^3x\left[ -\frac{3\Pi^2}{4A}+A(\partial_i\short)^2\right]. \label{eq:cq-app-j}
\end{equation}
Here $\Pi=2A\short'$ is the free short-field canonical momentum.
Let $\delta_3$ denote the correction generated by one cubic $H_{LSS}$
insertion, including its factor of $\zetal$. The initial time $\tau_i$
is taken to the adiabatic past. For short fields or canonical momenta
$X$ and $Y$, the unordered contraction receives the correction
\begin{align}
 \delta_3\langle X(\tau_1)Y(\tau_2)\rangle ={}&i\zetal\int_{\tau_i}^{\tau_1}d\eta\, \langle[J(\eta),X(\tau_1)]Y(\tau_2)\rangle\nonumber\\ &+i\zetal\int_{\tau_i}^{\tau_2}d\eta\, \langle X(\tau_1)[J(\eta),Y(\tau_2)]\rangle. \label{eq:cq-app-line-commutator}
\end{align}
The two terms insert the quadratic interaction at either endpoint.
We first evaluate the correction to an undifferentiated field,
\begin{equation}
 \delta_3\short(\tau) =i\zetal\int_{\tau_i}^{\tau}d\eta\, [J(\eta),\short(\tau)]. \label{eq:cq-app-field-insertion}
\end{equation}
Rather than integrate the mode products directly, differentiate this
equation. The canonical commutators give
\begin{align}
 (\delta_3\short)' &=\frac{\delta_3\Pi}{2A}-3\zetal\short', \label{eq:cq-app-canonical-field}\\ \left(\frac{\delta_3\Pi}{2A}\right)' &=-\frac{A'}A\frac{\delta_3\Pi}{2A} +\partial^2\delta_3\short +\zetal\partial^2\short. \label{eq:cq-app-canonical-momentum}
\end{align}
The last term in Eq.~\eqref{eq:cq-app-canonical-field} comes from the
upper endpoint of the insertion integral. Eliminating $\delta_3\Pi$
then gives, for each Fourier mode,
\begin{equation}
 \mathscr L_q\delta_3\short(\tau,\mathbf q) =2\zetal q^2\short(\tau,\mathbf q), \qquad \mathscr L_q=\partial_\tau^2+\frac{A'}A\partial_\tau+q^2. \label{eq:cq-app-forced-mode}
\end{equation}

The free mode satisfies $\mathscr L_qu_q=0$. Differentiating this
identity with respect to its momentum immediately gives
\begin{equation}
 \mathscr L_q(q\partial_qu_q)=-2q^2u_q. \label{eq:cq-app-momentum-particular}
\end{equation}
Thus the mode coefficient of the field correction has the form
$\delta_3u_q=\zetal[-q\partial_qu_q+\alpha_qu_q+\beta_qu_q^*]$.
The adiabatic positive-frequency prescription fixes $\beta_q=0$.
Preserving the canonical commutator fixes
$\operatorname{Re}\alpha_q=-3/2$: this follows by varying
$u_qu_q'^*-u_q^*u_q'=i/(2A)$ together with
Eq.~\eqref{eq:cq-app-canonical-field}. A time-independent phase drops
out of contractions. We may therefore take
\begin{equation}
 \delta_3u_q=-\zetal\left(\frac32+q\partial_q\right)u_q. \label{eq:cq-app-field-mode-result}
\end{equation}
This fixes the solution with the stated state prescription. A finite
starting time requires the corresponding state correction and does not
in general permit setting all insertion corrections to zero there.

Applying Eq.~\eqref{eq:cq-app-field-mode-result} to both endpoints of
$C_q^{00}(\tau_1,\tau_2)=u_q(\tau_1)u_q^*(\tau_2)$ gives
\begin{equation}
 \delta_3C_q^{00} =\delta_3u_q(\tau_1)u_q^*(\tau_2) +u_q(\tau_1)\delta_3u_q^*(\tau_2) =-\zetal(3+q\partial_q)C_q^{00}. \label{eq:cq-app-plain-line}
\end{equation}
This is the desired replacement of the time-integrated insertion by
a momentum derivative.

For a differentiated endpoint we must insert the interaction on
$\short'=\Pi/(2A)$, rather than simply differentiate the corrected
field. Equation~\eqref{eq:cq-app-canonical-field} gives
\begin{equation}
 \delta_3\short' =\frac{\delta_3\Pi}{2A} =(\delta_3\short)'+3\zetal\short'. \label{eq:cq-app-derivative-endpoint}
\end{equation}
Its mode coefficient is consequently
$\zetal(3/2-q\partial_q)u_q'$. Defining
\begin{equation}
 C_q^{mn}(\tau_1,\tau_2) =u_q^{(m)}(\tau_1)u_q^{(n)*}(\tau_2),\qquad u_q^{(0)}=u_q,\quad u_q^{(1)}=u_q',\quad m,n\in\{0,1\},
\end{equation}
we obtain
\begin{align}
 \delta_3C_q^{00}&=-\zetal(3+q\partial_q)C_q^{00},\nonumber\\ \delta_3C_q^{01}&=-\zetal q\partial_qC_q^{01},\nonumber\\ \delta_3C_q^{11}&=\zetal(3-q\partial_q)C_q^{11}, \qquad \delta_3C_q^{mn} =\zetal[3(m+n-1)-q\partial_q]C_q^{mn}. \label{eq:cq-line-insertion-general}
\end{align}
These are canonical contractions. In ordered products one retains
the in-in ordering of the fields and momenta. One does not generate
them by differentiating time-ordering step functions. The reversed
unordered contractions follow by complex conjugation. This prescription
retains the canonical endpoint terms in Eq.~\eqref{eq:cq-app-derivative-endpoint}.

All these insertions are evaluated with the original loop window held
fixed. For a same-vertex contraction, integration by parts in momentum
then gives
\begin{align}
 \frac{\delta_3T_\Lambda}{\zetal}&=-\mathscr R_\Lambda T_\Lambda, &\frac{\delta_3\Sigma_0}{\zetal}&=(6-\mathscr R_\Lambda)\Sigma_0, \nonumber\\ \frac{\delta_3M_\Lambda}{\zetal}&=(3-\mathscr R_\Lambda)M_\Lambda, &\frac{\delta_3V_\Lambda}{\zetal}&=(2-\mathscr R_\Lambda)V_\Lambda. \label{eq:cq-app-bubble-insertions}
\end{align}
Here the differentiated contractions contain the unconditioned canonical
operator $\short'$. They must not be confused with the contractions of
$(\short^{[\zetal]})'$ of the conditioned field $\short^{[\zetal]}$.
For example, the coefficient
$6$ in the second relation is consistent with the conditioned canonical
momentum $\Pi_s^{[\zetal]}=2Ae^{3\zetal}(\short^{[\zetal]})'$.

\subsection{Bispectrum contribution proportional to $g_4$}
\label{app:cq-g4}
The notation below corresponds to the main-text labels as
$B_{4,\mathrm{vertex}}^{(1,g_4)}=B_5$,
$B_{4,\mathrm{bubble}}^{(1,g_4)}=B_{34}^{\rm loop}$, and
$B_{4,\mathrm{external}}^{(1,g_4)}=B_{34}^{\rm hard}$.

\noindent\textbf{Direct vertex attachment.}
Using the single-insertion formula with Eq.~\eqref{eq:cq-app-hlssss-g4},
\begin{align}
 B_{4,\mathrm{vertex}}^{(1,g_4)} &=i\int^{\tau_f}d\tau\, \langle[H_{LSSSS}^{(g_4)}(\tau),Q_3]\rangle_{\Lambda}' \nonumber\\ &=\vcenter{\hbox{\quarticBSvertexboundary}}=-9P^{(0)}(p)P_{\Lambda,4}^{(1,g_4)}(k). \label{eq:cq-app-g4-vertex}
\end{align}
The factor $-9$ is the ratio of the quintic soft vertex to the quartic
hard vertex at fixed canonical momentum. It is not the $+3$ obtained
by differentiating the conditioned-velocity expression at fixed
$(\short^{[\zetal]})'$.

\medskip
\noindent\textbf{Line attachments.}
The required second-order term is
\begin{align}
 B_{4,\mathrm{lines}}^{(1,g_4)} =-\int^{\tau_f}d\tau_1\int^{\tau_1}d\tau_2\, \Big\langle &[H_{LSS}(\tau_2),[H_{SSSS}^{(g_4)}(\tau_1),Q_3]] \nonumber\\[-1mm] +{}&[H_{SSSS}^{(g_4)}(\tau_2),[H_{LSS}(\tau_1),Q_3]] \Big\rangle_{\Lambda}'. \label{eq:cq-app-g4-lines-inin}
\end{align}
We evaluate this sum by correcting the bubble and each external line
of Eq.~\eqref{eq:cq-power-integral}. Equation~\eqref{eq:cq-app-bubble-insertions}
replaces $\Sigma_0$ by $\zetal(6-\mathscr R_\Lambda)\Sigma_0$.
Each external contraction $G_k'$ has one differentiated endpoint, so
Eq.~\eqref{eq:cq-line-insertion-general} replaces it by
$-\zetal k\partial_kG_k'$. Hence
\begin{align}
 B_{4,\mathrm{bubble}}^{(1,g_4)} &=\vcenter{\hbox{\quarticBSloopboundary}}=P^{(0)}(p)(6-\mathscr R_\Lambda)P_{\Lambda,4}^{(1,g_4)}(k), \nonumber\\ B_{4,\mathrm{external}}^{(1,g_4)} &=\vcenter{\hbox{\quarticBShardboundary}}+(\mathbf k_1\leftrightarrow\mathbf k_2)=-P^{(0)}(p)k\partial_kP_{\Lambda,4}^{(1,g_4)}(k). \label{eq:cq-app-g4-line-results}
\end{align}
The additional insertion-time integration in
Eq.~\eqref{eq:cq-app-g4-lines-inin} has thus been performed; only the
original quartic-vertex integral remains. Adding the direct vertex
attachment gives
\begin{equation}
 B_{4,\mathrm{hom}}^{(1,g_4)} =P^{(0)}(p)[-9+6-k\partial_k-\mathscr R_\Lambda] P_{\Lambda,4}^{(1,g_4)} =-P^{(0)}(p)(D_k+\mathscr R_\Lambda)P_{\Lambda,4}^{(1,g_4)}.
\end{equation}
Varying the contraction in the quartic vertex gives the effective
cutoff insertion
\begin{equation}
 \delta_{\rm cut}\mathcal I_4^{(g_4)}(\tau)
 =\frac{g_4}{4}\mathscr R_\Lambda\Sigma_0(\tau)
   \int d^3x\,\zetal\short'^2.
 \label{eq:cq-cutoff-effective-g4}
\end{equation}
Its single-insertion contribution is
\begin{align}
 B_{4,\mathrm{cut}}^{(1,g_4)} &=i\int^{\tau_f}d\tau\, \langle[\delta_{\rm cut}\mathcal I_4^{(g_4)}(\tau),Q_3]\rangle' \nonumber\\ &=-g_4P^{(0)}(p)\int^{\tau_f}d\tau\, \mathscr R_\Lambda\Sigma_0(\tau)\operatorname{Im}[G_k'(\tau)^2] \nonumber\\ &=\vcenter{\hbox{\quarticBScutoffboundary}}=P^{(0)}(p)\mathscr R_\Lambda P_{\Lambda,4}^{(1,g_4)}(k).
\end{align}
This gives Eq.~\eqref{eq:cq-cutoff-bispectrum} directly in ordinary
in-in perturbation theory.

\subsection{Bispectrum contribution independent of $g_4$}
\label{app:cq-quartic-zero}
Let us first consider the quartic bispectrum contribution.
Separate the power-spectrum expression into kinetic and gradient
pieces, $P_{\Lambda,4}^{(1,g_4^0)}=P_{4,K}+P_{4,\partial}$, defined by
\begin{align}
 P_{4,K}&=-2\int^{\tau_f}d\tau\,A\, \operatorname{Im}[9T_\Lambda G_k'^2+9\Sigma_0G_k^2 +36M_\Lambda G_kG_k'],\nonumber\\ P_{4,\partial}&=-2\int^{\tau_f}d\tau\,A\, \operatorname{Im}[(V_\Lambda+k^2T_\Lambda)G_k^2]. \label{eq:cq-app-quartic-pieces}
\end{align}
These labels refer to the quartic vertex from which each term originates.

\noindent\textbf{Direct vertex attachment.}
The single-insertion expression is
\begin{align}
 B_{4,\mathrm{vertex}}^{(1,g_4^0)} &=i\int^{\tau_f}d\tau\, \langle[H_{LSSSS}^{(g_4^0)}(\tau),Q_3]\rangle_{\Lambda}' \nonumber\\ &=\vcenter{\hbox{\quarticBSvertexboundary}}=P^{(0)}(p)[-3P_{4,K}+P_{4,\partial}]. \label{eq:cq-app-zero-vertex}
\end{align}
The relative coefficients $-3$ and $+1$ follow directly from
Eq.~\eqref{eq:cq-app-hlssss-zero} and the quartic Hamiltonian.

\medskip
\noindent\textbf{Line attachments.}
The line drawings are
\begin{equation}
 B_{4,\mathrm{lines}}^{(1,g_4^0)}=\vcenter{\hbox{\quarticBSloopboundary}}+\vcenter{\hbox{\quarticBShardboundary}}+(\mathbf k_1\leftrightarrow\mathbf k_2).
\end{equation}
The two-insertion expression is
\begin{align}
 B_{4,\mathrm{lines}}^{(1,g_4^0)} =-\int^{\tau_f}d\tau_1\int^{\tau_1}d\tau_2\, \Big\langle &[H_{LSS}(\tau_2),[H_{SSSS}^{(g_4^0)}(\tau_1),Q_3]] \nonumber\\[-1mm] +{}&[H_{SSSS}^{(g_4^0)}(\tau_2),[H_{LSS}(\tau_1),Q_3]] \Big\rangle_{\Lambda}'. \label{eq:cq-app-zero-lines-inin}
\end{align}
Use Eq.~\eqref{eq:cq-app-bubble-insertions} for the contracted pair,
and Eq.~\eqref{eq:cq-line-insertion-general} for the external lines.
For $T_\Lambda G_k'^2$, $\Sigma_0G_k^2$, and $M_\Lambda G_kG_k'$, the
constant weights from the bubble and external lines sum to zero.
Their combined insertion is therefore
$-(k\partial_k+\mathscr R_\Lambda)P_{4,K}$.
For $V_\Lambda G_k^2$ the corresponding weight is $2-6=-4$.
The $k^2T_\Lambda G_k^2$ term gives the same result: expressing the
derivative as $k\partial_k$ of the entire term adds $+2$ to the external
weight $-6$. Consequently,
\begin{equation}
 B_{4,\mathrm{lines}}^{(1,g_4^0)} =P^{(0)}(p)\left[ -(k\partial_k+\mathscr R_\Lambda)P_{4,K} -(4+k\partial_k+\mathscr R_\Lambda)P_{4,\partial}\right]. \label{eq:cq-app-zero-line-results}
\end{equation}
Adding Eq.~\eqref{eq:cq-app-zero-vertex} gives $-P^{(0)}(p)(D_k+\mathscr R_\Lambda)P_{\Lambda,4}^{(1,g_4^0)}$. The cutoff insertion is obtained by varying only the contractions in $\mathcal I_4^{(g_4^0)}$:
\begin{equation}
 \delta_{\rm cut}\mathcal I_4^{(g_4^0)}=\frac A2\int d^3x\,\zetal\left[9\mathscr R_\Lambda T_\Lambda\short'^2+(9\mathscr R_\Lambda\Sigma_0+\mathscr R_\Lambda V_\Lambda)\short^2+36\mathscr R_\Lambda M_\Lambda\frac{\short\short'+\short'\short}{2}+\mathscr R_\Lambda T_\Lambda(\partial\short)^2\right].
\end{equation}
With $Q_3$ defined at the start of this appendix,
\begin{equation}
 B_{4,\mathrm{cut}}^{(1,g_4^0)}=\vcenter{\hbox{\quarticBScutoffboundary}}=i\int^{\tau_f}d\tau\,\langle[\delta_{\rm cut}\mathcal I_4^{(g_4^0)}(\tau),Q_3]\rangle'=P^{(0)}(p)\mathscr R_\Lambda(P_{4,K}+P_{4,\partial}).
\end{equation}
This cancels the remaining boundary term and gives Eq.~\eqref{eq:cq-cutoff-quartic-zero}. For the de Sitter modes, radial differentiation at $q=\cutoff(\tau)$ yields
\begin{equation}
 \begin{aligned} \mathscr R_\Lambda T_\Lambda&=\frac{H^2+\Lambda_{\rm UV}^2}{8\pi^2C},\qquad \mathscr R_\Lambda\Sigma_0=\frac{a^2\Lambda_{\rm UV}^4}{8\pi^2C},\\ \mathscr R_\Lambda M_\Lambda&=\frac{\Lambda_{\rm UV}^2}{8\pi^2C\tau},\qquad \mathscr R_\Lambda V_\Lambda=\frac{a^2\Lambda_{\rm UV}^2(H^2+\Lambda_{\rm UV}^2)}{8\pi^2C}. \end{aligned} \label{eq:cq-zero-ds-boundaries}
\end{equation}
These UV-shell quantities do not require assigning a finite value to the IR-divergent contraction $T_\Lambda$.

We now turn to the cubic-cubic bispectrum contribution.
The original hard loop contains two $H_{SSS}$ vertices and has two
vertex-time integrations. We distinguish direct vertex attachments
from line attachments before simplifying either contribution.

\noindent\textbf{Direct vertex attachments.}
The two possible vertex attachments are represented by
\begin{equation}
 B_{33,\mathrm{vertex}}^{(1)}=\vcenter{\hbox{\cubicBvertexboundary}}
\end{equation}
Their second-order expression is
\begin{align}
 B_{33,\mathrm{vertex}}^{(1)} =-\int^{\tau_f}d\tau_1\int^{\tau_1}d\tau_2\, \Big\langle &[H_{LSSS}^{(g_4^0)}(\tau_2),[H_{SSS}(\tau_1),Q_3]] \nonumber\\[-1mm] +{}&[H_{SSS}(\tau_2),[H_{LSSS}^{(g_4^0)}(\tau_1),Q_3]] \Big\rangle_{\Lambda}'\big|_{\mathrm{bubble}}. \label{eq:cq-app-33-vertex-inin}
\end{align}
Consider a term $P_{33}^{[n_K,n_\partial]}$ in the hard loop with
$n_K$ kinetic cubic vertices and $n_\partial$ gradient cubic vertices,
where $n_K+n_\partial=2$. The soft vertex replaces a kinetic vertex by
$-3$ times that vertex, or a gradient vertex by $+1$ times that vertex.
Summing the two possible attachments gives
\begin{equation}
 B_{33,\mathrm{vertex}}^{[n_K,n_\partial]} =P^{(0)}(p)(-3n_K+n_\partial) P_{33}^{[n_K,n_\partial]}. \label{eq:cq-app-33-vertex-result}
\end{equation}

\medskip
\noindent\textbf{Line attachments.}
The internal- and external-line drawings are
\begin{equation}
 B_{33,\mathrm{lines}}^{(1)}=\vcenter{\hbox{\cubicBloopboundary}}\vcenter{\hbox{\cubicBhardboundary}}
\end{equation}
These arise at third order:
\begin{equation}
 \begin{aligned}
 B_{33,\mathrm{lines}}^{(1)}
 &=-i\int^{\tau_f}d\tau_1\int^{\tau_1}d\tau_2
       \int^{\tau_2}d\tau_3
       \sum_{(V_1,V_2,V_3)\in\mathcal P}
 \left\langle
 [V_3(\tau_3),[V_2(\tau_2),[V_1(\tau_1),Q_3]]]
 \right\rangle_{\Lambda}'.
 \end{aligned}
 \label{eq:cq-app-33-lines-inin}
\end{equation}
Here $\mathcal P$ contains the three ordered placements
$(H_{LSS},H_{SSS},H_{SSS})$,
$(H_{SSS},H_{LSS},H_{SSS})$, and
$(H_{SSS},H_{SSS},H_{LSS})$
at the times $(\tau_1,\tau_2,\tau_3)$.
The short legs of $H_{LSS}$ attach to either internal or either
external hard line of the original cubic--cubic diagram.
Using Eq.~\eqref{eq:cq-line-insertion-general} performs its time
integration, leaving the original two integrations over the cubic
hard vertices.

For a fixed term with $n_K$ kinetic vertices there are $2n_K$
time-differentiated endpoints among the four short propagators.
Their constant insertion weights sum to $6n_K-12$.
The gradient vertices supply a momentum polynomial of total degree
$2n_\partial$. Including this polynomial, and integrating the loop
momentum derivative by parts, gives
\begin{equation}
 B_{33,\mathrm{lines}}^{[n_K,n_\partial]} =P^{(0)}(p) [6n_K-9+2n_\partial-k\partial_k-\mathscr R_\Lambda] P_{33}^{[n_K,n_\partial]}. \label{eq:cq-app-33-lines-result}
\end{equation}
In this step the two-line window must be kept. More explicitly, for a
loop integrand $F$ and a window satisfying
$(k\partial_k+\mathbf q\cdot\partial_{\mathbf q}
 +\mathscr R_\Lambda)W_\Lambda=0$, one uses
\begin{equation}
 \int \frac{d^3 q}{(2\pi)^3}W_\Lambda (k\partial_k+\mathbf q\cdot\partial_{\mathbf q})F =(k\partial_k-3+\mathscr R_\Lambda) \int \frac{d^3 q}{(2\pi)^3}W_\Lambda F. \label{eq:cq-app-window-integration-by-parts}
\end{equation}
The external derivative denotes a dilatation of the external vector.
For the rotationally invariant final spectrum it is $k\partial_k$.
Equation~\eqref{eq:cq-app-window-integration-by-parts} includes the
variation of the cutoff on the second internal momentum
$|\mathbf k-\mathbf q|$, as well as on $q$.

Adding the vertex and line coefficients gives
\begin{equation}
 (-3n_K+n_\partial)+(6n_K-9+2n_\partial) =3(n_K+n_\partial)-9=-3.
\end{equation}
Therefore each kinetic/gradient combination, and hence their sum, obeys
\begin{equation}
 B_{33,\mathrm{hom}}^{(1)} =-P^{(0)}(p)(D_k+\mathscr R_\Lambda)P_{\Lambda,33}^{(1)}(k). \label{eq:cq-app-33-hom-final}
\end{equation}
For the crossed-cutoff contribution, set
$Q_2=\short(\tau_f,\mathbf k)\short(\tau_f,-\mathbf k)$.
It differentiates the complete window in the original two-time kernel:
\begin{equation}
 \begin{aligned} B_{33,\mathrm{cut}}^{(1)}&=\vcenter{\hbox{\cubicBcutoffboundary}}\\ &=-P^{(0)}(p)\int^{\tau_f}d\tau_1\int^{\tau_1}d\tau_2\,\mathscr R_\Lambda\langle[H_{SSS}(\tau_2),[H_{SSS}(\tau_1),Q_2]]\rangle_{\Lambda}'\\ &=P^{(0)}(p)\mathscr R_\Lambda P_{\Lambda,33}^{(1)}(k). \end{aligned}
\end{equation}
Both internal momentum restrictions are varied after Wick contraction. Adding this term to Eq.~\eqref{eq:cq-app-33-hom-final} gives Eq.~\eqref{eq:cq-completed-cubic-cubic}. The calculation uses canonical insertion equations and momentum integration by parts throughout.

\subsection{An additional check: the constant-background soft source}
\label{app:cq-soft-lines}
This check is independent of the main-text proof and does not require evaluating the full soft power correction. Following the organization of Pimentel et al.~\cite{Pimentel:2012tw}, we separate the quartics paired with cubic diagrams from those paired with tadpole counterterms. In our model all $g_4^0$ terms originate in the chosen symmetry completion. The separation concerns their role in the cancellation.

Write $\mathcal H_4^{(g_4^0)}=\mathcal H_{4,\mathrm{can}}+\mathcal H_{4,\mathrm{dil}}+\mathcal H_{4,\mathrm{rem}}$, with
\begin{equation}
 \begin{aligned} \mathcal H_{4,\mathrm{can}}&=9A\zeta^2\zeta'^2,\qquad \mathcal H_{4,\mathrm{dil}}=-A\zeta^2(\partial\zeta)^2,\\ \mathcal H_{4,\mathrm{rem}}&=\frac{3A}{2}[-3\zeta^2\zeta'^2+\zeta^2(\partial\zeta)^2]. \end{aligned} \label{eq:cq-app-soft-quartic-split}
\end{equation}
The first term comes from the Legendre transform, and the second is the spatial-derivative partner used in reconstructing the consistency-relation integrand. To exhibit the coefficients, write $H_{LSS}=\zetal J$, with
$J(\tau)=\int d^3x\,j(\tau,\mathbf x)$ and
$j=A[-3\short'^2+(\partial\short)^2]$. Averaging the short fields,
denoted by $\langle\cdot\rangle_s$, defines the undifferentiated
canonical source coefficient
\begin{equation}
 \overline J_\Lambda\equiv A(-3\Sigma_0+V_\Lambda)
 =\langle j\rangle_s.
\end{equation}
The bar distinguishes this coefficient from the full tadpole source, which also includes variations of long-field derivative terms. In the following equations $K$ denotes the coefficient of a constant $\zetal$ in this canonical source, before attachment to the external long propagators. The corresponding insertions in a squeezed tree diagram have the topologies
\begin{equation}
 \begin{array}{c@{\qquad}c@{\qquad}c@{\qquad}c} \vcenter{\hbox{\seagullBsoftboundary}}&\vcenter{\hbox{\cubicBsoftboundary}}&\vcenter{\hbox{\countertermBsoftboundary}}&\vcenter{\hbox{\cutoffBsoftboundary}}\\[1mm] {\scriptstyle\text{quartic}}&{\scriptstyle\text{cubic--cubic}}&{\scriptstyle\text{quadratic counterterm}}&{\scriptstyle\text{cutoff}} \end{array}. \label{eq:cq-soft-diagrams}
\end{equation}

Contracting the short fields in the remaining quartic gives $\mathcal H_{LL,\mathrm{rem}}\supset\tfrac32\overline J_\Lambda\zetal^2$. In this sector the completed counterterm supplies $\mathcal H_{\mathrm{ct}}\supset-\overline J_\Lambda(e^{3\zetal}-1)/3$, whose linear term cancels the tadpole and whose quadratic term cancels that seagull. Thus
\begin{equation}
 K_{4,\mathrm{rem}}=3\overline J_\Lambda,\qquad K_{\mathrm{ct}}=-3\overline J_\Lambda,\qquad K_{4,\mathrm{rem}}+K_{\mathrm{ct}}=0.
\end{equation}

The canonical and spatial-dilatation quartics give $K_{4,\mathrm{can}}=18A\Sigma_0$ and $K_{4,\mathrm{dil}}=-2AV_\Lambda$. Inserting $H_{LSS}=\zetal J$ into the expectation value of $j$,
with the boundary fixed, gives from Eq.~\eqref{eq:cq-app-bubble-insertions}
\begin{equation}
 K_{33}(\tau)
 =i\int^\tau d\eta\,
 \langle[J(\eta),j(\tau,\mathbf x)]\rangle_{\Lambda}
 =A[-3(6-\mathscr R_\Lambda)\Sigma_0
       +(2-\mathscr R_\Lambda)V_\Lambda].
 \label{eq:cq-app-source-insertion}
\end{equation}
Spatial homogeneity makes the result independent of $\mathbf x$.
Their bulk terms cancel, leaving the momentum-boundary contribution. The moving cutoff supplies its opposite:
\begin{equation}
 K_{33}+K_{4,\mathrm{can}}+K_{4,\mathrm{dil}}=-\mathscr R_\Lambda\overline J_\Lambda,\qquad K_{\mathrm{cut}}=+\mathscr R_\Lambda\overline J_\Lambda. \label{eq:cq-soft-source-cancellation}
\end{equation}
This displays the two cancellations in the undifferentiated canonical sector. Terms containing long-field derivatives must be retained before varying the full action to establish the complete long equation. The displayed source algebra is not a calculation of the entire finite-momentum propagator. The separate $g_4\zeta'^4$ interaction produces $\mathcal H_{LL,\mathrm{loop}}^{(g_4)}=(g_4/4)\Sigma_0\zetal'^2$, which preserves the constant solution. None of these observations fixes the full $P_\Lambda^{(1)}(p)$ entering Eq.~\eqref{eq:bw-one-loop-consistency}.

\section{Summary of induced Hamiltonian}
\label{app:induced_hamiltonian}

In this Appendix, we summarize the induced Hamiltonian associated with $\zeta'$. 
The following analysis is based on Ref.~\cite{Huang:2006eha}, though we explicitly obtain the quintic induced Hamiltonian, which is not obtained in the reference.
We start with 
\begin{align}
    \mathcal L = \frac{E}{2}[v^2 - (\partial_i \zeta)^2] + \mathcal L_3[\zeta,v] + \mathcal L_4[\zeta,v] + \mathcal L_5[\zeta,v],
\end{align}
where $E \equiv 2a^2 \epsilon (=2A)$ and $v \equiv \zeta'$. 
We have neglected $\mathcal O(\zeta^6)$ because it contributes only as two- or higher-loop corrections to the bispectrum. 
We define 
\begin{align}
    h = - \mathcal L_3, \ l = -\mathcal L_4, \ m = - \mathcal L_5.
\end{align}
The conjugate momentum for $\zeta$ is given by 
\begin{align}
    P \equiv \frac{\delta \mathcal L}{\delta v} = E v - h_{,v} - l_{,v} - m_{,v},
    \label{eq:P_def}
\end{align}
where $h_{,v} = \delta h[\zeta,v]/\delta v$ and the same for $l_{,v}$ and $m_{,v}$.
Using this, we can express $v$ with $P$.
For convenience, we define $u \equiv P/E$. 
The difference between $v$ and $u$ is given by 
\begin{align}
    \tilde \delta \equiv v-u = \frac{h_{,v}[\zeta,u+\tilde\delta] + l_{,v}[\zeta, u+ \tilde\delta] + m_{,v}[\zeta, u+ \tilde\delta]}{E}.
\end{align}
Since $h_{,u}$ is quadratic and $l_{,u}$ is cubic in perturbations, we set 
\begin{align}
    \tilde\delta = \tilde\delta_2 + \tilde\delta_3 + \cdots. 
\end{align}
Matching field orders gives 
\begin{align}
    \tilde\delta_2 = \frac{h_{,u}}{E}, \ \tilde\delta_3 = \frac{h_{,uu} h_{,u}}{E^2} + \frac{l_{,u}}{E},
\end{align}
where $h_{,u} = \delta h[\zeta,u]/\delta u$, $h_{,uu} = \delta^2 h[\zeta,u]/\delta u^2$, and $l_{,u} = \delta l[\zeta,u]/\delta u$.
The Hamiltonian density is given by 
\begin{align}
    \mathcal H &= Pv - \mathcal L.
\end{align}
We reexpress $\mathcal H$ with $u$ instead of $v$ by using Eq.~(\ref{eq:P_def}),
\begin{align} 
    \mathcal H&= \mathcal H_\text{free}[\zeta,u] + \mathcal H_\text{int}[\zeta,u],
\end{align}
where the free Hamiltonian $\mathcal H_\text{free}$ is given by 
\begin{align}
    \mathcal H_\text{free}  
    &= \frac{E}{2}[u^2 + (\partial_i \zeta)^2],
\end{align}
and the interaction Hamiltonian $\mathcal H_\text{int}$ is given by
\begin{align}
    \mathcal H_\text{int} &= -\frac{E}{2} \tilde\delta^2 + h[\zeta, u+ \tilde\delta] + l[\zeta,u+\tilde\delta] + m[\zeta,u+ \tilde\delta].
\end{align} 
Then, we obtain
\begin{align}
    \mathcal H^{(3)} &= h, \\ 
    \mathcal H^{(4)} &= l + h_{,u} \tilde\delta_2 - \frac{E}{2} \tilde\delta_2^2 \nonumber \\ 
    &= l + \frac{h_{,u}^2}{2E}, \\ 
    \mathcal H^{(5)} &= m+ h_{,u} \tilde\delta_3 - E \tilde\delta_2 \tilde\delta_3 + \frac{1}{2} h_{,uu} \tilde\delta_2^2 + l_{,u} \tilde\delta_2 \nonumber \\ 
    &= m + \frac{h_{,uu} h_{,u}^2}{2E^2} + \frac{l_{,u} h_{,u}}{E}.
\end{align}
The induced Hamiltonian is defined as $\mathcal H^{(4)}_{\text{ind}} \equiv \mathcal H^{(4)} - l$ and $\mathcal H^{(5)}_{\ind} \equiv \mathcal H^{(5)} - m$.

\section{Induced Hamiltonian contributions for the contact terms}
\label{app:one_loop_bispectrum}

In this Appendix, we clarify the relation between the induced Hamiltonian and the contact terms that arise when the interactions include $\zeta'$.
The main goal of this Appendix is to obtain the explicit expressions of the induced contributions in Eqs.~(\ref{eq:q_3_c}) and (\ref{eq:q_3_4c}).
We focus on the bispectrum of curvature perturbations, $Q_3(\tau_f) \equiv \zeta_{\bp_1}(\tau_f) \zeta_{\bp_2}(\tau_f) \zeta_{\bp_3}(\tau_f)$, with the same notations as in Appendix~\ref{app:induced_hamiltonian} and omit the time argument $\tau_f$ for brevity in the following.

We begin with the contributions related to the induced quintic Hamiltonian,
\begin{align}
    B_{5,\ind} &= i \int^{\tau_f}_{\tau_\star} d \tau_1 \expval{[H_{5,\ind}(\tau_1),Q_3]}' \nonumber \\ 
    &= i \int^{\tau_f}_{\tau_\star} d \tau_1 \sum_{\text{cyc(1,2,3)}}\sum_A [\zeta^A_{-\bp_1}(\tau_1),\zeta_{\bp_1}]' \left\langle \frac{\partial \mathcal H_{5,\ind}}{\partial \zeta^A}\Big|_{\tau_1,\bp_1} \zeta_{\bp_2} \zeta_{\bp_3} \right\rangle' \nonumber \\ 
    &= i \int^{\tau_f}_{\tau_\star} d \tau_1 \sum_{\text{cyc(1,2,3)}}\sum_A [\zeta^A_{-\bp_1}(\tau_1),\zeta_{\bp_1}]' \nonumber \\
    &\qquad \times \left \langle \left(\frac{h_{,uuA} h_{,u}^2 + 2 h_{,uu} h_{,u} h_{,uA}}{2E^2} + \frac{l_{,uA}h_{,u} + l_{,u} h_{,uA}}{E} \right)\Big|_{\tau_1,\bp_1} \zeta_{\bp_2} \zeta_{\bp_3}\right\rangle',
\end{align}
where $h_{,uA} = \delta^2 h/\delta u \delta \zeta^A$, $h_{,uuA} = \delta^3 h/\delta u^2 \delta \zeta^A$, and $l_{,uA} = \delta^2 l/\delta u \delta \zeta^A$.
For later convenience, we split this contribution into two parts, $\expval{\cdots}_{5,\ind} = \expval{\cdots}_{5,\ind,a} + \expval{\cdots}_{5,\ind,b}$, with
\begin{align}
    \label{eq:3_5_ind_a}
        B_{5,\ind,a}   &= i \int^{\tau_f}_{\tau_\star} d \tau_1 \sum_{\text{cyc(1,2,3)}}\sum_A [\zeta^A_{-\bp_1}(\tau_1),\zeta_{\bp_1}]' \nonumber \\ 
        &\qquad \times \left \langle \left(\frac{h_{,uuA} h_{,u}^2 + 2 h_{,uu} h_{,u} h_{,uA}}{2E^2} + \frac{l_{,u} h_{,uA}}{E} \right)\Big|_{\tau_1,\bp_1} \zeta_{\bp_2} \zeta_{\bp_3} \right \rangle', \\ 
    \label{eq:3_5_ind_b}        
    B_{5,\ind,b}   &= i \int^{\tau_f}_{\tau_\star} d \tau_1 \sum_{\text{cyc(1,2,3)}} \sum_A [\zeta^A_{-\bp_1}(\tau_1),\zeta_{\bp_1}]' \left \langle \frac{l_{,uA} h_{,u}}{E} \Big|_{\tau_1,\bp_1} \zeta_{\bp_2} \zeta_{\bp_3} \right \rangle'.
\end{align}
The quartic induced Hamiltonian provides 
\begin{align}
    \label{eq:3_43_ind}    
    B_{34,\ind} &= -\int^{\tau_f}_{\tau_\star} d \tau_1 \int^{\tau_1}_{-\infty} d\tau_2 \expval{[H^{(3)}(\tau_2),[H_{4,\ind}(\tau_1),Q_3]]}' \nonumber \\ 
    &= -\int^{\tau_f}_{\tau_\star} d \tau_1 \int^{\tau_1}_{-\infty} d\tau_2\sum_{\text{cyc(1,2,3)}}\sum_A [\zeta^A_{-\bp_1}(\tau_1),\zeta_{\bp_1}]' \nonumber \\ 
    &\qquad \times \frac{1}{E(\tau_1)}\left\langle [H^{(3)}(\tau_2),h_{,u} h_{,uA}|_{\tau_1,\bp_1} \zeta_{\bp_2} \zeta_{\bp_3}] \right\rangle'.
\end{align}
We here show that the sum of some contributions becomes
\begin{align}
    \label{eq:q3_connect}
    B_{3,a}
    &\equiv i\int^{\tau_f}_{\tau_\star} d \tau_1   \sum_{\text{cyc(1,2,3)}} \sum_A [\zeta^A_{-\bp_1}(\tau_1),\zeta_{\bp_1}]' \expval{h_{,A}|_{\tau_1,\bp_1}\zeta_{\bp_2} \zeta_{\bp_3}}_{c}',
\end{align}
where this was defined in Eq.~\eqref{eq:q_3_c} and the subscript $c$ denotes the connected part. 
Note that $h_{,A}$ can include $\zeta'$ and $\zeta'_H = U^\dagger_I \left(\zeta'_I + i[H_{\text{int},I},\zeta_I] \right) U_I$, where we here explicitly put the subscripts $H$ and $I$ to clarify the quantities in the Heisenberg and interaction pictures, respectively, and $U_I$ satisfies $i U_I' = H_{\text{int},I} U_I$. 
This difference between $\zeta_H'$ and $\zeta_I'$ provides the so-called contact terms~\cite{Pimentel:2012tw}.

The contact-term-related contributions can be expressed as 
\begin{align}
    \label{eq:3_c_contact}
    \expval{h_{,A}|_{\tau_1,\bp_1} \zeta_{\bp_2} \zeta_{\bp_3}}_{c,\text{contact}}' &= 
     - [u,\zeta]'_{\tau_1} \int^{\tau_1}_{-\infty} d\tau_2\expval{ [H^{(3)}(\tau_2), h_{,u} h_{,Au}|_{\tau_1,\bp_1} \zeta_{\bp_2} \zeta_{\bp_3}]}'_c  \nonumber \\ 
     & \quad + i [u,\zeta]'_{\tau_1} \left \langle \left(l_{,u} h_{,Au} + \frac{h_{,u} h_{,uu} h_{,Au}}{E}\right)\Big|_{\tau_1,\bp_1} \zeta_{\bp_2} \zeta_{\bp_3} \right \rangle'_c \nonumber \\ 
     &\quad - \frac{1}{2} ([u,\zeta]'_{\tau_1})^2 \expval{ h_{,u}^2 h_{,Auu}|_{\tau_1,\bp_1} \zeta_{\bp_2} \zeta_{\bp_3}}'_c \nonumber \\ 
     &= \frac{i}{E(\tau_1)} \int^{\tau_1}_{-\infty} d \tau_2\expval{ [H^{(3)}(\tau_2),h_{,u} h_{,Au}|_{\tau_1,\bp_1} \zeta_{\bp_2} \zeta_{\bp_3}] }'_c \nonumber \\ 
     &\qquad + \left\langle \left(\frac{ h_{,u}^2 h_{,Auu} + 2h_{,uu} h_{,u} h_{,Au}}{2E^2} + \frac{l_{,u} h_{,Au}}{E} \right)\Big|_{\tau_1,\bp_1} \zeta_{\bp_2} \zeta_{\bp_3} \right\rangle'_c,
\end{align}
where we have neglected higher-order contributions and used the fact that the equal-time commutator is given by 
\begin{align}
    \label{eq:u_zeta_rel}
    [u,\zeta]'_{\tau} = [\zeta'_{-\bp}(\tau),\zeta_{\bp}(\tau)]' = -\frac{i}{2 \epsilon(\tau)a^2(\tau)} = -\frac{i}{E(\tau)}.
\end{align}
Note that the connected part projection with the subscript $c$ on the right-hand side of Eq.~(\ref{eq:3_c_contact}) inherits the left-hand side before expanding the time-derivative contact terms. 
For instance, $\expval{\cdots h_{,Au} \zeta_{\bp_2} \zeta_{\bp_3}}'_c$ does not allow the contraction with $h_{,Au}$ and $\zeta_{\bp_2}$ or $\zeta_{\bp_3}$. 
Also, $\expval{ h_{,u}^2 h_{,Auu}|_{\tau_1,\bp_1} \zeta_{\bp_2} \zeta_{\bp_3}}'_c$ excludes the disconnected part, $\expval{h_{,u}} \expval{ h_{,u} h_{,Auu}|_{\tau_1,\bp_1} \zeta_{\bp_2} \zeta_{\bp_3}}'$.
By substituting Eq.~(\ref{eq:3_c_contact}) into Eq.~(\ref{eq:q3_connect}) and comparing it with Eqs.~(\ref{eq:3_5_ind_a}) and (\ref{eq:3_43_ind}), we obtain 
\begin{align}
    \label{eq:q3_connect_contact}
    B_{c,\text{contact}}
    &\equiv i\int^{\tau_f}_{\tau_\star} d \tau_1   \sum_{\text{cyc(1,2,3)}} \sum_A [\zeta^A_{-\bp_1}(\tau_1),\zeta_{\bp_1}]' \expval{h_{,A}|_{\tau_1,\bp_1}\zeta_{\bp_2} \zeta_{\bp_3}}_{c,\text{contact}}' \nonumber \\ 
    &= B_{34,\ind,c} + B_{5,\ind,a,c},
\end{align}
where the subscript $c$ means the connected part inherited from the left-hand side of Eq.~(\ref{eq:3_c_contact}), as explained below Eq.~(\ref{eq:u_zeta_rel}).
Using this, we can reexpress Eq.~(\ref{eq:q3_connect}) as 
\begin{align}
    \label{eq:q_3_c_app}
    B_{3,a}= B_{333,a} + B_{43,\text{1PI}} + B_{34,\ind,c} + B_{5,\ind,a,c}.
\end{align}
This corresponds to Eq.~(\ref{eq:q_3_c}) in the main text. 
More specifically, $B_{\ind,X} = B_{34,\ind,c} + B_{5,\ind,a,c}$. 

Next, we focus on the contact term in Eq.~(\ref{eq:q_3_4c}),
\begin{align}
    \label{eq:q4_connect}
    B_{4}
    &\equiv
    i\int^{\tau_f}_{\tau_\star} d \tau_1  \sum_{\text{cyc}(1,2,3)}\sum_{A} [\zeta^A_{-\bp_1}(\tau_1),\zeta_{\bp_1}]' 
    \expval{l_{,A}|_{\tau_1,\bp_1}\zeta_{\bp_2} \zeta_{\bp_3}}',
\end{align}
where we do not restrict ourselves to the 1-PI contributions unlike the main text. 
Similarly to Eq.~(\ref{eq:3_c_contact}), the contact-term-related contribution at one-loop level is given by 
\begin{align}
    \label{eq:l_a_contact}
    \expval{l_{,A}|_{\tau_1,\bp_1} \zeta_{\bp_2} \zeta_{\bp_3}}'_\text{contact} 
     &= i [u,\zeta]'_{\tau_1} \expval{ h_{,u} l_{,Au}|_{\tau_1,\bp_1} \zeta_{\bp_2} \zeta_{\bp_3}}' \nonumber \\ 
     &= \left \langle \frac{l_{,uA}h_{,u}}{E}\Big|_{\tau_1,\bp_1} \zeta_{\bp_2} \zeta_{\bp_3} \right \rangle'.
\end{align}
Note that we here only consider the single contact terms, while we consider the double contact terms in Eq.~(\ref{eq:3_c_contact}). 
This is because the double contact terms in Eq.~(\ref{eq:l_a_contact}) are only relevant to two- or higher-order loop contributions. 
From this and Eq.~(\ref{eq:3_5_ind_b}), we obtain 
\begin{align}
    \label{eq:q_4_app}
    B_{4} = B_{34,\text{non-ind}} + B_{5,\ind,b}.
\end{align}
This corresponds to Eq.~(\ref{eq:q_3_4c}) in the main text with $B_{\text{ind},Y} = B_{5,\text{ind},b,\text{1PI}}$, where the subscript 1PI here means the same projection as $B_{34,\text{1PI}}$ before expanding the time-derivative contact terms. 
It is explicitly given by 
\begin{align}
    B_{5,\text{ind},b,\text{1PI}} &= i \int^{\tau_f}_{\tau_\star} d \tau_1 \sum_{\sigma \in S_3} \sum_{A,B} [\zeta^A_{-\bp_{\sigma(1)}}(\tau_1),\zeta_{\bp_{\sigma(1)}}]' \nonumber \\ 
    &\qquad \times \langle \zeta^B_{-\bp_{\sigma(2)}}(\tau_1) \zeta_{\bp_{\sigma(2)}} \rangle'  \left \langle \frac{l_{,uAB} h_{,u}}{E} \Big|_{\tau_1,\bp_{\sigma(1)} + \bp_{\sigma(2)}} \zeta_{\bp_{\sigma(3)}} \right \rangle'_\text{1PI},
\end{align}
where the 1-PI projection excludes the contribution proportional to $\expval{h_{,u}}\expval{l_{,uAB} |_{\tau_1,\bp_{\sigma(1)} + \bp_{\sigma(2)}} \zeta_{\bp_{\sigma(3)}}}'$.

\bibliographystyle{JHEP}
\bibliography{Biblio}

@article{Fang:2026off,
    author = "Fang, Cheng-Jun and Guo, Zong-Kuan",
    title = "{Fixing the renormalization of inflationary loops via Ward identities}",
    eprint = "2605.13325",
    archivePrefix = "arXiv",
    primaryClass = "gr-qc",
    doi = "10.1007/JHEP10(2026)121",
    journal = "JHEP",
    volume = "10",
    pages = "121",
    year = "2026"
}

@article{Benincasa:2024ptf,
    author = "Benincasa, Paolo and Brunello, Giacomo and Mandal, Manoj K. and Mastrolia, Pierpaolo and Vaz{\~a}o, Francisco",
    title = "{One-loop corrections to the Bunch-Davies wave function of the universe}",
    eprint = "2408.16386",
    archivePrefix = "arXiv",
    primaryClass = "hep-th",
    doi = "10.1103/PhysRevD.111.085016",
    journal = "Phys. Rev. D",
    volume = "111",
    number = "8",
    pages = "085016",
    year = "2025"
}

@article{Lee:2023jby,
    author = "Lee, Mang Hei Gordon and McCulloch, Ciaran and Pajer, Enrico",
    title = "{Leading loops in cosmological correlators}",
    eprint = "2305.11228",
    archivePrefix = "arXiv",
    primaryClass = "hep-th",
    doi = "10.1007/JHEP11(2023)038",
    journal = "JHEP",
    volume = "11",
    pages = "038",
    year = "2023"
}

@article{Jain:2025maa,
    author = "Jain, Diksha and Pajer, Enrico and Tong, Xi",
    title = "{Unitary and Analytic Renormalisation of Cosmological Correlators}",
    eprint = "2509.02696",
    archivePrefix = "arXiv",
    primaryClass = "hep-th",
    doi = "10.1007/JHEP03(2026)225",
    journal = "JHEP",
    volume = "03",
    pages = "225",
    year = "2026"
}

@article{Farren:2026hao,
    author = "Farren, Alexander and McCulloch, Ciaran and Pajer, Enrico and Tong, Xi",
    title = "{All-Loop Renormalization and the Phase of the de Sitter Wavefunction}",
    eprint = "2603.08794",
    archivePrefix = "arXiv",
    primaryClass = "hep-th",
    month = "3",
    year = "2026"
}

@article{Qin:2026yuo,
    author = "Qin, Zhehan and Xianyu, Zhong-Zhi",
    title = "{Cosmological Collider Signals From a Triangle Loop}",
    eprint = "2609.09288",
    archivePrefix = "arXiv",
    primaryClass = "hep-th",
    month = "9",
    year = "2026"
}

@article{Pimentel:2026kqc,
    author = "Pimentel, Guilherme L. and Westerdijk, Tom",
    title = "{On Cosmological Correlators at One Loop}",
    eprint = "2601.00952",
    archivePrefix = "arXiv",
    primaryClass = "hep-th",
    month = "1",
    year = "2026"
}

@article{Henn:2026lfz,
    author = "Henn, Johannes and Mei, Jiajie and Yang, Qinglin",
    title = "{A compact analytic formula for the one-loop triangle cosmological correlator}",
    eprint = "2608.17877",
    archivePrefix = "arXiv",
    primaryClass = "hep-th",
    reportNumber = "MPP-2026-138",
    month = "8",
    year = "2026"
}

@article{Qin:2023bjk,
    author = "Qin, Zhehan and Xianyu, Zhong-Zhi",
    title = "{Inflation correlators at the one-loop order: nonanalyticity, factorization, cutting rule, and OPE}",
    eprint = "2304.13295",
    archivePrefix = "arXiv",
    primaryClass = "hep-th",
    doi = "10.1007/JHEP09(2023)116",
    journal = "JHEP",
    volume = "09",
    pages = "116",
    year = "2023"
}

@article{Bodas:2025vpb,
    author = "Bodas, Arushi and Broadberry, Edward and Sundrum, Raman and Xu, Zhaohui",
    title = "{Charged loops at the cosmological collider with chemical potential}",
    eprint = "2507.22978",
    archivePrefix = "arXiv",
    primaryClass = "hep-ph",
    reportNumber = "FERMILAB-PUB-25-0519-V",
    doi = "10.1007/JHEP01(2026)083",
    journal = "JHEP",
    volume = "01",
    pages = "083",
    year = "2026"
}

@article{Braglia:2026fzp,
    author = "Braglia, Matteo and Franciolini, Gabriele and Ijaz, Nadir and Peloso, Marco",
    title = "{Mean-Field Backreaction in Single-Field Inflation}",
    eprint = "2610.26669",
    archivePrefix = "arXiv",
    primaryClass = "astro-ph.CO",
    reportNumber = "CERN-TH-2026-239",
    month = "10",
    year = "2026"
}

@article{Ballesteros:2026fwo,
    author = "Ballesteros, Guillermo and Gamb{\'\i}n Egea, Jes{\'u}s and Riccardi, Flavio",
    title = "{Background-independent one-loop renormalization of tensor and scalar primordial spectra}",
    eprint = "2607.06042",
    archivePrefix = "arXiv",
    primaryClass = "hep-th",
    doi = "10.1007/JHEP11(2026)054",
    journal = "JHEP",
    volume = "11",
    pages = "054",
    year = "2026"
}

@article{Inomata:2022yte,
    author = "Inomata, Keisuke and Braglia, Matteo and Chen, Xingang and Renaux-Petel, S{\'e}bastien",
    title = "{Questions on calculation of primordial power spectrum with large spikes: the resonance model case}",
    eprint = "2211.02586",
    archivePrefix = "arXiv",
    primaryClass = "astro-ph.CO",
    doi = "10.1088/1475-7516/2023/04/011",
    journal = "JCAP",
    volume = "04",
    pages = "011",
    year = "2023",
    note = "[Erratum: JCAP 09, E01 (2023)]"
}

@article{Chen:2016hrz,
    author = "Chen, Xingang and Wang, Yi and Xianyu, Zhong-Zhi",
    title = "{Standard Model Mass Spectrum in Inflationary Universe}",
    eprint = "1612.08122",
    archivePrefix = "arXiv",
    primaryClass = "hep-th",
    doi = "10.1007/JHEP04(2017)058",
    journal = "JHEP",
    volume = "04",
    pages = "058",
    year = "2017"
}

@article{Tanaka:2026zew,
    author = "Tanaka, Takahiro and Urakawa, Yuko",
    title = "{Locality in effective field theory for inflationary soft modes}",
    eprint = "2605.19331",
    archivePrefix = "arXiv",
    primaryClass = "gr-qc",
    reportNumber = "KEK-TH-2839, KEK-Cosmo-0422",
    month = "5",
    year = "2026"
}

@article{Ema:2026dop,
    author = "Ema, Yohei and Hong, Muzi and Jinno, Ryusuke and Mukaida, Kyohei",
    title = "{Cancellation of loop corrections to soft scalar power spectrum}",
    eprint = "2603.01961",
    archivePrefix = "arXiv",
    primaryClass = "astro-ph.CO",
    month = "3",
    year = "2026"
}

@article{Braglia:2025qrb,
    author = "Braglia, Matteo and Pinol, Lucas",
    title = "{Freezing of the renormalized one-loop primordial scalar power spectrum}",
    eprint = "2504.13136",
    archivePrefix = "arXiv",
    primaryClass = "astro-ph.CO",
    doi = "10.1103/5jkv-mj3w",
    journal = "Phys. Rev. D",
    volume = "113",
    number = "6",
    pages = "L061302",
    year = "2026"
}

@article{Inomata:2025pqa,
    author = "Inomata, Keisuke",
    title = "{Role of the counterterms in the conservation of superhorizon curvature perturbations at one loop}",
    eprint = "2502.12112",
    archivePrefix = "arXiv",
    primaryClass = "astro-ph.CO",
    doi = "10.1103/r8bh-s48f",
    journal = "Phys. Rev. D",
    volume = "111",
    number = "12",
    pages = "123517",
    year = "2025"
}

@article{Fang:2025vhi,
    author = "Fang, Cheng-Jun and Lyu, Zhen-Hong and Chen, Chao and Guo, Zong-Kuan",
    title = "{Incorporating backreaction in one-loop corrections in ultraslow-roll inflation}",
    eprint = "2502.09555",
    archivePrefix = "arXiv",
    primaryClass = "gr-qc",
    doi = "10.1103/nkrq-3d39",
    journal = "Phys. Rev. D",
    volume = "112",
    number = "2",
    pages = "023547",
    year = "2025"
}

@article{Fumagalli:2023zzl,
    author = "Fumagalli, Jacopo",
    title = "{Absence of one-loop effects on large scales from small scales in non-slow-roll dynamics}",
    eprint = "2305.19263",
    archivePrefix = "arXiv",
    primaryClass = "astro-ph.CO",
    doi = "10.1007/JHEP05(2025)162",
    journal = "JHEP",
    volume = "05",
    pages = "162",
    year = "2025"
}

@article{Frolovsky:2025qre,
    author = "Frolovsky, Daniel and Ketov, Sergei V.",
    title = "{One-loop corrections to the E-type {\ensuremath{\alpha}}-attractor models of inflation and primordial black hole production}",
    eprint = "2502.00628",
    archivePrefix = "arXiv",
    primaryClass = "gr-qc",
    reportNumber = "IPMU25-0003",
    doi = "10.1103/PhysRevD.111.083533",
    journal = "Phys. Rev. D",
    volume = "111",
    number = "8",
    pages = "083533",
    year = "2025"
}

@article{Sheikhahmadi:2024peu,
    author = "Sheikhahmadi, Haidar and Nassiri-Rad, Amin",
    title = "{RG-Flow Renormalized One-Loop Corrections to the Power Spectrum in USR Inflation}",
    eprint = "2411.18525",
    archivePrefix = "arXiv",
    primaryClass = "astro-ph.CO",
    month = "11",
    year = "2024"
}

@article{Inomata:2026csq,
    author = "Inomata, Keisuke",
    title = "{Cancellation of one-loop time dependence in superhorizon curvature perturbations from all scales}",
    eprint = "2606.28247",
    archivePrefix = "arXiv",
    primaryClass = "astro-ph.CO",
    doi = "10.1103/vzh3-jxpm",
    journal = "Phys. Rev. D",
    volume = "114",
    number = "10",
    pages = "103514",
    year = "2026"
}

@article{Ballesteros:2024cef,
    author = "Ballesteros, Guillermo and Gamb{\'\i}n Egea, Jes{\'u}s and Riccardi, Flavio",
    title = "{Finite parts of inflationary loops}",
    eprint = "2411.19674",
    archivePrefix = "arXiv",
    primaryClass = "hep-th",
    reportNumber = "IFT UAM-CSIC 24-168",
    doi = "10.1007/JHEP06(2025)098",
    journal = "JHEP",
    volume = "06",
    pages = "098",
    year = "2025"
}

@article{Fang:2026qhu,
    author = "Fang, Cheng-Jun and Guo, Zong-Kuan",
    title = "{Nonlinear gauge maps and the structure of Goldstone loops in canonical inflation}",
    eprint = "2611.03571",
    archivePrefix = "arXiv",
    primaryClass = "hep-th",
    month = "11",
    year = "2026"
}

@article{Green:2024fsz,
    author = "Green, Daniel and Gupta, Kshitij",
    title = "{Soft metric fluctuations during inflation}",
    eprint = "2410.11973",
    archivePrefix = "arXiv",
    primaryClass = "hep-th",
    doi = "10.1103/9ywk-xccj",
    journal = "Phys. Rev. D",
    volume = "112",
    number = "4",
    pages = "043526",
    year = "2025"
}

@article{Braglia:2026fle,
    author = "Braglia, Matteo and C{\'e}spedes, Sebasti{\'a}n and Pinol, Lucas",
    title = "{Scale-dependent loop corrections to the inflationary power spectrum}",
    eprint = "2603.12216",
    archivePrefix = "arXiv",
    primaryClass = "astro-ph.CO",
    reportNumber = "CERN-TH-2026-038",
    doi = "10.1007/JHEP07(2026)250",
    journal = "JHEP",
    volume = "07",
    pages = "250",
    year = "2026"
}

@article{Ghosh:2023agt,
    author = "Ghosh, Diptimoy and Panchal, Kushan and Ullah, Farman",
    title = "{Mixed graviton and scalar bispectra in the EFT of inflation: Soft limits and Boostless Bootstrap}",
    eprint = "2303.16929",
    archivePrefix = "arXiv",
    primaryClass = "hep-th",
    doi = "10.1007/JHEP07(2023)233",
    journal = "JHEP",
    volume = "07",
    pages = "233",
    year = "2023"
}

@article{Wang:2021qez,
    author = "Wang, Lian-Tao and Xianyu, Zhong-Zhi and Zhong, Yi-Ming",
    title = "{Precision calculation of inflation correlators at one loop}",
    eprint = "2109.14635",
    archivePrefix = "arXiv",
    primaryClass = "hep-ph",
    doi = "10.1007/JHEP02(2022)085",
    journal = "JHEP",
    volume = "02",
    pages = "085",
    year = "2022"
}

@article{Xianyu:2022jwk,
    author = "Xianyu, Zhong-Zhi and Zhang, Hongyu",
    title = "{Bootstrapping one-loop inflation correlators with the spectral decomposition}",
    eprint = "2211.03810",
    archivePrefix = "arXiv",
    primaryClass = "hep-th",
    doi = "10.1007/JHEP04(2023)103",
    journal = "JHEP",
    volume = "04",
    pages = "103",
    year = "2023"
}

@article{Borinsky:2026roj,
    author = "Borinsky, Michael and Herderschee, Aidan and Lu, Qianshu",
    title = "{On the Numerical Integration of One-Loop Cosmological Collider Signals}",
    eprint = "2609.10673",
    archivePrefix = "arXiv",
    primaryClass = "hep-ph",
    reportNumber = "MIT-CTP/6107",
    month = "9",
    year = "2026"
}

@article{Cohen:2021fzf,
    author = "Cohen, Timothy and Green, Daniel and Premkumar, Akhil and Ridgway, Alexander",
    title = "{Stochastic Inflation at NNLO}",
    eprint = "2106.09728",
    archivePrefix = "arXiv",
    primaryClass = "hep-th",
    doi = "10.1007/JHEP09(2021)159",
    journal = "JHEP",
    volume = "09",
    pages = "159",
    year = "2021"
}

@article{Ota:2022hvh,
    author = {Ota, Atsuhisa and Sasaki, Misao and Wang, Yi},
    title = {{Scale-invariant enhancement of gravitational waves during inflation}},
    journal = {Mod. Phys. Lett. A},
    volume = {38},
    number = {12-13},
    pages = {2350063},
    year = {2023},
    doi = {10.1142/S0217732323500633},
    eprint = {2209.02272},
    archivePrefix = {arXiv},
    primaryClass = {astro-ph.CO}
}

@article{Ota:2022xni,
    author = {Ota, Atsuhisa and Sasaki, Misao and Wang, Yi},
    title = {{One-loop tensor power spectrum from an excited scalar field during inflation}},
    journal = {Phys. Rev. D},
    volume = {108},
    number = {4},
    pages = {043542},
    year = {2023},
    doi = {10.1103/PhysRevD.108.043542},
    eprint = {2211.12766},
    archivePrefix = {arXiv},
    primaryClass = {astro-ph.CO}
}

@article{Knizhnik:1988ak,
    author = {Knizhnik, V. G. and Polyakov, Alexander M. and Zamolodchikov, A. B.},
    title = {{Fractal Structure of 2D Quantum Gravity}},
    journal = {Mod. Phys. Lett. A},
    volume = {3},
    pages = {819--826},
    year = {1988},
    doi = {10.1142/S0217732388000982}
}

@article{David:1988hj,
    author = {David, F.},
    title = {{Conformal Field Theories Coupled to 2D Gravity in the Conformal Gauge}},
    journal = {Mod. Phys. Lett. A},
    volume = {3},
    pages = {1651--1656},
    year = {1988},
    doi = {10.1142/S0217732388001975}
}

@article{Distler:1988jt,
    author = {Distler, Jacques and Kawai, Hikaru},
    title = {{Conformal Field Theory and 2D Quantum Gravity}},
    journal = {Nucl. Phys. B},
    volume = {321},
    pages = {509--527},
    year = {1989},
    doi = {10.1016/0550-3213(89)90354-4}
}

@article{Cotler:2026flq,
    author = {Cotler, Jordan and Ivo, Victor and Maldacena, Juan},
    title = {{Logarithmic gravity from a very slow roll limit of inflation}},
    eprint = {2609.38052},
    archivePrefix = {arXiv},
    primaryClass = {hep-th},
    year = {2026}
}

@article{Seo:2026ajv,
    author = {Seo, Min-Seok},
    title = {{Reheating surface volume renormalization and eternal inflation under broken scale invariance}},
    eprint = {2610.07807},
    archivePrefix = {arXiv},
    primaryClass = {hep-th},
    year = {2026}
}

@article{Iacconi:2023ggt,
    author        = "Iacconi, Laura and Mulryne, David and Seery, David",
    title         = "{Loop corrections in the separate universe picture}",
    eprint        = "2312.12424",
    archivePrefix = "arXiv",
    primaryClass  = "astro-ph.CO",
    doi           = "10.1088/1475-7516/2024/06/062",
    journal       = "JCAP",
    volume        = "06",
    pages         = "062",
    year          = "2024"
}

@article{Iacconi:2026uzo,
    author        = "Iacconi, Laura and Mulryne, David and Seery, David",
    title         = "{Decoupling of large-scale, adiabatic inflationary perturbations from enhanced small-scale modes at one-loop}",
    eprint        = "2601.14229",
    archivePrefix = "arXiv",
    primaryClass  = "astro-ph.CO",
    doi           = "10.1088/1475-7516/2026/08/034",
    journal       = "JCAP",
    volume        = "08",
    pages         = "034",
    year          = "2026"
}

@article{Hinterbichler:2013dpa,
    author        = "Hinterbichler, Kurt and Hui, Lam and Khoury, Justin",
    title         = "{An Infinite Set of Ward Identities for Adiabatic Modes in Cosmology}",
    eprint        = "1304.5527",
    archivePrefix = "arXiv",
    primaryClass  = "hep-th",
    doi           = "10.1088/1475-7516/2014/01/039",
    journal       = "JCAP",
    volume        = "01",
    pages         = "039",
    year          = "2014"
}

@article{Senatore:2012ya,
    author        = "Senatore, Leonardo and Zaldarriaga, Matias",
    title         = "{The constancy of $\zeta$ in single-clock Inflation at all loops}",
    eprint        = "1210.6048",
    archivePrefix = "arXiv",
    primaryClass  = "hep-th",
    doi           = "10.1007/JHEP09(2013)148",
    journal       = "JHEP",
    volume        = "09",
    pages         = "148",
    year          = "2013"
}

@article{Hinterbichler:2012nm,
    author        = "Hinterbichler, Kurt and Hui, Lam and Khoury, Justin",
    title         = "{Conformal Symmetries of Adiabatic Modes in Cosmology}",
    eprint        = "1203.6351",
    archivePrefix = "arXiv",
    primaryClass  = "hep-th",
    doi           = "10.1088/1475-7516/2012/08/017",
    journal       = "JCAP",
    volume        = "08",
    pages         = "017",
    year          = "2012"
}

@article{tHooft:1972tcz,
    author  = "'t Hooft, Gerard and Veltman, M. J. G.",
    title   = "{Regularization and Renormalization of Gauge Fields}",
    doi     = "10.1016/0550-3213(72)90279-9",
    journal = "Nucl. Phys. B",
    volume  = "44",
    pages   = "189--213",
    year    = "1972"
}

@article{Wilson:1973jj,
    author  = "Wilson, Kenneth G. and Kogut, John B.",
    title   = "{The Renormalization Group and the Epsilon Expansion}",
    doi     = "10.1016/0370-1573(74)90023-4",
    journal = "Phys. Rept.",
    volume  = "12",
    pages   = "75--200",
    year    = "1974"
}

@article{Polchinski:1983gv,
    author  = "Polchinski, Joseph",
    title   = "{Renormalization and Effective Lagrangians}",
    doi     = "10.1016/0550-3213(84)90287-6",
    journal = "Nucl. Phys. B",
    volume  = "231",
    pages   = "269--295",
    year    = "1984"
}

@article{Chen:2017ryl,
    author        = {Chen, Xingang and Wang, Yi and Xianyu, Zhong-Zhi},
    title         = {{Schwinger-Keldysh Diagrammatics for Primordial Perturbations}},
    journal       = {JCAP},
    volume        = {12},
    pages         = {006},
    year          = {2017},
    doi           = {10.1088/1475-7516/2017/12/006},
    eprint        = {1703.10166},
    archivePrefix = {arXiv},
    primaryClass  = {hep-th}
}

@article{Joyce:2014aqa,
    author = "Joyce, Austin and Khoury, Justin and Simonovi{\'c}, Marko",
    title = "{Multiple Soft Limits of Cosmological Correlation Functions}",
    eprint = "1409.6318",
    archivePrefix = "arXiv",
    primaryClass = "hep-th",
    doi = "10.1088/1475-7516/2015/01/012",
    journal = "JCAP",
    volume = "01",
    pages = "012",
    year = "2015"
}

@article{Creminelli:2004yq,
    author = "Creminelli, Paolo and Zaldarriaga, Matias",
    title = "{Single field consistency relation for the 3-point function}",
    eprint = "astro-ph/0407059",
    archivePrefix = "arXiv",
    reportNumber = "HUTP-04-A032",
    doi = "10.1088/1475-7516/2004/10/006",
    journal = "JCAP",
    volume = "10",
    pages = "006",
    year = "2004"
}

@article{Senatore:2012wy,
    author = "Senatore, Leonardo and Zaldarriaga, Matias",
    title = "{A Note on the Consistency Condition of Primordial Fluctuations}",
    eprint = "1203.6884",
    archivePrefix = "arXiv",
    primaryClass = "astro-ph.CO",
    doi = "10.1088/1475-7516/2012/08/001",
    journal = "JCAP",
    volume = "08",
    pages = "001",
    year = "2012"
}

@article{Ganc:2010ff,
    author = "Ganc, Jonathan and Komatsu, Eiichiro",
    title = "{A new method for calculating the primordial bispectrum in the squeezed limit}",
    eprint = "1006.5457",
    archivePrefix = "arXiv",
    primaryClass = "astro-ph.CO",
    doi = "10.1088/1475-7516/2010/12/009",
    journal = "JCAP",
    volume = "12",
    pages = "009",
    year = "2010"
}

@article{Assassi:2012et,
    author = "Assassi, Valentin and Baumann, Daniel and Green, Daniel",
    title = "{Symmetries and Loops in Inflation}",
    eprint = "1210.7792",
    archivePrefix = "arXiv",
    primaryClass = "hep-th",
    reportNumber = "SITP-12-36",
    doi = "10.1007/JHEP02(2013)151",
    journal = "JHEP",
    volume = "02",
    pages = "151",
    year = "2013"
}

@article{Goldberger:2013rsa,
    author = "Goldberger, Walter D. and Hui, Lam and Nicolis, Alberto",
    title = "{One-particle-irreducible consistency relations for cosmological perturbations}",
    eprint = "1303.1193",
    archivePrefix = "arXiv",
    primaryClass = "hep-th",
    doi = "10.1103/PhysRevD.87.103520",
    journal = "Phys. Rev. D",
    volume = "87",
    number = "10",
    pages = "103520",
    year = "2013"
}

@article{Senatore:2009cf,
    author = "Senatore, Leonardo and Zaldarriaga, Matias",
    title = "{On Loops in Inflation}",
    eprint = "0912.2734",
    archivePrefix = "arXiv",
    primaryClass = "hep-th",
    doi = "10.1007/JHEP12(2010)008",
    journal = "JHEP",
    volume = "12",
    pages = "008",
    year = "2010"
}

@article{Artigas:2026vfi,
    author = "Artigas, Danilo and Pi, Shi and Tanaka, Takahiro and Urakawa, Yuko",
    title = "{How Dilatation Invariance Suppresses Loop Corrections to Curvature Perturbations on CMB Scales}",
    eprint = "2608.03725",
    archivePrefix = "arXiv",
    primaryClass = "astro-ph.CO",
    reportNumber = "KEK-TH-2859, KEK-Cosmo-0428",
    month = "8",
    year = "2026"
}

@article{Ballesteros:2025nhz,
    author = "Ballesteros, Guillermo and Egea, Jes{\'u}s Gamb{\'\i}n and Riccardi, Flavio",
    title = "{Finite parts of inflationary loops II: A streamlined UV in-in algorithm and distinguishable signatures}",
    eprint = "2512.20467",
    archivePrefix = "arXiv",
    primaryClass = "hep-th",
    reportNumber = "IFT UAM-CSIC 25-168",
    doi = "10.1007/JHEP07(2026)059",
    journal = "JHEP",
    volume = "07",
    pages = "059",
    year = "2026"
}

@article{Bhowmick:2024kld,
    author = "Bhowmick, Supritha and Ghosh, Diptimoy and Ullah, Farman",
    title = "{Bispectrum at 1-loop in the Effective Field Theory of Inflation}",
    eprint = "2405.10374",
    archivePrefix = "arXiv",
    primaryClass = "hep-th",
    doi = "10.1007/JHEP10(2024)057",
    journal = "JHEP",
    volume = "10",
    pages = "057",
    year = "2024"
}

@article{Cotler:2026lna,
    author = "Cotler, Jordan and Ivo, Victor and Maldacena, Juan",
    title = "{Renormalization in Liouville gravity and stochastic inflation}",
    eprint = "2608.26244",
    archivePrefix = "arXiv",
    primaryClass = "hep-th",
    month = "8",
    year = "2026"
}

@article{Fang:2025hid,
    author = "Fang, Cheng-Jun and Hu, Han-Wen and Guo, Zong-Kuan",
    title = "{One-loop corrections to infrared GWs is forbidden by symmetries}",
    eprint = "2509.00420",
    archivePrefix = "arXiv",
    primaryClass = "gr-qc",
    doi = "10.1088/1475-7516/2026/06/023",
    journal = "JCAP",
    volume = "06",
    pages = "023",
    year = "2026"
}

@article{Ema:2025ftj,
    author = "Ema, Yohei and Hong, Muzi and Jinno, Ryusuke and Mukaida, Kyohei",
    title = "{Cancellation of one-loop correction to soft tensor power spectrum}",
    eprint = "2506.15780",
    archivePrefix = "arXiv",
    primaryClass = "astro-ph.CO",
    reportNumber = "CERN-TH-2025-117, RESCEU-14/25, IPMU25-0030, KOBE-COSMO-25-13, KEK-TH-2731",
    doi = "10.1088/1475-7516/2026/01/040",
    journal = "JCAP",
    volume = "01",
    pages = "040",
    year = "2026"
}

@article{Braglia:2025cee,
    author = "Braglia, Matteo and Pinol, Lucas",
    title = "{One-loop renormalization of the effective field theory of inflationary fluctuations from gravitational interactions}",
    eprint = "2504.07926",
    archivePrefix = "arXiv",
    primaryClass = "astro-ph.CO",
    doi = "10.1103/f6q6-5jxb",
    journal = "Phys. Rev. D",
    volume = "113",
    number = "6",
    pages = "063513",
    year = "2026"
}

@article{Inomata:2025bqw,
    author = "Inomata, Keisuke",
    title = "{Conservation of superhorizon curvature perturbations at one loop: Backreaction in the in-in formalism and renormalization}",
    eprint = "2502.08707",
    archivePrefix = "arXiv",
    primaryClass = "astro-ph.CO",
    doi = "10.1103/PhysRevD.111.103504",
    journal = "Phys. Rev. D",
    volume = "111",
    number = "10",
    pages = "103504",
    year = "2025"
}

@article{Urakawa:2010it,
    author = "Urakawa, Yuko and Tanaka, Takahiro",
    title = "{IR divergence does not affect the gauge-invariant curvature perturbation}",
    eprint = "1007.0468",
    archivePrefix = "arXiv",
    primaryClass = "hep-th",
    doi = "10.1103/PhysRevD.82.121301",
    journal = "Phys. Rev. D",
    volume = "82",
    pages = "121301",
    year = "2010"
}

@article{Urakawa:2009my,
    author = "Urakawa, Yuko and Tanaka, Takahiro",
    title = "{Influence on Observation from IR Divergence during Inflation. I.}",
    eprint = "0902.3209",
    archivePrefix = "arXiv",
    primaryClass = "hep-th",
    doi = "10.1143/PTP.122.779",
    journal = "Prog. Theor. Phys.",
    volume = "122",
    pages = "779--803",
    year = "2009"
}

@article{Fumagalli:2024jzz,
    author = "Fumagalli, Jacopo",
    title = "{Absence of one-loop effects on large scales from small scales in non-slow-roll dynamics. Part 2. Quartic interactions and consistency relations}",
    eprint = "2408.08296",
    archivePrefix = "arXiv",
    primaryClass = "astro-ph.CO",
    doi = "10.1007/JHEP01(2025)108",
    journal = "JHEP",
    volume = "01",
    pages = "108",
    year = "2025"
}

@article{Tada:2023rgp,
    author = "Tada, Yuichiro and Terada, Takahiro and Tokuda, Junsei",
    title = "{Cancellation of quantum corrections on the soft curvature perturbations}",
    eprint = "2308.04732",
    archivePrefix = "arXiv",
    primaryClass = "hep-th",
    reportNumber = "CTPU-PTC-23-31",
    doi = "10.1007/JHEP01(2024)105",
    journal = "JHEP",
    volume = "01",
    pages = "105",
    year = "2024"
}

@article{Ballesteros:2024zdp,
    author = "Ballesteros, Guillermo and Egea, Jes\'us Gamb\'\i{}n",
    title = "{One-loop power spectrum in ultra slow-roll inflation and implications for primordial black hole dark matter}",
    eprint = "2404.07196",
    archivePrefix = "arXiv",
    primaryClass = "astro-ph.CO",
    doi = "10.1088/1475-7516/2024/07/052",
    journal = "JCAP",
    volume = "07",
    pages = "052",
    year = "2024"
}

@article{Inomata:2024lud,
    author = "Inomata, Keisuke",
    title = "{Superhorizon Curvature Perturbations Are Protected against One-Loop Corrections}",
    eprint = "2403.04682",
    archivePrefix = "arXiv",
    primaryClass = "astro-ph.CO",
    doi = "10.1103/PhysRevLett.133.141001",
    journal = "Phys. Rev. Lett.",
    volume = "133",
    number = "14",
    pages = "141001",
    year = "2024"
}

@article{Davies:2023hhn,
    author = "Davies, Matthew W. and Iacconi, Laura and Mulryne, David J.",
    title = "{Numerical 1-loop correction from a potential yielding ultra-slow-roll dynamics}",
    eprint = "2312.05694",
    archivePrefix = "arXiv",
    primaryClass = "astro-ph.CO",
    doi = "10.1088/1475-7516/2024/04/050",
    journal = "JCAP",
    volume = "04",
    pages = "050",
    year = "2024"
}

@article{Maity:2023qzw,
    author = "Maity, Suvashis and Ragavendra, H. V. and Sethi, Shiv K. and Sriramkumar, L.",
    title = "{Loop contributions to the scalar power spectrum due to quartic order action in ultra slow roll inflation}",
    eprint = "2307.13636",
    archivePrefix = "arXiv",
    primaryClass = "astro-ph.CO",
    doi = "10.1088/1475-7516/2024/05/046",
    journal = "JCAP",
    volume = "05",
    pages = "046",
    year = "2024"
}

@article{Wands:2000dp,
    author = "Wands, David and Malik, Karim A. and Lyth, David H. and Liddle, Andrew R.",
    title = "{A New approach to the evolution of cosmological perturbations on large scales}",
    eprint = "astro-ph/0003278",
    archivePrefix = "arXiv",
    doi = "10.1103/PhysRevD.62.043527",
    journal = "Phys. Rev. D",
    volume = "62",
    pages = "043527",
    year = "2000"
}

@article{Weinberg:2003sw,
    author = "Weinberg, Steven",
    title = "{Adiabatic modes in cosmology}",
    eprint = "astro-ph/0302326",
    archivePrefix = "arXiv",
    reportNumber = "UTTG-12-02",
    doi = "10.1103/PhysRevD.67.123504",
    journal = "Phys. Rev. D",
    volume = "67",
    pages = "123504",
    year = "2003"
}

@article{Weinberg:2005vy,
    author = "Weinberg, Steven",
    title = "{Quantum contributions to cosmological correlations}",
    eprint = "hep-th/0506236",
    archivePrefix = "arXiv",
    reportNumber = "UTTG-01-05",
    doi = "10.1103/PhysRevD.72.043514",
    journal = "Phys. Rev. D",
    volume = "72",
    pages = "043514",
    year = "2005"
}

@article{Fumagalli:2023loc,
    author = "Fumagalli, Jacopo and Bhattacharya, Sukannya and Peloso, Marco and Renaux-Petel, S\'ebastien and Witkowski, Lukas T.",
    title = "{One-loop infrared rescattering by enhanced scalar fluctuations during inflation}",
    eprint = "2307.08358",
    archivePrefix = "arXiv",
    primaryClass = "astro-ph.CO",
    doi = "10.1088/1475-7516/2024/04/029",
    journal = "JCAP",
    volume = "04",
    pages = "029",
    year = "2024"
}

@article{Kawaguchi:2024rsv,
    author = "Kawaguchi, Ryodai and Tsujikawa, Shinji and Yamada, Yusuke",
    title = "{Proving the absence of large one-loop corrections to the power spectrum of curvature perturbations in transient ultra-slow-roll inflation within the path-integral approach}",
    eprint = "2407.19742",
    archivePrefix = "arXiv",
    primaryClass = "hep-th",
    reportNumber = "WUCG-24-07",
    doi = "10.1007/JHEP12(2024)095",
    journal = "JHEP",
    volume = "12",
    pages = "095",
    year = "2024"
}

@article{Pimentel:2012tw,
    author = "Pimentel, Guilherme L. and Senatore, Leonardo and Zaldarriaga, Matias",
    title = "{On Loops in Inflation III: Time Independence of zeta in Single Clock Inflation}",
    eprint = "1203.6651",
    archivePrefix = "arXiv",
    primaryClass = "hep-th",
    doi = "10.1007/JHEP07(2012)166",
    journal = "JHEP",
    volume = "07",
    pages = "166",
    year = "2012"
}

@article{Motohashi:2023syh,
    author = "Motohashi, Hayato and Tada, Yuichiro",
    title = "{Squeezed bispectrum and one-loop corrections in transient constant-roll inflation}",
    eprint = "2303.16035",
    archivePrefix = "arXiv",
    primaryClass = "astro-ph.CO",
    doi = "10.1088/1475-7516/2023/08/069",
    journal = "JCAP",
    volume = "08",
    pages = "069",
    year = "2023"
}

@article{Firouzjahi:2023ahg,
    author = "Firouzjahi, Hassan and Riotto, Antonio",
    title = "{Primordial Black Holes and loops in single-field inflation}",
    eprint = "2304.07801",
    archivePrefix = "arXiv",
    primaryClass = "astro-ph.CO",
    doi = "10.1088/1475-7516/2024/02/021",
    journal = "JCAP",
    volume = "02",
    pages = "021",
    year = "2024"
}

@article{Kristiano:2023scm,
    author = "Kristiano, Jason and Yokoyama, Jun'ichi",
    title = "{Note on the bispectrum and one-loop corrections in single-field inflation with primordial black hole formation}",
    eprint = "2303.00341",
    archivePrefix = "arXiv",
    primaryClass = "hep-th",
    reportNumber = "RESCEU-3/23",
    doi = "10.1103/PhysRevD.109.103541",
    journal = "Phys. Rev. D",
    volume = "109",
    number = "10",
    pages = "103541",
    year = "2024"
}

@article{Riotto:2023hoz,
    author = "Riotto, Antonio",
    title = "{The Primordial Black Hole Formation from Single-Field Inflation is Not Ruled Out}",
    eprint = "2301.00599",
    archivePrefix = "arXiv",
    primaryClass = "astro-ph.CO",
    month = "1",
    year = "2023"
}

@article{Firouzjahi:2023aum,
    author = "Firouzjahi, Hassan",
    title = "{One-loop corrections in power spectrum in single field inflation}",
    eprint = "2303.12025",
    archivePrefix = "arXiv",
    primaryClass = "astro-ph.CO",
    doi = "10.1088/1475-7516/2023/10/006",
    journal = "JCAP",
    volume = "10",
    pages = "006",
    year = "2023"
}

@article{Kristiano:2022maq,
    author = "Kristiano, Jason and Yokoyama, Jun'ichi",
    title = "{Constraining Primordial Black Hole Formation from Single-Field Inflation}",
    eprint = "2211.03395",
    archivePrefix = "arXiv",
    primaryClass = "hep-th",
    reportNumber = "RESCEU-20/22",
    doi = "10.1103/PhysRevLett.132.221003",
    journal = "Phys. Rev. Lett.",
    volume = "132",
    number = "22",
    pages = "221003",
    year = "2024"
}

@article{Huang:2006eha,
	Archiveprefix = {arXiv},
	Author = {Chen, Xingang and Huang, Min-xin and Shiu, Gary},
	Doi = {10.1103/PhysRevD.74.121301},
	Eprint = {hep-th/0610235},
	Journal = {Phys. Rev.},
	Pages = {121301},
	Slaccitation = {%%CITATION = HEP-TH/0610235;%%},
	Title = {{The inflationary trispectrum for models with large non- Gaussianities}},
	Volume = {D74},
	Year = {2006}}

@article{Sasaki:1998ug,
	Archiveprefix = {arXiv},
	Author = {Sasaki, Misao and Tanaka, Takahiro},
	Doi = {10.1143/PTP.99.763},
	Eprint = {gr-qc/9801017},
	Journal = {Prog. Theor. Phys.},
	Pages = {763-782},
	Slaccitation = {%%CITATION = GR-QC/9801017;%%},
	Title = {{Super-horizon scale dynamics of multi-scalar inflation}},
	Volume = {99},
	Year = {1998}}

@article{Chen:2016uwp,
	Archiveprefix = {arXiv},
	Author = {Chen, Xingang and Wang, Yi and Xianyu, Zhong-Zhi},
	Doi = {10.1103/PhysRevLett.118.261302},
	Eprint = {1610.06597},
	Journal = {Phys. Rev. Lett.},
	Number = {26},
	Pages = {261302},
	Primaryclass = {hep-th},
	Slaccitation = {%%CITATION = ARXIV:1610.06597;%%},
	Title = {{Standard Model Background of the Cosmological Collider}},
	Volume = {118},
	Year = {2017}}

@article{Maldacena:2002vr,
	Archiveprefix = {arXiv},
	Author = {Maldacena, Juan Martin},
	Doi = {10.1088/1126-6708/2003/05/013},
	Eprint = {astro-ph/0210603},
	Journal = {JHEP},
	Pages = {013},
	Primaryclass = {astro-ph},
	Slaccitation = {%%CITATION = ASTRO-PH/0210603;%%},
	Title = {{Non-Gaussian features of primordial fluctuations in single field inflationary models}},
	Volume = {05},
	Year = {2003}}

@article{Arkani-Hamed:2015bza,
	Archiveprefix = {arXiv},
	Author = {Arkani-Hamed, Nima and Maldacena, Juan},
	Eprint = {1503.08043},
	Primaryclass = {hep-th},
	Slaccitation = {%%CITATION = ARXIV:1503.08043;%%},
	Title = {{Cosmological Collider Physics}},
	Year = {2015}}

@article{BM,
	Archiveprefix = {arXiv},
	Author = {Baumann, Daniel and McAllister, Liam},
	Doi = {10.1103/PhysRevD.75.123508},
	Eprint = {hep-th/0610285},
	Journal = {Phys. Rev.},
	Pages = {123508},
	Slaccitation = {%%CITATION = HEP-TH/0610285;%%},
	Title = {{A Microscopic Limit on Gravitational Waves from D-brane Inflation}},
	Volume = {D75},
	Year = {2007}}

@article{Sasaki:1995aw,
	Archiveprefix = {arXiv},
	Author = {Sasaki, Misao and Stewart, Ewan D.},
	Doi = {10.1143/PTP.95.71},
	Eprint = {astro-ph/9507001},
	Journal = {Prog. Theor. Phys.},
	Pages = {71-78},
	Slaccitation = {%%CITATION = ASTRO-PH/9507001;%%},
	Title = {{A General analytic formula for the spectral index of the density perturbations produced during inflation}},
	Volume = {95},
	Year = {1996}}

@article{Assassi:2012zq,
	Archiveprefix = {arXiv},
	Author = {Assassi, Valentin and Baumann, Daniel and Green, Daniel},
	Doi = {10.1088/1475-7516/2012/11/047},
	Eprint = {1204.4207},
	Journal = {JCAP},
	Pages = {047},
	Primaryclass = {hep-th},
	Slaccitation = {%%CITATION = ARXIV:1204.4207;%%},
	Title = {{On Soft Limits of Inflationary Correlation Functions}},
	Volume = {1211},
	Year = {2012}}
\end{document}